%% file: amplitudesfinalv4.tex
\documentclass[11pt, letter]{article}

\usepackage{stolenstyle}

\usepackage{amsmath,epsf,amssymb,latexsym,amsthm,setspace,bbm,array,pifont,color}

\usepackage[matrix,arrow,frame,import,curve,color]{xy}

\input{basedefs2}

\def\mm{{{\mathsf{m}}}}

\def\mms{\mm\slash}

\def\bJ{{{\boldsymbol{J}}}}
\def\bU{{{\boldsymbol{U}}}}
\def\bV{{{\boldsymbol{V}}}}

\def\xis{{{\xi\slash}}}
\def\kiss{{{k\slash}}}

\def\TYM{{{\text{YM}}}}
\def\AYM{{A_{\text{YM}}}}

\def\ed{\mathrm{d}}

\def\DC#1{{C_{#1}}}

\def\preprint{ {{MI-TH-1515}}}

\title{Some tree-level string amplitudes in the NSR formalism}
\author[a] {Katrin Becker,}
\author[a] {Melanie Becker,}
\author[b] {Ilarion V.~Melnikov,}
\author[a] {Daniel Robbins}
\author[a] {\\ and Andrew B.~Royston}

\affiliation[a]{ George P. and Cynthia W. Mitchell Institute for Fundamental Physics and Astronomy,
Texas A\&M University,
College Station, TX 77843, USA}
\affiliation[b]{Department of Mathematics \\ Harvard University, Cambridge, MA 02138, USA}
\emailAdd{kbecker@physics.tamu.edu}
\emailAdd{mbecker@physics.tamu.edu}
\emailAdd{ilarion@math.harvard.edu}
\emailAdd{drobbins@physics.tamu.edu}
\emailAdd{aroyston@physics.tamu.edu}

\abstract{
We calculate tree level scattering amplitudes for open strings using the NSR formalism. We present a streamlined symmetry-based and pedagogical approach to the computations, which we first develop by checking two-, three-, and four-point functions involving bosons and fermions. We calculate the five-point amplitude for massless gluons and find agreement with an earlier result by Brandt, Machado and Medina.  
We then compute the five-point amplitudes involving two and four fermions respectively, the general form of which has not been previously obtained in the NSR formalism. The results nicely confirm expectations from the supersymmetric $F^4$ effective action.  Finally we use the prescription of Kawai, Lewellen and Tye  (KLT) to compute the amplitudes for the closed string sector.}
\begin{document}

\maketitle

\date{today}

\section{Introduction}\label{s:intro}

Superstring theory is free of ultraviolet divergences, which is one important reason this theory is the leading candidate to unify gravity with quantum field theory. The low energy limit of superstring theory contains gauge theory as well as gravitational interactions.  In the mid 1980s the tree-level string corrections to Yang--Mills theory \cite{Gross:1986iv, Tseytlin:1986ti} and the Einstein--Hilbert theory \cite{Gross:1986iv} were computed from tree-level scattering amplitudes. These corrections emerge at order $\alpha^{\prime 2}$ and $\alpha^{\prime 3}$ respectively, both of which originate from a four-point amplitude. About one year later the (partial) five-point amplitude for gluons was computed in \cite{Kitazawa:1987xj} using open string theory and an effective action was proposed based on this result, which included $F^5$ as well as $D^2F^4$ terms.

Renewed interest in calculating higher order effective actions for open string theory emerged around the year 2000. There were basically three different groups working on this problem using three different methods \cite{Refolli:2001df}, \cite{Koerber:2001uu,Koerber:2001hk} and \cite{Medina:2002nk,Brandt:2003wu}. In \cite{Refolli:2001df} four-dimensional ${\cal N}=4$ supersymmetric Yang--Mills theory in superspace was used to calculate correlation functions and the corresponding effective action, including $F^5$ terms.  In \cite{Koerber:2001uu,Koerber:2001hk} deformations of BPS solutions of $D=10$ super Yang--Mills theory were used to calculate the bosonic non-abelian effective action to order $\alpha'^3$.  The third group \cite{Medina:2002nk,Brandt:2003wu} uses the complete five gluon scattering amplitude to calculate the bosonic (non-abelian) effective action for open strings. Their result for the $F^5$ term in the effective action does not agree with the earlier calculations in \cite{Kitazawa:1987xj,Refolli:2001df} but agrees with the work of \cite{Koerber:2001uu,Koerber:2001hk}.  A supersymmetric version of the bosonic action appearing in \cite{Koerber:2001uu,Medina:2002nk} was presented in \cite{Collinucci:2002ac}.  This action was successfully tested in \cite{deRoo:2002ap}.  

$N$-point gluon tree amplitudes for $N \geq 5$ were investigated systematically starting in \cite{Stieberger:2006bh,Stieberger:2006te}, following a {\it tour de force} analysis of the six gluon amplitude in \cite{Oprisa:2005wu}.  Supersymmetric Ward identities were developed in \cite{Stieberger:2007jv,Stieberger:2007am}, in order to infer relations between gluon amplitudes and amplitudes involving gauginos and scalars.  With the exception of \cite{Oprisa:2005wu}, which was carried out in a $10D$ covariant language, these references work in a four-dimensional set-up where one can take advantage of the spinor helicity formalism to write compact expressions for MHV and NMHV amplitudes, generalizing known $4D$ field theory results.  Five-point amplitudes involving matter fermions on D-brane intersections were considered in this context in \cite{Lust:2009pz}.

In recent times diverse tree level $N$-point functions have been calculated using the pure spinor approach of Berkovits \cite{Berkovits:2000fe}. In particular the five-point amplitude was calculated in \cite{Mafra:2009bz}, while tree level $N$-point functions for $N>5$ were presented in \cite{Mafra:2010gj,Mafra:2011nv} (see also the references therein). The results were  expressed in terms of $10D$ superspace variables (or in terms of partial amplitudes of the SYM field theory limit which were identified from their representation in terms of $10D$ superspace variables described in \cite{Mafra:2010jq}).  Hence by expanding in components one can recover amplitudes involving bosons and fermions.  Indeed a large number of such results are available on the website \cite{website}.  (See also \cite{Mafra:2010pn}.)    
%%The correspondence between the NSR formalism and the pure spinor formalism has been verified for correlation functions involving up to four fermions \cite{Berkovits:2000ph}. 
%The pure spinor formalism and the NSR formalism were shown to be equivalent for tree-level correlation functions involving up to four fermions and arbitrary numbers of bosons in~\cite{Berkovits:2000ph}.  However, explicit component expansions of amplitudes involving five or more external particles are quite complicated, regardless of which formalism is used, so it is good to have an independent calculation.
% %Also, it would be interesting to check if the results of both formalisms agree and if the correspondence works for an arbitrary number of fermions.  
% %For this purpose \cite{Kostelecky:1986ab,Hartl:2010ks} might be relevant.  
 A nice review of NSR and pure spinor approaches for open string amplitudes can be found in~\cite{Schlotterer:2012zz}.
 
Calculations of five-point tree-level amplitudes involving fermions have not been previously carried out directly in the 10$D$-covariant NSR formalism.  It has however been shown in \cite{Berkovits:2000ph} that the pure spinor and NSR formalisms give equivalent prescriptions for obtaining tree-level amplitudes with an arbitrary number of bosons and up to four fermions.  Therefore results for tree-level five-point amplitudes obtained via NSR should agree with the corresponding pure spinor results available in the literature cited above.  Nevertheless, we feel that an independent computation within the NSR formalism is valuable.  The amplitudes are quite nontrivial and yet take a remarkably elegant form when expressed in terms of color-ordered field theory sub-amplitudes.  

Our first goal is to see how this structure comes about when carrying out a standard NSR analysis, following the lead of \cite{Medina:2002nk,Barreiro:2005hv} in the five gluon case.  To that end we begin by using the NSR formalism to reproduce results for open-string tree-level $N$-point amplitudes involving gauge bosons and gauginos for $N=2,3,4$. These have previously appeared in the literature and we generally find agreement. However, our expression for the four-point amplitude containing two bosons and two fermions with a particular color ordering differs by a relative sign as compared to the standard references \cite{Green:1981xx,Schwarz:1982jn}. This typo has propagated into some standard literature \cite{Kawai:1985xq,Green:1987mn}, so we give an additional check of the result.  Next we verify the complete five-point amplitude involving gauge bosons that was originally obtained by Brandt, Machado and Medina \cite{Medina:2002nk} and then simplified in \cite{Barreiro:2005hv}. We find agreement with their result, which has also been independently checked in \cite{Oprisa:2005wu}.  

We then explicitly compute, for the first time in the literature, the five-point amplitudes involving two and four fermionic fields respectively in a manifestly $10D$-covariant NSR formalism.  Our calculation agrees with the proposal based on supersymmetry given in \cite{Medina:2006uf}, which was expressed in terms of color-ordered field theory amplitudes.  To show this, we provide some details of the necessary field theory calculation, decomposing the field theory amplitudes into a sum of diagrams, and providing a comparison of the pole structure with the string amplitude result.

Finally, we generalize our calculations to the closed string sector using the KLT formalism~\cite{Kawai:1985xq}.

Our second goal is to illustrate a pedagogical approach to computing amplitudes in the NSR formalism.  Such computations quickly become opaque if carried out by brute force.  Fortunately, however, the structure is greatly clarified by symmetries.  We systematically use the Ward identities of the world-sheet current algebra corresponding to the space-time Lorentz symmetry to organize the computations.  The requisite technology, the principles of which were laid out in classic references, {\it e.g.}~\cite{Ward:1950xp,Friedan:1985ge,Kostelecky:1986xg}, is developed in the next section.

\section{The world-sheet set up} \label{s:worldsheet}
In this section we provide some basic world-sheet conventions as well as tools that we will use for the computation of the amplitudes.  Essentially we work with the conventions of~\cite{Polchinski:1998rr}.  We have the following fields on a Euclidean world-sheet with complex coordinates $(z,\zb)$:
\begin{enumerate}
\item the matter fields are the bosons $X^\mu(z,\zb)$, and the fermions $\psi^\mu(z)$, $\psit^\mu(\zb)$, where $\mu = 0,\ldots, 9$ is a space-time vector index;
\item the ghosts are the usual $c(z)$ and $\ct (\zb)$ (we have no need of the $b$ ghosts), as well as the $\vphi(z),\vphit(\zb)$ bosons from bosonizing the $\beta\gamma$ ghosts: $\beta\gamma \sim \partial\varphi$;
 \item we also have the spin fields $\Theta_a(z)$ and $\Thetat_a(\zb)$ that have weights  $(5/8,0)$ and $(0,5/8)$ respectively and transform as Majorana spinors under $\spin(1,9)$.
\end{enumerate}
The basic fields have the familiar OPEs:
\begin{align}
X^\mu(z,\zb) X^\nu(0) &\sim -\frac{\alpha'}{2} \eta^{\mu\nu} \log |z|^2~,&
\psi^\mu(z) \psi^\nu(0) &\sim \frac{\eta^{\mu\nu}}{z}~,&\nonumber\\
c(z) c(0) &\sim z~,&
e^{s\vphi} (z) e^{t\vphi}(0) & = z^{-st} : e^{s\vphi}(z) e^{t\vphi}(0) :~.
\end{align}
Here $\eta_{\mu\nu}$ is the ten-dimensional Minkowski metric.
Note that $e^{s\vphi}$ then has weight $(-s-s^2/2,0)$ and carries a ghost number $s$.  Throughout the computations we will leave off normal ordering symbols from the fields whenever it is not likely to cause confusion.

\subsection{Open string vertex operators}
We will, for the most part, work in the open string sector, where we insert the vertex operators on the real line in the complex plane, $z=\zb = t$.   It is then convenient to work in units of $\alpha' = 1/2$, as the physical vertex operators for massless fields take a simple form in terms of the (1,0) holomorphic currents
\begin{align}
 \cV_a & \equiv e^{-\vphi/2} \Theta_a~,&
 \cU^\mu &\equiv e^{-\vphi} \psi^\mu~,&
 J^{\alpha\beta} & \equiv \psi^\alpha\psi^\beta~.
\end{align}
The $\cV_a$ are fermionic, while the $\cU^\mu$ and $J^{\alpha\beta}$ are bosonic operators.  With these currents the gaugino vertex operator with polarization $\zeta_a$ and momentum $k$ is given by
\begin{align}\label{eq:gVO}
V_g^{-1/2}[\zeta,k] &= \DC{1} \bV[\zeta] e^{i k \cdot X}~,&
\bV[\zeta] &\equiv \zeta_a \cV_a~,
\end{align}
while the vector boson vertex operator with polarization $\xi$ and momentum $k$ takes one of two forms:
\begin{align}
V_v^{-1}[\xi,k] & = \DC{2} \bU[\xi] e^{i k\cdot X}~,&  \bU[\xi] &\equiv  \xi_\mu \cU^\mu~,&
\end{align}
and
\begin{align}
V_v^{0}[\xi,k] & =  \DC{2} ( i \xi_\mu \dot{X}^\mu + k_\alpha \xi_\beta J^{\alpha\beta}) e^{ik \cdot X}~.
\end{align}
Here $\dot X^\mu = \p_t X^\mu$.  As we will see below, the normalization constants $\DC{i}$ are determined by unitarity of the amplitude to be
\begin{align}
\label{eq:vernorm}
\DC{1}^2 & = -\frac{g^2}{\sqrt{2}}~,&
\DC{2} & = - g~,
\end{align}
where $g$ is the Yang--Mills coupling constant.\footnote{Since the amplitude always involves an even number of fermions we need only determine $\DC{1}^2$.}

These massless vertex operators are physical provided that the polarizations and momenta obey the physical state conditions:
\begin{align}\label{eq:physstate}
k\cdot \xi & = 0~,&
k\slash \zeta = k_\mu \Gamma^\mu \zeta& = 0~,&
k^2 &=0.
\end{align}
We also have gauge transformations on the vector polarizations,
\begin{equation}
\xi_\mu\sim\xi_\mu+\lambda k_\mu,
\end{equation}
for arbitrary $\lambda$.  In particular, gauge invariance of an amplitude implies that if we take any vector polarization to be proportional to the corresponding momentum, then the entire amplitude must vanish.

We should also say a word about our Dirac--Majorana matrices $\Gamma^\mu$.  These are real and obey the usual Clifford algebra $\AC{\Gamma^\mu}{\Gamma^\nu} = 2 \eta^{\mu\nu} \iden$, and we have an anti-symmetric charge conjugation matrix $\cC$ that satisfies $\cC^2 = \iden$ and $\cC \Gamma^\mu \cC = -~^t\Gamma^\mu$;  the Dirac conjugate spinor is given by $\zetab = ~^t\zeta \cC$.  The GSO projection requires us to take $\zeta$ to be a Majorana--Weyl spinor, i.e. $\Gamma \zeta = \zeta$, where $\Gamma$ is the ten-dimensional chirality matrix.  Note that $\AC{\cC}{\Gamma} = 0$ and $~^t\Gamma = \Gamma$.

Finally, we recall that in order to fix the super-diffeomorphisms on the plane we must fix the positions of three vertex operators and arrange the total.  The operator $e^{s \vphi}$ carries picture charge $s$, and we indicated the picture charges of the vertex operators by a superscript.

\subsection{The open-string Koba--Nielsen amplitude}
The correlators for the bosonic fields are easily evaluated from the general formula for products of exponentials and derivatives of $X$.\footnote{See, for instance, chapter 6 of~\cite{Polchinski:1998rq}.}  We will quote the results we will need for our computation and in the process introduce some convenient notation.

Let us introduce a useful short-hand:  we will often abbreviate the operator and its insertion by the same label; e.g. $e^{i k_1 \cdot X}(z_1,\zb_1)$ will simply be $e^{i k_1\cdot X}$.  This will reduce the clutter and with any luck will not increase confusion.   Let
\begin{align}
\label{eq:bosoinvariants}
q_{ij} & \equiv k_i\cdot k_j~,&
\Delta^i_j & \equiv \xi^i \cdot k_j~,&
\omega^{ij} & \equiv \xi^i \cdot \xi^j~.
\end{align}
 We then have the following familiar results for the open bosonic amplitudes.  First, we have the ``$s$-point tachyon amplitude''
\begin{align}
\cT_s \equiv \la e^{i k_1 \cdot X} \cdots e^{i k_s \cdot X} \ra =  \DC{3} (2\pi)^{10} \delta^{10}(\textstyle\sum_i k_i) \prod_{i<j} |z_{ij}|^{q_{ij}}~,
\end{align}
where unitarity fixes
\begin{align}
\label{eq:tachnorm}
\DC{3} = \frac{2i}{g^2}~.
\end{align}
In what follows we will leave off the factor of $(2\pi)^{10} \delta^{10}(\textstyle\sum_i k_i)$ when writing the amplitude.  Next, we have
\begin{align}
\cT_{s}^{(2)} \equiv \la e^{i k_1 \cdot X} [ i \xi^2 \cdot \dot X e^{i k_2 \cdot X}]_2 \cdots e^{i k_s \cdot X}\ra
= -\cT_s \sum_{l \neq 2} \frac{\Delta^2_l}{z_{l2} }~.
\end{align}
We will also have use for
\begin{align}
\cT_{s}^{(2,3)} \equiv \la e^{i k_1 \cdot X} [ i \xi^2 \cdot \dot X e^{i k_2 \cdot X}]_2  [ i \xi^3 \cdot \dot X e^{i k_3 \cdot X}]_3\cdots e^{i k_s \cdot X}\ra
= \cT_s \left[\frac{\omega^{23}}{z_{23}^2}+ \sum_{l \neq 2,m\neq 3} \frac{\Delta^2_l\Delta^3_m}{z_{l2} z_{m3} }\right]~,
\end{align}
and
\begin{align}
\cT_s^{(2,3,4)}&\equiv \la e^{i k_1 \cdot X} [ i \xi^2 \cdot \dot X e^{i k_2 \cdot X}]_2  [ i \xi^3 \cdot \dot X e^{i k_3 \cdot X}]_3[ i \xi^4 \cdot \dot X e^{i k_4 \cdot X}]_4\cdots e^{i k_s \cdot X}\ra \nonumber\\
& = \left[
 \frac{\omega^{23}}{z_{23}^2} \sum_{l\neq 4} \frac{\Delta^4_l}{z_{4l}}
+ \frac{\omega^{24}}{z_{24}^2} \sum_{l\neq 3} \frac{\Delta^3_l}{z_{1l}}
+ \frac{\omega^{34}}{z_{34}^2} \sum_{l\neq 2} \frac{\Delta^2_l}{z_{2l}}
~+ \!\!\!\sum_{l\neq 2,m\neq 3,n\neq 4} \frac{\Delta^2_l \Delta^3_m \Delta^4_n}{z_{2l}z_{3m}z_{4n}}
 \right] \cT_s~.
\end{align}
It is useful to keep in mind that the physical state conditions and momentum conservation imply that for every $i$,
\begin{align}\label{eq:Deltanull}
\Delta^i_i &= 0~,&\sum_j \Delta^i_j &= 0~.
\end{align}

\subsection{A little current algebra}
The world-sheet correlation functions that we will need can be computed in terms of the free-field presentation.  However, such computations quickly become unwieldy, and to tame them it is useful to introduce the structure of a Kac--Moody (KM) algebra.

Recall that a holomorphic KM algebra is generated by $(1,0)$ conserved currents $J^a(z)$, where $a$ runs over the adjoint of a Lie algebra $\Lg$.\footnote{A pedagogical presentation may be found in Chapter 11 of~\cite{Polchinski:1998rr}, as well as in~\cite{Ginsparg:1988ui}.}  The currents have the OPE
\begin{align}
J^a(z) J^b(0) \sim \frac{k^{ab}}{z^2} +  \frac{ f^{ab}_{~~c}}{z} J^c(0)~,
\end{align}
where $k^{ab}$ is a metric on $\Lg$ and $f^{ab}_{~~c}$ are the structure constants.  The remaining fields of the CFT assemble into KM-primary fields $\Phi$ in representations of $\Lg$, as well as their descendants.  The primary fields in representation $\rep{r}$ are characterized by the OPE
\begin{align}
J^a(z) \Phi(0) \sim \frac{ t^a_{\rep{r}} \cdot \Phi}{z}~,
\end{align}
where $t^a_{\rep{r}}$ denote the (anti-Hermitian) generators of $\Lg$ in representation $\rep{r}$.

The correlation functions of KM currents and KM primaries satisfy well-known Ward identities, and we will use these to reduce some complicated correlators to simpler ones.  For instance, we have
\begin{align}
\label{eq:1current}
\la J^a(z) \Phi_1(w_1) \Phi_2(w_2) \cdots \Phi_n(w_n) \ra =
\sum_{i=1}^n \frac{1}{z-w_i} \la \Phi_1(w_1)\cdots t^a_{\rep{r_i}} \cdot \Phi_i(w_i) \cdots \Phi_n(w_n)\ra~.
\end{align}

The situation is slightly modified in the presence of a quasi-primary KM field like a current, but the Ward identity is still determined by the OPE.  For example, we find
\begin{align}
\label{eq:2current}
\la J^a(z) J^b(u) \Phi_1 \Phi_2 \cdots \Phi_n \ra
&= \frac{f^{ab}_{~~c}}{z-u} \la J^c(u) \Phi_1 \Phi_2\cdots \Phi_n\ra
+\frac{k^{ab}}{(z-u)^2} \la  \Phi_1 \Phi_2 \cdots \Phi_n \ra \nonumber\\
&\quad+\sum_{i=1}^n \frac{1}{z-w_i} \la J^b(u) \Phi_1\cdots t^a_{\rep{r_i}} \cdot \Phi_i \cdots \Phi_n\ra~.
\end{align}
In our case, the relevant currents are the Lorentz currents $J^{\alpha\beta} = \psi^\alpha\psi^\beta$ with OPE
\begin{align*}
J^{\gamma\delta}(z) J^{\alpha\beta}(0)  \sim \frac{\eta^{\alpha\delta} \eta^{\beta\gamma} -\eta^{\alpha\gamma}\eta^{\beta\delta}}{z^2} +
\frac{1}{z} \left[ \eta^{\alpha\delta} J^{\gamma\beta}(0) -\eta^{\beta\delta} J^{\gamma\alpha}(0)
-\eta^{\alpha\gamma} J^{\delta\beta}(0) +\eta^{\gamma\beta} J^{\delta\alpha}(0) \right]~.
\end{align*}
The $J^{\gamma\delta}$ generate the action of the global Lorentz symmetry $\so(1,9)$ on the fermions.  To see this, let $\omega_{\alpha\beta}$ be any anti-symmetric matrix.  Then
\begin{align}
-\frac{1}{2} \omega_{\alpha\beta} J^{\alpha\beta}(z) \psi^\mu(0) \sim
\frac{1}{z}  \omega^\mu_{~\nu} \psi^\nu(0)~.
\end{align}
The Lorentz symmetry then fixes the OPE with the spin-field operators $\cV_a$ as well:
\begin{align}
%-\frac{1}{2}\omega_{\alpha\beta} J^{\alpha\beta}(z)  \cV_a(0)
%\sim
%\frac{1}{z} \left[ -\frac{i}{2} \omega_{\alpha\beta} \cV_b(0) \Sigma^{\alpha\beta}_{ba} \right]~,
-\frac{1}{2}\omega_{\alpha\beta} J^{\alpha\beta}(z)  \cV_a(0)
\sim
\frac{1}{z} \left[ -\frac{1}{2} \omega_{\alpha\beta} \cV_b(0) \Sigma^{\alpha\beta}_{ba} \right]~,
\end{align}
where
\begin{align}\label{eq:spingens}
\Sigma^{\alpha\beta} \equiv \frac{1}{4} \CO{\Gamma^\alpha}{\Gamma^\beta}
\end{align}
is the Lorentz generator in the Majorana representation.\footnote{The Lorentz transformation of a spinor, such as the polarization $\zeta$, is given by $\delta_{\omega}\zeta = \ff{1}{2} \omega_{\alpha\beta} \Sigma^{\alpha\beta} \zeta$.  The spin fields then naturally transform as $\zetab = ~^t \zeta \cC$ since they are contracted with $\zeta$ to form a Lorentz scalar.}  This presentation still suffers from an over-abundance of indices.  To cure this, we observe that for our applications the $J^{\alpha\beta}$ are always contracted into
\begin{align}
\label{eq:mmdef}
\mm^i_{\alpha\beta} \equiv \ff{1}{2} \left( k^i_\alpha \xi^i_\beta - k^i_\beta\xi^i_\alpha\right)~,
\end{align}
and so we define
\begin{align}
\label{eq:wiseJ}
\bJ[\mm](z) \equiv \mm_{\alpha\beta} J^{\alpha\beta}(z)~,
\end{align}
which have the OPEs
\begin{align}
\label{eq:wiseJOPE}
\bJ [\mm^i] (z_i)\bJ[\mm^j](z_j)
%& \sim \mm^i_{\alpha_i \mu_i} \mm^j_{\alpha_j \mu_j}
%\left[
%\frac{2}{z_{ij}^2} \eta^{\mu_i\alpha_j} \eta^{\mu_j\alpha_i} +\frac{2}{z_{ij}} \left( \eta^{\mu_i \alpha_j} J^{\alpha_i \mu_j}_j -\eta^{\mu_i\mu_j} J^{\alpha_i\alpha_j}_j\right)
%\right] \nonumber\\
&\sim \frac{2\tr(\mm^i \mm^{j})}{z_{ij}^2} +\frac{4}{z_{ij}} \bJ[\mm^{ij} ] (z_j) ~,
\end{align}
where we use $\eta$ to contract indices and anti-symmetrize:
\begin{align}
\tr(\mm^i \mm^j) &\equiv \eta^{\alpha\beta} \mm^i_{\alpha\gamma} \eta^{\gamma\delta} \mm^j_{\delta\beta}~, &
(\mm^{ij})_{\alpha \beta} &\equiv \frac{1}{2} \left[\mm^i_{\alpha\gamma}\eta^{\gamma\delta} \mm^j_{\delta \beta} - (\alpha\leftrightarrow \beta)\right]~.
\end{align}
In applying the Ward identities we use the OPEs
\begin{align}\label{eq:wiseJUVOPE}
\bJ[\mm^i] (z_i) \bU[\xi^k] (z_k) & \sim
\frac{2}{z_{ik}} \bU[\mm^i \xi^k] (z_k)~,&
\bJ[\mm^i] (z_i) \bV[\zeta_k] (z_k) & \sim
\frac{1}{2z_{ik}} \bV[\mms\zeta](z_k)~,
\end{align}
where
\begin{align}
\mms^i = m^i_{\alpha\beta} \Gamma^{\alpha\beta} = \kiss^i\xis^i = - \xis^i \kiss^i~,
\end{align}
and as usual $\Gamma^{\alpha\beta} = \ff{1}{2} (\Gamma^\alpha\Gamma^\beta-\Gamma^\beta\Gamma^\alpha)$.  Note that the $\bJ \bV$ OPE determines
\begin{align}
\label{eq:psiThetacheat}
\psi^{\mu}(z_1) \Theta_a(z_2) \sim \pm \frac{1}{\sqrt{2 z_{12}}} \Theta_b(z_2) \Gamma^{\mu}_{ba}~.
\end{align}

\subsection{Two-point and three-point correlation functions}
Consider the sector of the world-sheet CFT generated by the $\psi^\mu$ and the $\beta\gamma$ ghost $\vphi$.  Almost all correlation functions in this sector that are relevant to our computations can be reduced by the Lorentz KM Ward identities to two- and three-point functions.  (The one exception is the $4$-$\cV$ correlator given in formula \eqref{eq:VVVVcor} below which does not involve any currents.)  We will present these in this section.

As a first step, we compute the two-point functions of $\cU$ and $\cV$ currents.  First, up to overall normalization the two-point function of the $\cU^\mu$ is determined by conformal invariance and representation theory:
\begin{align}
\label{eq:2points}
%\la \cV_{a}(z_1) \cV_{b}(z_2) \ra &= 0~, &
\la \cU^{\mu} (z_1) \cU^{\nu} (z_2) \ra & = -\frac{\eta^{\mu\nu}}{z_{12}^2}~,
\end{align}
where $z_{ij} \equiv z_i - z_j$.  Conformal invariance fixes the position dependence, and Lorentz invariance determines the form of the coefficients:  $\Sym^2 (\rep{10})$ contains a unique trivial representation corresponding to $\eta$.  The remaining overall constant can be absorbed into normalization constants of the vertex operators, which are in turn fixed by unitarity of the S-matrix.  Note that the two-point function of the $\cV_a$ is necessarily zero on the disk or sphere, since it has a total ghost number $-1 \neq -2$.

Next, we consider the three-point function
\begin{align}\label{eq:VUV3pt}
\la \cV_{a}(z_1) \cU^\mu(z_2) \cV_{b}(z_3) \ra = \pm \frac{ (\cC\Gamma^\mu)_{ab}}{\sqrt{2}z_{12} z_{13} z_{23}}~.
\end{align}
The position-dependence is once again fixed by conformal invariance, while the coefficient is constrained by the OPE in~(\ref{eq:psiThetacheat}); we absorb the sign ambiguity into a sign of the $\Gamma^\mu$ and obtain
\begin{align}\label{eq:UUcor}
\la \bU[\xi^i] \bU[\xi^j] \ra &= -\frac{ \omega^{ij}}{z_{ij}^2}~, &
\la \bV[\zeta^1] \bU[\xi^2] \bV[\zeta^3]\ra & = \frac{\zetab^3 \xis^2 \zeta^1}{\sqrt{2} z_{12} z_{13} z_{23}}~.
\end{align}
Finally, we consider the $\cV_a(z_1) \cV_b(z_2)$ OPE.  By splitting the operators into the ghost $e^{-\vphi/2}$ and spin field contributions, we arrive at
%Conformal invariance and ghost number conservation fix it to be
%\begin{align}
%\cV_a(z_1) \cV_b(z_2) \sim \frac{K_{ba}e^{-\vphi}(z_2) }{z_{12}^{3/2}} + \frac{1}{z_{12}} J_{ba}(z_2)~,
%\end{align}
%where $K_{ba} = -K_{ab}$ and $J_{ba} = J_{ab}$ are bosonic, respectively of weight $(0,0)$ and $(1,0)$, and $J_{ab}$ carries ghost number $-1$.  Space-time Lorentz invariance and~(\ref{eq:2points}, \ref{eq:UUcor}) then lead to
\begin{align}
\label{eq:VVOPE}
\cV_a(z_1) \cV_b(z_2) \sim \frac{\cC_{ba} e^{-\vphi}(z_2)}{z_{12}^{3/2}} + \frac{(\cC\Gamma_\mu)_{ba}}{\sqrt{2} z_{12}} \cU^\mu(z_2)~.
\end{align}
The second term is consistent with the three-point function~(\ref{eq:VUV3pt}) and the two-point function~(\ref{eq:2points}), as it should be. Note that since $\AC{\Gamma}{\cC} = 0$, the first term in the OPE will vanish when contracted with a pair of Majorana--Weyl spinors of the same chirality.  This will play an important role in computations involving 4 gaugino vertex operators.

\subsection{Three-point super-Yang--Mills amplitudes}
Armed with this knowledge, we embark on our first set of open string amplitudes:  the three-point functions.  We begin with the color-ordered three-vector amplitude, which we write as
\begin{align}
A(1,2,3) = \la c_1 V_v^{-1} [\xi^1,k_1] c_2 V_v^0[\xi^2,k_2] c_3 V_v^{-1}[\xi^3,k_3] \ra.
\end{align}
Factorizing the amplitude into $c$, $X$, and current algebra sectors, we obtain
\begin{align}
A(1,2,3)= \DC{2}^3  \la c_1 c_2 c_3 \ra \left[
\la \bU[\xi^1]\bU[\xi^3] \ra\cT_3^{(2)}+
\la \bU[\xi^1]\bJ[\mm_2] \bU[\xi^3]\ra \cT_3
\right]~.
\end{align}
The Ward identity yields the only so-far undetermined correlator:
\begin{align}\label{eq:JUUcor}
\la \bU[\xi^1]\bJ[\mm_2] \bU[\xi^3]\ra =
\frac{2}{z_{21}} \la\bU_1[\mm^2 \xi^1] \bU[\xi^3]\ra
+
\frac{2}{z_{23}} \la \bU_3[\mm^2 \xi^3] \bU[\xi^1]\ra
= \frac{2 ~^t \xi^3 \mm^2 \xi^1}{z_{12} z_{13} z_{23}}~.
\end{align}
Here the short-hand notation is
\begin{align}
~^t \xi^3 \mm_2 \xi^1 = \xi^3_{\mu} \eta^{\mu\alpha} \mm^{2}_{\alpha\beta} \eta^{\beta\nu} \xi^1_{\nu}~.
\end{align}
Using $\la c_1 c_2 c_3 \ra = z_{12} z_{13} z_{23}$~, we obtain
\begin{align}
A(1,2,3) &= z_{12} z_{13} z_{23}\times\left[ -\frac{\omega^{13}}{z_{13}^2} \left(-\frac{\Delta^2_1}{z_{12}} -\frac{\Delta^2_3}{z_{32}}\right) + \frac{ 2 ~^t\xi^3 \mm^2 \xi^1}{z_{12} z_{13} z_{23}} \right]\times \cT_3.
\end{align}
Since there are no non-vanishing Lorentz-invariants for massless $3$ particle kinematics, we have $\cT_3 = \DC{3}$; finally, using $\Delta^2_3 + \Delta^2_1 = 0$, and setting $k_{ij} \equiv k_i -k_j$, we find
\begin{align}
A(1,2,3) & = \DC{2}^3 \DC{3} \left[ \omega^{13} \Delta^2_1 + 2 ~^t\xi^3 \mm^2 \xi^1 \right] \nonumber \\&=
-\frac{\DC{2}^3 \DC{3}}{2} \left[ \xi^1 \cdot k_{23} \xi^2\cdot \xi^3 + \xi^2 \cdot k_{31} \xi^3\cdot \xi^1 + \xi^3 \cdot k_{12} \xi^1 \cdot \xi^2 \right]~.
\end{align}
This is indeed the familiar color-ordered $3$-point Yang--Mills amplitude with Yang--Mills coupling $g$ provided we set
\begin{align}
\DC{2}^3 \DC{3} = -2ig~
\end{align}
and obtain
\begin{align}
\label{eq:3vAmplitude}
A(1,2,3) = A_{\TYM}(1,2,3) = +ig \left[ \xi^1 \cdot k_{23} \xi^2\cdot \xi^3 + \xi^2 \cdot k_{31} \xi^3\cdot \xi^1 + \xi^3 \cdot k_{12} \xi^1 \cdot \xi^2 \right]~.
\end{align}
This agrees with \cite{Green:1987mn} equation 7.4.40 once momentum conservation and physical state conditions are taken into account.
All of this notation is a very heavy-handed way to derive this familiar result, but it will be of great help in organizing the computation of higher point amplitudes.

Next, we compute the amplitude with two gauginos and one vector boson.  We will distinguish the gauginos by a $\tilde{~}$ over the corresponding index in the ordered amplitude, so that we are interested in $A(\tilde{1},2,\tilde{3})$.  All the work is already done:
\begin{align}
A(\tilde{1},2,\tilde{3}) & = \la c_1 V_g^{-1/2}[\zeta^1,k_1] c_2 V_v^{-1}[\xi^2,k_2] c_3 V_g^{-1/2}[\zeta^3,k_3]\ra \nonumber \\
& = \DC{2}\DC{1}^2\la c_1 c_2 c_3\ra \la \bV[\zeta^1] \bU[\xi^2]\bV[\zeta^3] \ra \cT_3
=\frac{\DC{2}\DC{1}^2\DC{3}}{\sqrt{2}} \zetab^1 \xis^2 \zeta^3~.
\end{align}
We set
\begin{align}
\DC{2}\DC{1}^2 \DC{3} = i\sqrt{2} g~,
\end{align}
so that the computation yields the standard $3$-point super Yang--Mills (sYM) amplitude:
\begin{align}
\label{eq:v2fAmplitude}
A(\tilde{1},2,\tilde{3}) & = A_{\TYM} (\tilde{1},2,\tilde{3}) = +ig \zetab^1 \xis^2 \zeta^3~,
\end{align}
see e.g.\ \cite{Green:1987mn} formula 7.4.31.
%

%%%%%%%%%%%%%%%%%%%%%%%%
%%%%%%%%%%%%%%%%%%%%%%%%
\section{Four-point open superstring amplitudes} \label{s:4point}
%%%%%%%%%%%%%%%%%%%%%%%%
%%%%%%%%%%%%%%%%%%%%%%%%

In this section we recall the tree-level four-point amplitudes.  The purpose is two-fold.  First, we will gain familiarity with the condensed notation and techniques.  Second, the explicit results will be useful in the following.

Before we jump into the details, we mention the massless kinematics in terms of the Mandelstam variables
\begin{align}
\label{eq:kinematics4}
s & = -2q_{12} = -2q_{34}~,&
t  & = -2q_{23} = -2q_{14}~,&
u & = -2q_{13} = -2q_{24}~,
\end{align}
which satisfy $s+t+u = 0$.  We will often find it convenient to work with $q_{12}$ and $q_{23}$ as independent kinematic variables.

%%%%%%%%%%%%%%%%%%%%%%%%
\subsection{Four vectors}
%%%%%%%%%%%%%%%%%%%%%%%%

We write the color-ordered four-vector amplitude as
\begin{equation}\label{A4v1}
A(1,2,3,4) = \int_{z_1}^{z_3} \ed z_2 \langle c_1 V_{v}^{-1}[\xi^1,k_1] V_{v}^0[\xi^2,k_2] c_3 V_{v}^0[\xi^3,k_3] c_4 V_{v}^{-1}[\xi^4, k_4] \rangle~.
\end{equation}
Some comments are due regarding the distribution of ghost insertions and picture choices.  Three ghost insertions are required on the half-plane to soak up the residual $\PSL(2,\mathbb{R})$ of ${\rm Diff}\times{\rm Weyl}$.  In \eqref{A4v1} we have chosen to fix positions $z_{1,3,4}$.  In general, for a tree-level $n$-point correlator we will follow the common convention of fixing $z_1,z_{n-1},z_n$.  The remaining positions will be integrated over, subject to the condition that the specified ordering is preserved.  In \eqref{A4v1} for example, we integrate $z_2$ over the range $z_1 \le z_2 \le  z_3$.  The picture charges can be distributed arbitrarily, subject to the constraint that they sum to $-2$~\cite{Polchinski:1998rr}.  The amplitude is independent of this choice.  Comparing different choices can lead to interesting integral identities, especially for higher point functions, which can in turn be helpful in simplifying expressions.  See e.g.\ \cite{Oprisa:2005wu}.  We will make choices that streamline the computation as much as possible.

Writing the correlator as a sum of terms, factorized into ghost, bosonic, and current algebra sectors, we have
\begin{align}\label{A4v2}
A(1,2,3,4) =&~ \DC{2}^4 \int_{z_1}^{z_3} \ed z_2 \la c_1 c_3 c_4 \ra \bigg[ \la \bU[\xi^1] \bU[\xi^4] \ra \cT_{4}^{(2,3)} + \la \bU[\xi^1] \bJ[\mm^2] \bU[\xi^4] \ra \cT_{4}^{(3)} + \cr
&~ \qquad \qquad + \la \bU[\xi^1] \bJ[\mm^3] \bU[\xi^4] \ra \cT_{4}^{(2)}  + \la \bU[\xi^1] \bJ[\mm^2] \bJ[\mm^3] \bU[\xi^4] \ra \cT_4 \bigg]~.
\end{align}
Making use of the Ward identity \eqref{eq:2current} together with the OPE's \eqref{eq:wiseJOPE}, \eqref{eq:wiseJUVOPE}, the new correlator appearing in the last term of \eqref{A4v2} is
\begin{align}
\la \bU[\xi^1] \bJ[\mm^2] \bJ[\mm^3] \bU[\xi^4] \ra =&~ \frac{2 \tr(\mm^2 \mm^3)}{z_{23}^2} \la \bU[\xi^1] \bU[\xi^4] \ra + \frac{4}{z_{23}} \la \bU[\xi^1] \bJ_3[\mm^{23}]  \bU[\xi^4] \ra + \cr
& + \frac{2}{z_{21}} \la \bU_1[\mm^2 \xi^1] \bJ[\mm^3] \bU[\xi^4] \ra + \frac{2}{z_{24}} \la \bU[\xi^1] \bJ[\mm^3] \bU_4[\mm^2 \xi^4] \ra~. \qquad
\end{align}
Then with \eqref{eq:UUcor} and \eqref{eq:JUUcor} this is simplified to
\begin{align}
\la \bU[\xi^1] \bJ[\mm^2] \bJ[\mm^3] \bU[\xi^4] \ra =&~ - \frac{2 \tr(\mm^2 \mm^3) \omega^{14}}{z_{23}^2 z_{14}^2} + \frac{8 ~^t\xi^4 \mm^{23} \xi^1}{z_{23} z_{13} z_{14} z_{34}} - \frac{4 ~^t\xi^4 \mm^3 \mm^2 \xi^1}{z_{12} z_{13} z_{14} z_{34}} + \cr
&~ - \frac{4 ~^t\xi^4 \mm^2 \mm^3 \xi^1}{z_{24} z_{13} z_{14} z_{34}} \cr
=&~  - \frac{2 \tr(\mm^2\mm^3) \omega^{14}}{z_{23}^2 z_{14}^2} + ~^t\xi^4 \left[ \frac{4 \mm^2 \mm^3}{z_{13} z_{14} z_{23} z_{24}} - \frac{4 \mm^3 \mm^2}{z_{12} z_{14} z_{23} z_{34}} \right] \xi^1 . \qquad
\end{align}
In the last line we wrote the result in a way that makes the expected symmetry under the separate exchanges $1\leftrightarrow 4$ and $2\leftrightarrow 3$ apparent.
Collecting the remaining pieces, the full amplitude is
\begin{align}\label{A4v3}
A(1,2,3,4) =&~ \DC{2}^4 \int_{z_1}^{z_3} \ed z_2 \bigg\{ -\frac{\omega^{14}}{z_{41}^2} \bigg[ \frac{\omega^{23}}{z_{32}^2} + \left(- \frac{\Delta_{1}^2}{z_{21}} + \frac{\Delta_{3}^2}{z_{32}} + \frac{\Delta_{4}^2}{z_{42}}\right) \left( -\frac{\Delta_{1}^3}{z_{31}} - \frac{\Delta_{2}^3}{z_{32}} + \frac{\Delta_{4}^3}{z_{43}} \right) \bigg] +  \cr
& + \frac{2 ~^t\xi^4 \mm^2 \xi^1}{z_{21} z_{41} z_{42}} \left( -\frac{\Delta_{1}^3}{z_{31}} - \frac{\Delta_{2}^3}{z_{32}} + \frac{\Delta_{4}^3}{z_{43}} \right) + \frac{2 ~^t\xi^4 \mm^3 \xi^1}{z_{31} z_{41} z_{43}} \left(- \frac{\Delta_{1}^2}{z_{21}} + \frac{\Delta_{3}^2}{z_{32}} + \frac{\Delta_{4}^2}{z_{42}}\right) + \cr
& - \frac{2 \tr(\mm^2\mm^3) \omega^{14}}{z_{32}^2 z_{41}^2} + \frac{4 ~^t\xi^4 \mm^2 \mm^3 \xi^1}{z_{31} z_{41} z_{32} z_{42}} - \frac{4 ~^t\xi^4 \mm^3 \mm^2 \xi^1}{z_{21} z_{41} z_{32} z_{43}} \bigg\} \times (-z_{31} z_{41} z_{43}) \times \cT_4 ~. \qquad \raisetag{22pt}
\end{align}

Worldsheet diffeomorphism invariance guarantees that the amplitude is independent of our choice for $z_{1,3,4}$.  At this point there is nothing to gain by leaving them arbitrary, so we make the conventional choice that brings the integral into a well-known form:
\begin{equation}\label{4ptzs}
z_1= 0~, \qquad z_3 = 1~, \qquad  z_4 \to \infty~.
\end{equation}
We also set $z_2 \equiv x \in (0,1)$.  Using the kinematic relations in~(\ref{eq:kinematics4}), we find that the four-tachyon correlator is finite as $z_4\to \infty$ and given by
\begin{equation}\label{fixed4tach}
\cT_4 = \DC{3} x^{q_{12}} (1-x)^{q_{23}} \left( 1 + O(1/z_{4}) \right)~.
\end{equation}
The integrand involves four types of denominator structure:
\begin{align}\label{4vnumden}
A(1,2,3,4) =&~ \DC{2}^4 \DC{3} \int_{0}^1 \ed x \, t^{q_{12}} (1-x)^{q_{23}} \left\{ \frac{N_1}{(1-x)^2} + \frac{N_2}{x(1-x)} + \frac{N_3}{1-x} + \frac{N_4}{x} \right\} \cr
\equiv &~ \DC{2}^4 \DC{3} \sum_{i=1}^4 N_i I_i~,
\end{align}
where the numerators are
\begin{align}
N_1 =&~ \omega^{14} \omega^{23} (1-q_{23})~, \cr
N_2 =&~ -\omega^{12} \omega^{34} q_{23} - \omega^{12} \Delta_{2}^3 \Delta_{1}^4 - \omega^{23} \Delta_{2}^1 \Delta_{3}^4 + \omega_{34} \Delta_{2}^1 \Delta_{3}^2 + \omega^{41} \Delta_{1}^2 \Delta_{2}^3 - \omega^{24} \Delta_{2}^1 \Delta_{2}^3~, \cr
N_3 =&~ \omega^{13} \omega^{24} q_{23} + \omega^{23} \Delta_{3}^1 \Delta_{2}^4 + \omega^{34} \Delta_{3}^1 \Delta_{3}^2 - \omega^{41} \Delta_{3}^2\Delta_{1}^3 + \omega^{13} \Delta_{3}^2 \Delta_{1}^4 - \omega^{24} \Delta_{3}^1 \Delta_{2}^3 ~, \cr
N_4 =&~ \omega^{12} \Delta_{1}^3 \Delta_{2}^4 - \omega^{34} \Delta_{3}^1 \Delta_{1}^2 + \omega^{41} \Delta_{1}^2 \Delta_{1}^3 + \omega^{13} \Delta_{1}^2 \Delta_{3}^4 - \omega^{24} \Delta_{2}^1 \Delta_{1}^3~.
\end{align}
In order to obtain these results we replaced the pieces of \eqref{A4v3} involving $\mm$'s with equivalent expressions in terms of the kinematic invariants \eqref{eq:bosoinvariants}; e.g.
\begin{align}
2 ~^t\xi^4 \mm^2 \xi^1 =&~ \omega^{12} \Delta_{2}^4 - \omega^{24} \Delta_{2}^1~, \cr
2 \tr(\mm^2\mm^3) =&~ \Delta_{2}^3 \Delta_{3}^2 - \omega^{23} q_{23}~, \cr
4 ~^t\xi^4 \mm^2 \mm^3 \xi^1 =&~ \omega^{13} \Delta_{3}^2 \Delta_{2}^4 - \omega^{23} \Delta_{3}^1 \Delta_{2}^4 - \omega^{13} \omega^{24} q_{23} + \omega^{24} \Delta_{3}^1 \Delta_{2}^3 ~.
\end{align}

The integral $I_2$ is the standard Euler beta function
\begin{align}
I_2 = B(q_{12},q_{23}) = \frac{\Gamma(q_{12}) \Gamma(q_{23})}{\Gamma(q_{12} + q_{23})}~,
\end{align}
and the other integrals are simply related to it:
\begin{equation}
I_1 = \frac{q_{12}}{q_{23}-1} I_2~, \qquad I_3 = \frac{q_{12}}{q_{12} + q_{23}} I_2~, \qquad I_4 = \frac{q_{23}}{q_{12} + q_{23}} I_2~.
\end{equation}
Hence the amplitude can be put in the form of a kinematic factor times the beta function $I_2$.   Momentum conservation and the mass shell condition imply $q_{12} + q_{23} = -q_{24} = - q_{13}$.  By using~(\ref{eq:kinematics4}), along with \eqref{eq:Deltanull}, we express the kinematic factor in the manifestly crossing-symmetric form given by Green and Schwarz \cite{Green:1981xx,Schwarz:1982jn}, or \cite{Green:1987mn} formula 7.4.44:
\begin{align}\label{4vstrings}
A(1,2,3,4) =&~ \frac{\DC{2}^4 \DC{3}}{q_{13}} B(q_{12},q_{23}) \times K(1,2,3,4)~,
\end{align}
with
\begin{align}
K(1,2,3,4) \equiv&~ - \bigg\{ \omega^{12} \omega^{34} q_{23} q_{13} + \omega^{23} \omega^{41} q_{12} q_{13} + \omega^{13} \omega^{24} q_{12} q_{23} + \cr
&~  \qquad + q_{12} \left[ \omega^{23} \Delta_{2}^4 \Delta_{3}^1 + \omega^{41} \Delta_{4}^2 \Delta_{1}^3  + \omega^{13} \Delta_{1}^4 \Delta_{3}^2 + \omega^{24} \Delta_{2}^3 \Delta_{4}^1 \right] + \cr
&~ \qquad + q_{23} \left[ \omega^{12} \Delta_{1}^3 \Delta_{2}^4 + \omega^{34} \Delta_{3}^1 \Delta_{4}^2 + \omega^{13} \Delta_{1}^2 \Delta_{3}^4 + \omega^{24} \Delta_{2}^1 \Delta_{4}^3 \right] + \cr
&~ \qquad + q_{13} \left[  \omega^{12} \Delta_{1}^4 \Delta_{2}^3 + \omega^{34} \Delta_{3}^2 \Delta_{4}^1 + \omega^{23} \Delta_{2}^1 \Delta_{3}^4 + \omega^{41} \Delta_{4}^3 \Delta_{1}^2 \right] \bigg\}~,
\end{align}
or equivalently
\begin{align}\label{K4vec}
K(1,2,3,4) =&~ - \frac{1}{4} \left( t u \, \omega^{12} \omega^{34} + s u \, \omega^{23} \omega^{41} + s t \, \omega^{13}\omega^{24} \right) + \cr
&~  + \frac{s}{2} \left[ \omega^{23} \Delta_{2}^4 \Delta_{3}^1 + \omega^{41} \Delta_{4}^2 \Delta_{1}^3  + \omega^{13} \Delta_{1}^4 \Delta_{3}^2 + \omega^{24} \Delta_{2}^3 \Delta_{4}^1 \right] + \cr
&~  + \frac{t}{2} \left[ \omega^{12} \Delta_{1}^3 \Delta_{2}^4 + \omega^{34} \Delta_{3}^1 \Delta_{4}^2 + \omega^{13} \Delta_{1}^2 \Delta_{3}^4 + \omega^{24} \Delta_{2}^1 \Delta_{4}^3 \right] + \cr
&~ + \frac{u}{2} \left[  \omega^{12} \Delta_{1}^4 \Delta_{2}^3 + \omega^{34} \Delta_{3}^2 \Delta_{4}^1 + \omega^{23} \Delta_{2}^1 \Delta_{3}^4 + \omega^{41} \Delta_{4}^3 \Delta_{1}^2 \right]  ~. \qquad
\end{align}

It is not difficult to see that the amplitude has the $\Z_4$ cyclic symmetry that we expect from the color-ordered open string amplitude.  The polarization factor, \eqref{K4vec}, and the stringy factor $q_{13}^{-1} B(q_{12},q_{23})$ independently carry the symmetry.  One must make use of the mass shell constraints, \eqref{eq:kinematics4}, to see the latter.   Of course the full amplitude is obtained by summing all cyclically-inequivalent permutations of the color-ordered amplitudes and will have the full $S_4$ permutation symmetry on the gauge boson lines.

In fact the polarization factor \eqref{K4vec} itself carries the full $S_4$ symmetry, as pointed out by \cite{Green:1981xx,Schwarz:1982jn}.  While a priori this needn't have been the case, it plays an important role in the KLT relations for four-point amplitudes.

%%%%%%%%%%%%%%%%%%%
\subsection{Two vectors and two gauginos}
%%%%%%%%%%%%%%%%%%%

Next we consider the case where particles $1$ and $4$ are gauginos.  We use vertex operators \eqref{eq:gVO} with picture charge $-1/2$.  Particles $2$ and $3$ are gauge bosons, and we take their vertex operators in the $0$ and $-1$ pictures respectively.  Hence the color-ordered amplitude we consider is
\begin{align}\label{A2g2vA1}
A(\tilde{1},2,3,\tilde{4}) =&~ \DC{1}^2 \DC{2}^2 \int_{z_{1}}^{z_3} \ed z_2 \langle c_1 V_{g}^{-1/2}[\zeta^1,k_1] V_{v}^{0}[\xi^2,k_2] c_3 V_{v}^{-1}[\xi^3,k_3] c_4 V_{g}^{-1/2}[\zeta^4,k_4] \rangle \cr
=&~ \DC{1}^2 \DC{2}^2 \int_{z_1}^{z_3} \ed z_2 \langle c_1 c_3 c_4 \rangle \bigg[ \langle \bV[\zeta^1] \bU[\xi^3] \bV[\zeta^4] \rangle \cT_{4}^{(2)} + \cr
&~ \qquad\qquad\qquad\qquad\qquad\qquad + \langle \bV[\zeta^1] \bJ[\mm^2] \bU[\xi^3] \bV[\zeta^4]\rangle \cT_4 \bigg]~.
\end{align}
We use the Ward identity with \eqref{eq:wiseJUVOPE} and \eqref{eq:UUcor} to evaluate the new correlator:
\begin{align}
\langle \bV[\zeta^1] \bJ[\mm^2] \bU[\xi^3] \bV[\zeta^4]\rangle =&~ \frac{1}{2z_{21}} \langle \bV_1[\mms^2 \zeta^1] \bU[\xi^3] \bV[\zeta^4] \rangle + \frac{2}{z_{23}} \langle \bV[\zeta^1] \bU_3[\mm^2 \xi^3] \bV[\zeta^4] \rangle + \cr
&~ + \frac{1}{2z_{24}} \langle \bV[\zeta^1] \bU[\xi^3] \bV_4[\mms^2 \zeta^4] \rangle \cr
=&~ - \frac{\zetab^1 \mms^2 \xis^3 \zeta^4}{2\sqrt{2} z_{21} z_{13} z_{14} z_{34}} + \frac{\zetab^1 (\omega^{23} \kiss_2  - \Delta_{2}^3 \xis^2 ) \zeta^4}{\sqrt{2} z_{23} z_{13} z_{14} z_{34}} + \frac{\zetab^1 \xis^3 \mms^2 \zeta^4}{2\sqrt{2} z_{24} z_{13} z_{14} z_{34}}~ . \cr \raisetag{16pt}
\end{align}
Note in order to use formula \eqref{eq:UUcor} in the case of the first term, one uses $\overline{\mms \zeta} \equiv ~^t(\mms \zeta) \cC = - ~^t\zeta \cC \mms = - \zetab \mms$.  Hence the amplitude is
\begin{align}\label{A2g2vA2}
A(\tilde{1},2,3,\tilde{4}) =&~ \frac{\DC{1}^2 \DC{2}^2}{\sqrt{2}} \int_{z_{1}}^{z_3} \ed z_2 \bigg[ \frac{\zetab^1 \xis^3 \zeta^4}{z_{31} z_{41} z_{43}} \left( -\frac{\Delta_{1}^2}{z_{21}} + \frac{\Delta_{3}^2}{z_{32}} + \frac{\Delta_{4}^2}{z_{42}} \right) + \frac{\zetab^1 \mms^2 \xis^3 \zeta^4}{2z_{21} z_{31} z_{41} z_{43}} + \cr
&~ +  \frac{\zetab^1 (\omega^{23} \kiss_2  - \Delta_{2}^3 \xis^2 ) \zeta^4}{z_{32} z_{31} z_{41} z_{43}} + \frac{\zetab^1 \xis^3 \mms^2 \zeta^4}{2z_{42} z_{31} z_{41} z_{43}} \bigg] \times (-z_{31} z_{41} z_{43}) \times \cT_4~.
\end{align}

We fix $z_{1,3,4}$ to the values \eqref{4ptzs} and use \eqref{fixed4tach}.  In this case there are only two denominator types, corresponding to the integrals $I_{3,4}$ of \eqref{4vnumden}.  We find
\begin{align}
A(\tilde{1},2,3,\tilde{4}) =&~ -\frac{\DC{1}^2 \DC{2}^2 \DC{3}}{\sqrt{2}} \int_{0}^1 \ed x \, x^{q_{12}} (1-x)^{q_{23}} \left\{ \frac{\widetilde{N}_3}{1-x} + \frac{\widetilde{N}_4}{x} \right\} \cr
=&~ \frac{\DC{1}^2 \DC{2}^2 \DC{3}}{\sqrt{2}} B(q_{12},q_{23}) \times \left\{ \frac{q_{12}}{q_{13}} \widetilde{N}_3 + \frac{q_{23}}{q_{13}} \widetilde{N}_4 \right\}~,
\end{align}
with
\begin{equation}
\widetilde{N}_3 := \zetab^1 \left[ \Delta_{3}^2 \xis^3 + \omega^{23} \kiss_2 - \Delta_{2}^3 \xis^2 \right] \zeta^4~, \qquad \widetilde{N}_4 := \zetab^1 \left[ -\Delta_{1}^2 \xis^3 + \ff{1}{2} \mms^2 \xis^3 \right] \zeta^4~.
\end{equation}
Therefore we have
\begin{align}
A(\tilde{1},2,3,\tilde{4}) =&~ \frac{\DC{1}^2 \DC{2}^2 \DC{3}}{2\sqrt{2}} \frac{B(q_{12},q_{23})}{q_{13}} K(\tilde{1},2,3,\tilde{4})~, \qquad \textrm{with} \cr
K(\tilde{1},2,3,\tilde{4}) \equiv&~ 2 q_{12} \zetab^1 \left[ \Delta_{3}^2 \xis^3 - \Delta_{2}^3 \xis^2 + \omega^{23} \kiss_2 \right] \zeta^4 - 2 q_{23} \zetab^1 \left[ \Delta_{1}^2 \xis^3 - \half \kiss_2 \xis^2 \xis^3 \right] \zeta^4~. \qquad
\end{align}
We used $\mms^2 = \kiss_2 \xis^2$, which follows from the physical state conditions and~(\ref{eq:mmdef}).

In order to compare this result with \cite{Schwarz:1982jn}, we use momentum conservation and the physical state conditions to write
\begin{align}\label{pstrick1}
& \zetab^1 \kiss_2 \zeta^4 =  - \zetab^1 \kiss_3 \zeta^4~, \cr
& \zetab^1 \kiss_2 \xis^2 \xis^3 = - \zetab^1 (\kiss_3 + \kiss_4) \xis^2 = \zetab^1 \left[ \xis^2 (\kiss_3 + \kiss_4) - 2 (\Delta_{3}^2 + \Delta_{4}^2) \right]~.
\end{align}
In particular, the second relation leads to some cancelation of terms in $K(\tilde{1},2,3,\tilde{4})$ upon using $\Delta_{1}^2 + \Delta_{3}^2 + \Delta_{4}^2 = 0$.  Setting $q_{12} = -s/2$ and $q_{23} = -t/2$ as well, we have
\begin{equation}\label{ourK2v2g}
K(\tilde{1},2,3,\tilde{4}) = -s \, \zetab^1 \left[ \Delta_{3}^2 \xis^3 - \Delta_{2}^3 \xis^2 - \omega^{23}\kiss_3 \right] \zeta^4 - \frac{t}{2} \, \zetab^1 \xis^2 (\kiss_3 + \kiss_4) \xis^3 \zeta^4 ~,
\end{equation}
This differs from equation (4.33) of reference \cite{Schwarz:1982jn}, (following from equation (4.13) of reference \cite{Green:1981xx}), by the relative sign of the $s$ and $t$ terms.  The overall prefactor of $K$ is of course a matter of definition since there is an ambiguity in how we factorize $A(\tilde{1},2,3,\tilde{4})$ into a ``polarization factor'' and the remaining stringy content.  The relative sign between the two terms in $K$ however is physical.

We claim there is a typo in \cite{Green:1981xx,Schwarz:1982jn} and that \eqref{ourK2v2g} is the correct expression.  Since this typo has propagated into some standard references, e.g. \cite{Kawai:1985xq} and 7.6.46 of \cite{Green:1987mn}, we give a quick check of the result.  Longitudinal gauge bosons should decouple from the color-ordered amplitude, and this requires the relative plus sign between the $s$ and $t$ terms.  To see this, set $\xi^2 = k_2$ and $\xi^3 = k_3$.  Using the physical state conditions, the two terms in~(\ref{ourK2v2g}) reduce to, respectively, $-\frac{st}{2} \zetab^1 \kiss_2 \zeta^4$ and $\frac{st}{2} \zetab^1 \kiss_2 \zeta^4$;  gauge invariance only holds for the relative plus sign as in~(\ref{ourK2v2g}).

With this corrected sign, the kinematic factor is actually even under exchange of the two bosons, and odd under exchange of the fermions, i.e.
\begin{equation}
\label{eq:2v2fSymmetries}
K(\tilde{1},3,2,\tilde{4})=K(\tilde{1},2,3,\tilde{4}),\qquad K(\tilde{4},2,3,\tilde{1})=-K(\tilde{1},2,3,\tilde{4}).
\end{equation}

\subsubsection*{The second color ordering}
In the two-vector, two-gaugino case we are considering there is a second disk amplitude that cannot be obtained from the previous computation by either cyclically permuting the particle labels or reflecting them about a bisecting line.  In the previous computation the ordering around the disc was $bffb$ where ``$b$'' stands for boson and ``$f$'' for fermion.  The second inequivalent amplitude corresponds to the ordering $fbfb$ and is given by
\begin{align}\label{A2g2vB1}
A(\tilde{1},2,\tilde{3},4) =&~ \DC{1}^2 \DC{2}^2 \int_{z_{1}}^{z_3} \ed z_2 \langle c_1 V_{g}^{-1/2}[\zeta^1,k_1] V_{v}^{0}[\xi^2,k_2] c_3 V_{g}^{-1/2}[\zeta^3,k_3]  c_4 V_{v}^{-1}[\xi^4,k_4] \rangle \cr
=&~ -\DC{1}^2 \DC{2}^2 \int_{z_1}^{z_3} \ed z_2 \langle c_1 c_3 c_4 \rangle \bigg[ \langle \bV[\zeta^1] \bV[\zeta^3] \bU[\xi^4] \rangle \cT_{4}^{(2)} + \cr
&~ \qquad\qquad\qquad\qquad\qquad\qquad + \langle \bV[\zeta^1] \bJ[\mm^2] \bV[\zeta^3] \bU[\xi^4]\rangle \cT_4 \bigg]~.
\end{align}

In comparing \eqref{A2g2vB1} with \eqref{A2g2vA1}, we see that the only differences are an overall minus sign and that particle labels 3 and 4 are exchanged in the $\psi$ matter correlators, while the ghost and Koba--Nielson amplitudes remain the same.  The minus sign is due to the odd number of exchanges of the worldsheet fermions $c_i$ and $\bV[\zeta^j]$ necessary to factorize the ghost and matter sectors.  Since $\bV[\zeta]$ and $\bU[\xi]$ commute,
\begin{align}
A(\tilde{1},2,\tilde{3},4) =&~- \frac{\DC{1}^2 \DC{2}^2}{\sqrt{2}} \int_{z_{1}}^{z_3} \ed z_2 \bigg[ \frac{\zetab^1 \xis^4 \zeta^3}{z_{14} z_{13} z_{34}} \left( -\frac{\Delta_{1}^2}{z_{21}} + \frac{\Delta_{3}^2}{z_{32}} + \frac{\Delta_{4}^2}{z_{42}} \right) + \frac{\zetab^1 \mms^2 \xis^4 \zeta^3}{2z_{21} z_{14} z_{13} z_{34}} + \cr
&~ +  \frac{\zetab^1 (\omega^{24} \kiss_2  - \Delta_{2}^4 \xis^2 ) \zeta^3}{z_{42} z_{14} z_{13} z_{34}} + \frac{\zetab^1 \xis^4 \mms^2 \zeta^3}{2z_{32} z_{14} z_{13} z_{34}} \bigg] \times (z_{13} z_{14} z_{34}) \times \cT_4~.
\end{align}
Plugging in \eqref{4ptzs} and \eqref{fixed4tach} we have
\begin{align}
A(\tilde{1},2,\tilde{3},4) =&~ \frac{\DC{1}^2 \DC{2}^2 \DC{3}}{\sqrt{2}} \int_{0}^1 \ed x \, x^{q_{12}} (1-x)^{q_{23}} \left\{ \frac{\widetilde{N}_{3}'}{1-x} + \frac{\widetilde{N}_{4}'}{x} \right\} \cr
=&~ -\frac{\DC{1}^2 \DC{2}^2 \DC{3}}{2\sqrt{2}} B(q_{12},q_{23}) \times \left\{ \frac{q_{12}}{q_{13}} \widetilde{N}_{3}' + \frac{q_{23}}{q_{13}} \widetilde{N}_{4}' \right\}~,
\end{align}
with
\begin{equation}
\widetilde{N}_{3}' := -\zetab^1 \left[2 \Delta_{3}^2 \xis^4 + \xis^4 \mms^2 \right] \zeta^3~, \qquad \widetilde{N}_{4}' := \zetab^1 \left[ 2\Delta_{1}^2 \xis^4 - \mms^2 \xis^4 \right] \zeta^3~.
\end{equation}
Hence
\begin{align}
A(\tilde{1},2,\tilde{3},4) =&~  -\frac{\DC{1}^2 \DC{2}^2 \DC{3}}{2\sqrt{2}} \frac{B(q_{12},q_{23})}{q_{13}} K(\tilde{1},2,\tilde{3},4)~, \qquad \textrm{with} \cr
K(\tilde{1},2,\tilde{3},4) \equiv &~ - q_{12} \zetab^1 \xis^4\left[ 2\Delta_{3}^2 +  \kiss_2 \xis^2 \right] \zeta^3 + q_{23} \zetab^1 \left[ 2\Delta_{1}^2  -  \kiss_2 \xis^2  \right] \xis^4\zeta^3~.
\end{align}
Via the same sort of manipulations as in \eqref{pstrick1} we find that the polarization tensor can also be written as
\begin{equation}\label{K2v2gB}
K(\tilde{1},2,\tilde{3},4) = \frac{s}{2} \, \zetab^1 \xis^4 (\kiss_2 + \kiss_3) \xis^2 \zeta^3 + \frac{t}{2} \, \zetab^1 \xis^2 (\kiss_3 + \kiss_4) \xis^4 \zeta^3~,
\end{equation}
in agreement with \cite{Schwarz:1982jn} and 7.4.47 of \cite{Green:1987mn}.  In fact, this kinematic factor matches the factor that appeared in the first ordering; with some now-familiar manipulations using momentum conservation and physical state conditions we can show that
\begin{equation}
\label{eq:KequalsK}
K(\tilde{1},2,\tilde{3},4)=K(\tilde{1},2,4,\tilde{3}).
\end{equation}
As such, it also inherits the symmetries (\ref{eq:2v2fSymmetries}), i.e.\ it is even under exchange of the vectors and odd under exchange of the gauginos.  
The amplitudes have prefactors proportional to $B(q_{12},q_{23})/q_{13}$ and $B(q_{12},q_{24})/q_{14}$ respectively.  Since these are different, the amplitudes themselves still disagree,
\begin{equation}
A(\tilde{1},2,\tilde{3},4)\ne A(\tilde{1},2,4,\tilde{3}).
\end{equation}

%%%%%%%%%%%%%%%%%%%
\subsection{Four gauginos}
%%%%%%%%%%%%%%%%%%%

Our final four-point amplitude is the four-gaugino amplitude,
\begin{align}
A(\tilde{1},\tilde{2},\tilde{3},\tilde{4}) =&~ -\DC{1}^4 \int_{z_{1}}^{z_3} \ed z_2 \langle c_1 V_{g}^{-1/2}[\zeta^1,k_1] V_{g}^{-1/2}[\zeta^2,k_2] c_3 V_{g}^{-1/2}[\zeta^3,k_3]  c_4 V_{g}^{-1/2}[\zeta^4,k_4] \rangle \cr
=&~- \DC{1}^4 \int_{z_1}^{z_3} \ed z_2 \langle c_1 c_3 c_4\rangle \langle \bV[\zeta^1] \bV[\zeta^2] \bV[\zeta^3] \bV[\zeta^4] \rangle \cT_4 ~. \raisetag{20pt}
\end{align}
We need to compute the correlator of four $\cV$ currents.  This is an easy task because the correlator is holomorphic, and thus completely determined by its singularities and their residues.  These, in turn, are fixed by the OPE~(\ref{eq:VVOPE}) and the three-point function in~(\ref{eq:UUcor}).  Restricting the external spinors to be Majorana--Weyl, which excludes the $O(1/z_{ij}^2)$ term in \eqref{eq:VVOPE}, we obtain
\begin{align}
\label{eq:VVVVcor}
\la \cV_{a_1} \cV_{a_2} \cV_{a_3} \cV_{a_4} \ra & = -\frac{ (\cC\Gamma_\mu)_{a_2 a_1} (\cC\Gamma^\mu)_{a_4 a_3}}{2 z_{12}z_{23} z_{24} z_{34}}
-\frac{ (\cC\Gamma_\mu)_{a_3 a_1} (\cC\Gamma^\mu)_{a_4 a_2}}{2 z_{13}z_{23}  z_{24}z_{34}}
-\frac{ (\cC\Gamma_\mu)_{a_4 a_1} (\cC\Gamma^\mu)_{a_2 a_3}}{2 z_{14} z_{23} z_{24}z_{34}}~.
\end{align}
As a check, we can see that the operator satisfies the $\Lsl(2,\R)$ conformal Ward identities.  Dilatation covariance and translation invariance hold trivially, while invariance under special conformal transformations requires the right-hand side to be annihilated by $\sum_{i=1}^4( z_i^2 \p_{z_i} + 2z_i)$; this holds as a consequence of the following famous ten-dimensional Fierz identity for Majorana--Weyl spinors\footnote{A proof is given in the appendix of chapter 4 of~\cite{Green:1987mn}.}~:
\begin{align}
\label{eq:Fierz}
(\cC\Gamma_\mu)_{a_2 a_1} (\cC\Gamma^\mu)_{a_4 a_3}
+(\cC\Gamma_\mu)_{a_3 a_1} (\cC\Gamma^\mu)_{a_4 a_2}
+(\cC\Gamma_\mu)_{a_4 a_1} (\cC\Gamma^\mu)_{a_2 a_3} = 0~.
\end{align}
We emphasize that without the GSO projection to Majorana--Weyl spinors,~(\ref{eq:VVVVcor}) and~(\ref{eq:Fierz}) do not hold.

Contracting \eqref{eq:VVVVcor} with the external spinors we obtain
\begin{equation}\label{4Vcor}
\langle \bV[\zeta^1] \bV[\zeta^2] \bV[\zeta^3] \bV[\zeta^4] \rangle = - \frac{ \zetab^1 \Gamma^\mu \zeta^2 \,\zetab^3 \Gamma_\mu \zeta^4 }{2 z_{21} z_{32} z_{43} z_{42}} - \frac{\zetab^1 \Gamma^\mu \zeta^3 \,\zetab^2 \Gamma_\mu \zeta^4 }{2 z_{31} z_{32} z_{42} z_{43}} - \frac{\zetab^1 \Gamma^\mu \zeta^4 \,\zetab^2 \Gamma_\mu \zeta^3  }{2 z_{41} z_{42} z_{32} z_{43}} ~.
\end{equation}
Setting $\langle c_1 c_3 c_4 \rangle = -z_{31} z_{41} z_{43}$ and using \eqref{4ptzs} and \eqref{fixed4tach}  leads to
\begin{align}
A(\tilde{1},\tilde{2},\tilde{3},\tilde{4}) =&~ -\frac{\DC{1}^4 \DC{3}}{2} \int_{0}^1 \ed x \, x^{q_{12}} (1-x)^{q_{13}} \left\{  \frac{\zetab^1 \Gamma^\mu \zeta^2 \,\zetab^3 \Gamma_\mu \zeta^4  }{x(1-x)} + \frac{\zetab^1 \Gamma^\mu \zeta^3 \,\zetab^2 \Gamma_\mu \zeta^4 }{1-x} \right\} \cr
=&~ -\frac{\DC{1}^4 \DC{3}}{2} \frac{B(q_{12}, q_{23})}{q_{13}} K(\tilde{1},\tilde{2},\tilde{3},\tilde{4})~, \qquad \textrm{where}~, \cr
K(\tilde{1},\tilde{2},\tilde{3},\tilde{4}) \equiv &~ q_{13} \zetab^1 \Gamma^\mu \zeta^2 \,\zetab^3 \Gamma_\mu \zeta^4 - q_{12} \zetab^1 \Gamma^\mu \zeta^3 \,\zetab^2 \Gamma_\mu \zeta^4~.
\end{align}
The mass shell condition $q_{12} + q_{13} + q_{23} = 0$ together with the Fierz identity \eqref{eq:Fierz} allow us to write this polarization tensor in different ways.  For example, eliminating $q_{13} = -u/2$ in favor of $q_{12} = -s/2$ and $q_{23} = -t/2$ leads to
\begin{equation}
K(\tilde{1},\tilde{2},\tilde{3},\tilde{4}) = -\frac{s}{2} \, \zetab^1 \Gamma^\mu \zeta^4 \, \zetab^2 \Gamma_\mu \zeta^3 + \frac{t}{2} \, \zetab^1 \Gamma^\mu \zeta^2 \, \zetab^3 \Gamma_\mu \zeta^4~,
\end{equation}
which is the form given in \cite{Schwarz:1982jn} and 7.4.48 of \cite{Green:1987mn}.  Indeed one may show with such manipulations that $K(\tilde{1},\tilde{2},\tilde{3},\tilde{4})$ is odd under the full $S_4$ permutation group.

\section{Comparison to ordered super-Yang--Mills amplitudes} \label{s:YM}
We now compare the previous results to tree-level computations in gauge theory.  This will allow us to match the low energy limit, verify unitarity at the massless level, and fix the normalization constants $\DC{i}$ as advertised in~(\ref{eq:vernorm}) and (\ref{eq:tachnorm}).

\subsection{The ordered Feynman rules}
We begin by writing down the color-ordered Feynman rules for our super-Yang--Mills theory.  For a nice introduction to these ideas the reader may consult one of the recent reviews~\cite{Elvang:2013cua,Henn:2014yza}.  We write our amplitudes with all momenta incoming (as we already did tacitly above), and we fix Feynman gauge.  We construct color-ordered diagrams with external labels increasing in a counter-clockwise fashion, and we add all cyclically inequivalent diagrams with desired external states using the following rules:
\begin{align}\label{Feynrules1}
\xymatrix@R=4mm@C=15mm{
%%%%
%%%%  vector propagator
\begin{xy} <1.0mm,0mm>:
(0,8)*{~};
(0,-8)*{~};
(9,0)*{\mu_1}="e1",
(-9,0)*{\mu_2}="e2",
\ar@{~}"e1";"e2"
\end{xy}  &  \displaystyle{-\frac{i}{k^2}\eta^{\mu_1\mu_2} } ~, \\
%%%%
%%%%   3-vector vertex
\begin{xy} <1.0mm,0mm>:
(0,0)*{\bullet}="v",
(10,0)*{1}="e1",
(0,7)*{2}="e2",
(-10,0)*{3}="e3",
%(0,-7)*{4}="e4",
\ar@{~}"e1";"v"
\ar@{~}"v";"e2"
\ar@{~}"v";"e3"
%\ar@{~}"v";"e4"
\end{xy}  &  +ig\left[ (\xi^1\cdot k_{23}) (\xi^2\cdot \xi^3)+ (\xi^2\cdot k_{31}) (\xi^3\cdot \xi^1)+(\xi^3\cdot k_{12}) (\xi^1\cdot \xi^2) \right] ~, \\
%%%%
%%%%   4-vector vertex
\begin{xy} <1.0mm,0mm>:
(0,0)*{\bullet}="v",
(10,0)*{1}="e1",
(0,7)*{2}="e2",
(-10,0)*{3}="e3",
(0,-7)*{4}="e4",
\ar@{~}"e1";"v"
\ar@{~}"v";"e2"
\ar@{~}"v";"e3"
\ar@{~}"v";"e4"
\end{xy}  &  +i g^2 \left[ 2(\xi^1\cdot\xi^3)(\xi^2\cdot\xi^4) - (\xi^1\cdot \xi^2)(\xi^3\cdot \xi^4) -(\xi^1\cdot\xi^4)(\xi^2\cdot \xi^3) \right] ~.
}
\end{align}
We wrote the vertices in a condensed notation by dotting in with dummy external polarizations.  To obtain the vertices we of course need to expand these as, say, $\xi^1_{\mu_1} \xi^2_{\mu_2} \xi^3_{\mu_3} V^{\mu_1 \mu_2\mu_3}$.  We will typically express these vertices by amplitudes with some polarizations stripped off.  For instance,
\begin{align}
\AYM(1,2,\cdot)_\mu &= ig \left[ k_{12\mu} (\xi^1\cdot\xi^2) + 2(\xi^1\cdot k_2 \xi^2_\mu -\xi^2\cdot k_1 \xi^1_\mu) \right]~, \nonumber\\
\AYM(\cdot,3,4)_\mu &= ig \left[ k_{34\mu} (\xi^3\cdot\xi^4) + 2(\xi^3\cdot k_4 \xi^{4}_{\mu} -\xi^4\cdot k_3 \xi^{3}_{\mu}) \right]~,
\end{align}
where momentum conservation and the physical state conditions on the external polarizations were used.  We then have a key property: as long as $\xi^{1,2}$ obey physical conditions, then
\begin{align}
\AYM(1,2,\cdot)_\mu (k_1+k_2)^\mu = 0~.
\end{align}
In other words, the longitudinal photon does not show up as an intermediate state.

To include the gauginos we use the rules (written in the same condensed fashion)
\begin{align}\label{Feynrules2}
\xymatrix@R=2mm{
\begin{xy}<1.0mm,0mm>:
(-10,0)*{c}="c";
(10,0)*{b} ="b";
(0,-6)*{p}
\ar@{-}|{<}"c";"b"
\ar@{<-}(-7,-4);(7,-4)
\end{xy}
&
\begin{xy} <1.0mm,0mm>:
%(10,0)*{5}="v1",
(20,0)*{3}="e3",
(30,0)*{\bullet}="v",
(40,0)*{1}="e1",
%(20,9)*{3}="e3",
(30,9)*{2}="e2",
%\ar@{-}|{>}"v1";"v2"
\ar@{-}|{<}"e3";"v"
\ar@{-}|{<}"v";"e1"
%\ar@{~}"v2";"e3"
\ar@{~}"v";"e2"
\end{xy}
&
\begin{xy} <1.0mm,0mm>:
%(10,0)*{5}="v1",
(20,0)*{2}="e3",
(30,0)*{\bullet}="v",
(40,0)*{1}="e1",
%(20,9)*{3}="e3",
(30,-9)*{3}="e2",
%\ar@{-}|{>}"v1";"v2"
\ar@{-}|{<}"e3";"v"
\ar@{-}|{<}"v";"e1"
%\ar@{~}"v2";"e3"
\ar@{~}"v";"e2"
\end{xy}
\\
\frac{i (p\slash)_{cb}}{p^2} & ig\zetab^3 \xis^2\zeta^1 &  ig\zetab^2 \xis^3\zeta^1
}
\end{align}
As in QED, to completely specify the rules we must describe some additional signs due to the Fermi statistics; we will only need these in our discussion of the $4$-gaugino amplitude, and so we will not give a complete description here.

\subsection{The amplitudes}
The three-point ``amplitudes'' can be read off directly from the vertices, while the four-point amplitudes are constructed by applying the rules.  For starters we consider
\begin{align}
\label{eq:YMbbbb}
\AYM(1,2,3,4)~ & =~
%%%%%%%%%%%%
%%%%%%%%%%%%
\begin{xy} <1.0mm,0mm>:
(10,-7)*{4}="e6",
(10,0)*{\bullet}="v1",
(20,0)*{\bullet}="v2",
(0,0)*{3}="e4",
(20,7)*{2}="e3",
(30,0)*{1}="e2"
\ar@{~}"v1";"v2"
\ar@{~}"v2";"e2"
\ar@{~}"v2";"e3"
\ar@{~}"v1";"e4"
\ar@{~}"v1";"e6"
\end{xy}  ~+~
%%%%%%%%%%%%
%%%%%%%%%%%%
\begin{xy} <1.0mm,0mm>:
(10,7)*{2}="e6",
(10,0)*{\bullet}="v1",
(20,0)*{\bullet}="v2",
(0,0)*{3}="e4",
(20,-7)*{4}="e3",
(30,0)*{1}="e2"
\ar@{~}"v1";"v2"
\ar@{~}"v2";"e2"
\ar@{~}"v2";"e3"
\ar@{~}"v1";"e4"
\ar@{~}"v1";"e6"
\end{xy}  ~+~
\begin{xy} <1.0mm,0mm>:
(10,-7)*{4}="e6",
(10,0)*{\bullet}="v1",
(0,0)*{3}="e4",
(10,7)*{2}="e3",
(20,0)*{1}="e2"
\ar@{~}"e6";"v1"
\ar@{~}"v1";"e2"
\ar@{~}"v1";"e3"
\ar@{~}"v1";"e4"
\end{xy} ~ \nonumber\\
& = D_1 + D_2 + D_3~,
\end{align}
with
\begin{align}
D_1 &= \frac{ i g^2}{ 2 q_{12}}\times
\left[ k_{12\mu} \omega^{12} + 2 (\Delta^1_2 \xi^2_\mu - \Delta^2_1\xi^1_\mu) \right]
\times
\left[ k_{34}^{\mu} \omega^{34} + 2 (\Delta^3_4 \xi^{4\mu}-\Delta^4_3 \xi^{3\mu}) \right]~,\nonumber\\
D_2 & = \frac{ i g^2}{ 2 q_{41}}\times
\left[ k_{41\mu} \omega^{41} + 2 (\Delta^4_1 \xi^1_\mu - \Delta^1_4\xi^4_\mu) \right]
\times
\left[ k_{23}^{\mu} \omega^{23} + 2 (\Delta^2_3 \xi^{3\mu}-\Delta^3_2 \xi^{2\mu}) \right]~, \nonumber\\
D_3 & =ig^2\left[ 2\omega^{13}\omega^{24}-\omega^{12}\omega^{34}-\omega^{14}\omega^{23} \right] ~.
\end{align}
We can easily verify gauge invariance by setting $\xi^1 = k_1$, where we find
\begin{align}
D_1 &\to ig^2\left[ (\Delta^2_3-\Delta^2_4) \omega^{34} + 2 (\Delta^3_4\omega^{24}-\Delta^4_3\omega^{23}) \right]~,\nonumber\\
D_2 & \to -ig^2\left[ (\Delta^4_2-\Delta^4_3) \omega^{23} + 2(\Delta^2_3\omega^{34}-\Delta^3_2\omega^{24})\right]~,\nonumber\\
D_3 & \to
ig^2\left[2\Delta^3_1\omega^{24}-\Delta^2_1\omega^{34} -\Delta^4_1 \omega^{23}\right]~,
\end{align}
so that the sum vanishes in the limit.

Next, we tackle the amplitudes with two fermions.
\begin{align}
\label{eq:YMfbbf}
%\xymatrix@R=0mm@C=10mm{
%%%%%%%%%%%%%%
%%%%%%%%%%%%%%
\AYM(\tilde{1},2,3,\tilde{4}) &=~
\begin{xy} <1.0mm,0mm>:
(10,-3)*{4}="v1",
(20,-3)*{\bullet}="v2",
(30,-3)*{1}="e1",
(20,4)*{\bullet}="v3",
(10,4)*{3}="e3",
(30,4)*{2}="e2",
\ar@{-}|{<}"v1";"v2"
\ar@{-}|{<}"v2";"e1"
\ar@{~}"v2";"v3"
\ar@{~}"v3";"e3"
\ar@{~}"v3";"e2"
\end{xy} ~+~
%%%%%%%%%%%%%%
%%%%%%%%%%%%%%
\begin{xy} <1.0mm,0mm>:
(10,-3)*{4}="v1",
(20,-3)*{\bullet}="v2",
(30,-3)*{\bullet}="v3",
(40,-3)*{1}="e1",
(20,6)*{3}="e3",
(30,6)*{2}="e2",
\ar@{-}|{<}"v1";"v2"
\ar@{-}|{<}"v2";"v3"
\ar@{-}|{<}"v3";"e1"
\ar@{~}"v2";"e3"
\ar@{~}"v3";"e2"
\end{xy} \qquad ,\nonumber\\[3mm]
%%%%%%%%%%%%%%
%%%%%%%%%%%%%%
\AYM(\tilde{1},2,\tilde{3},4) & =
\begin{xy} <1.0mm,0mm>:
(10,0)*{3}="v1",
(20,0)*{\bullet}="v2",
(30,0)*{\bullet}="v3",
(40,0)*{1}="e1",
(20,-9)*{4}="e3",
(30,9)*{2}="e2",
\ar@{-}|{<}"v1";"v2"
\ar@{-}|{<}"v2";"v3"
\ar@{-}|{<}"v3";"e1"
\ar@{~}"v2";"e3"
\ar@{~}"v3";"e2"
\end{xy} ~+~
%%%%%%%%%%%%%%
%%%%%%%%%%%%%%
\begin{xy} <1.0mm,0mm>:
(10,0)*{3}="v1",
(20,0)*{\bullet}="v2",
(30,0)*{\bullet}="v3",
(40,0)*{1}="e1",
(20,9)*{2}="e3",
(30,-9)*{4}="e2",
\ar@{-}|{<}"v1";"v2"
\ar@{-}|{<}"v2";"v3"
\ar@{-}|{<}"v3";"e1"
\ar@{~}"v2";"e3"
\ar@{~}"v3";"e2"
\end{xy}
\end{align}
From the rules we obtain
\begin{align}
\label{eq:YMffff}
\AYM(\tilde{1},2,3,\tilde{4}) & = \frac{i g^2}{q_{23}} \zetab^4 \left[\omega^{23} \kiss_2 + \Delta^2_3 \xis^3 -\Delta^3_2\xis^2\right] \zeta^1~ -\frac{i g^2}{2q_{12}} \zetab^4 \xis^3(\kiss_1+\kiss_2) \xis^2 \zeta^1~,\nonumber\\[2mm]
\AYM(\tilde{1},2,\tilde{3},4) &= -\frac{ig^2}{2} \frac{1}{q_{12}}  \zetab^3 \xis^4 (\kiss_1+\kiss_2) \xis^2 \zeta^1
-\frac{ig^2}{2} \frac{1}{q_{23}}  \zetab^3\xis^2(\kiss_1+\kiss_4) \xis^4 \zeta^1~.
\end{align}
Finally, we have the four gaugino amplitude
\begin{align}
\AYM(\tilde{1},\tilde{2},\tilde{3},\tilde{4}) =
\begin{xy} <1.0mm,0mm>:
(10,-3)*{4}="v1",
(20,-3)*{\bullet}="v2",
(30,-3)*{1}="e1",
(20,4)*{\bullet}="v3",
(10,4)*{3}="e3",
(30,4)*{2}="e2",
\ar@{-}|{<}"v1";"v2"
\ar@{-}|{<}"v2";"e1"
\ar@{~}"v2";"v3"
\ar@{-}|{<}"v3";"e3"
\ar@{-}|{<}"v3";"e2"
\end{xy} ~~+~~
%%%%%%%%%%%%%
%%%%%%%%%%%%%
\begin{xy} <1.0mm,0mm>:
(10,-3)*{3}="v1",
(20,-3)*{\bullet}="v2",
(30,-3)*{4}="e1",
(20,4)*{\bullet}="v3",
(10,4)*{2}="e3",
(30,4)*{1}="e2",
\ar@{-}|{<}"v1";"v2"
\ar@{-}|{<}"v2";"e1"
\ar@{~}"v2";"v3"
\ar@{-}|{<}"v3";"e3"
\ar@{-}|{<}"v3";"e2"
\end{xy}~~.
\end{align}
The amplitude is then
\begin{align}\label{AYM4gaugino}
\AYM(\tilde{1},\tilde{2},\tilde{3},\tilde{4}) & =
ig^2 \frac{ (\zetab^1\Gamma_\mu \zeta^4)(\zetab^2\Gamma^\mu\zeta^3)}{2 q_{23}}
~+~ (-1) \times ig^2 \frac{ (\zetab^1\Gamma_\mu\zeta^2)  (\zetab^3\Gamma^\mu\zeta^4)}{2 q_{12}}~.
\end{align}
The relative sign between the two diagrams is due to Fermi statistics.  The simplest way to obtain it is to compare the two sets of contractions that lead to the different contributions; the result of this comparison is an extra sign due to an exchange of two fermions.

\subsection{Comparison to the string theory results}
We can now compare the string theory computations of the previous section with the super-YM results.  From matching the three-point computations, we obtain the normalizations
\begin{align}
\DC{2}^3\DC{3} &= -2ig~,&
\DC{1}^2 \DC{2}\DC{3} & = i\sqrt{2} g~.
\end{align}
To compare the four-point amplitudes we need to take the momenta to zero and use
\begin{align}
\frac{B(q_{12},q_{23})}{q_{13}} = -\frac{1}{q_{12}q_{23}} + \frac{\pi^2}{6} + O(q)~.
\end{align}
Taking the leading term, we find the string amplitudes reduce to
\begin{align}
A(1,2,3,4) & \to -\DC{2}^4\DC{3} \frac{1}{q_{12}q_{23}} K(1,2,3,4)~,\nonumber\\[2mm]
A(\tilde{1},2,3,\tilde{4}) & \to -\frac{\DC{1}^2\DC{2}^2\DC{3}}{2\sqrt{2}} \frac{1}{q_{12}q_{23}} K(\tilde{1},2,3,\tilde{4})~,\nonumber\\[2mm]
A(\tilde{1},2,\tilde{3},4) & \to \frac{\DC{1}^2\DC{2}^2\DC{3}}{2\sqrt{2}} \frac{1}{q_{12}q_{23}} K(\tilde{1},2,\tilde{3},{4})~,\nonumber\\[2mm]
A(\tilde{1},\tilde{2},\tilde{3},\tilde{4})  &\to \frac{\DC{1}^4 \DC{3}}{2} \frac{1}{q_{12}q_{23}} K(\tilde{1},\tilde{2},\tilde{3},\tilde{4})~.
\end{align}
By comparing the $\AYM$ amplitudes with these terms we find that they are matched by the corresponding limit of a string amplitude if and only if
\begin{align}
\DC{2}^4 \DC{3} & = 2ig^2~,&
\DC{1}^2 \DC{2}^2 \DC{3} & = -i\sqrt{2}g^2~,&
\DC{1}^4\DC{3} & = ig^2~.
\end{align}
The only case where the comparison is not completely straightforward is in the 4-vector amplitude.  To see that the claim is sensible, we can consider the $\omega^2$ terms in the two amplitudes.  We have
\begin{align}
\left.\AYM(1,2,3,4)\right|_{{\omega^2}}& = 2ig^2 \times\left[- \frac{q_{23}}{q_{12}} \omega^{12}\omega^{34} - \frac{q_{12}}{q_{23}} \omega^{14}\omega^{23} -\omega^{12}\omega^{34}-\omega^{14}\omega^{23}+\omega^{13}\omega^{24}\right]~,  \nonumber\\
\left. -K(1,2,3,4)\right|_{{\omega^2}}& =
 \omega^{12} \omega^{34} q_{23} q_{13} + \omega^{23} \omega^{41} q_{12} q_{13} + \omega^{13} \omega^{24} q_{12} q_{23} \nonumber\\[2mm]
& =- \omega^{12} \omega^{34} q^2_{23}  - \omega^{23} \omega^{41} q^2_{12}  + (\omega^{13} \omega^{24}-\omega^{12} \omega^{34} - \omega^{23} \omega^{41}) q_{12} q_{23}~.
\end{align}
Once we have matched these Yang--Mills results unitarity is of course obvious, and we see that the $\DC{i}$ take the values claimed in~(\ref{eq:vernorm}) and (\ref{eq:tachnorm}).

We can then finally summarize all of the color-ordered four-point amplitudes as follows:
\begin{align}
\label{eq:FourPointSummary}
A(1,2,3,4) & = 2ig^2 \frac{B(q_{12},q_{23})}{q_{13}} K(1,2,3,4)~,\nonumber\\[2mm]
A(\tilde{1},2,3,\tilde{4}) & =-\frac{ig^2}{2} \frac{B(q_{12},q_{23})}{q_{13}} K(\tilde{1},2,3,\tilde{4})~,\nonumber\\[2mm]
A(\tilde{1},2,\tilde{3},4) & = \frac{ig^2}{2} \frac{B(q_{12},q_{23})}{q_{13}} K(\tilde{1},2,\tilde{3},{4})~,\nonumber\\[2mm]
A(\tilde{1},\tilde{2},\tilde{3},\tilde{4})  & =-\frac{ig^2}{2} \frac{B(q_{12},q_{23})}{q_{13}} K(\tilde{1},\tilde{2},\tilde{3},\tilde{4})~.
\end{align}
The leading terms in the momentum expansion are just the Yang--Mills amplitudes, while the higher order terms arise from the supersymmetric form of the $\tr F^4$ coupling in the effective Lagrangian~\cite{Metsaev:1987by,Cederwall:2001td}.  Indeed, we could explicitly subtract off the Yang--Mills contributions from the amplitude to derive the additional vertices
\begin{align}
\label{eq:newvertices4}
\xymatrix@C=4mm{
\begin{xy} <1.0mm,0mm>:
(10,-7)*{}="v1",
(10,0)*{\blacksquare}="v2",
(0,0)*{}="e4",
(10,7)*{}="e3",
(20,0)*{}="e2"
\ar@{~}"v1";"v2"
\ar@{~}"v2";"e2"
\ar@{~}"v2";"e3"
\ar@{~}"v2";"e4"
\end{xy} 
& &
\begin{xy} <1.0mm,0mm>:
(10,0)*{}="v1",
(20,0)*{\blacksquare}="v2",
(30,0)*{}="e1",
(14,7)*{}="e4",
(27,7)*{}="e2",
%(10,7)*{3}="e3",
%(30,7)*{2}="e2",
\ar@{-}|{>}"v1";"v2"
\ar@{-}|{>}"v2";"e1"
\ar@{~}"v2";"e4"
\ar@{~}"v2";"e2"
%\ar@{~}"v3";"e3"
%\ar@{~}"v3";"e2"
\end{xy}  
&&
\begin{xy} <1.0mm,0mm>:
(10,-7)*{}="e1",
(10,0)*{\blacksquare}="v",
(0,0)*{}="e4",
(10,7)*{}="e3",
(20,0)*{}="e2"
\ar@{~}"e1";"v"
%\ar@{~}"v";"e2"
\ar@{-}|{>}"v";"e2"
\ar@{~}"v";"e3"
\ar@{-}|{>}"e4";"v"
%\ar@{~}"v";"e4"
\end{xy} 
&&
\begin{xy} <1.0mm,0mm>:
(10,-7)*{}="e1",
(10,0)*{\blacksquare}="v",
(0,0)*{}="e4",
(10,7)*{}="e3",
(20,0)*{}="e2"
\ar@{-}|{\vee}"e1";"v"
\ar@{-}|{\vee}"v";"e3"
%\ar@{~}"v1";"v2"
\ar@{-}|{>}"v";"e2"
\ar@{-}|{>}"e4";"v"
\end{xy} ~~.
%&
%\begin{xy} <1.0mm,0mm>:
%(10,0)*{}="v1",
%(20,0)*{\blacksquare}="v2",
%(30,0)*{}="e1",
%(14,7)*{}="e4",
%(20,7)*{}="e3",
%(27,7)*{}="e2",
%\ar@{-}|{>}"v1";"v2"
%\ar@{-}|{>}"v2";"e1"
%\ar@{~}"v2";"e4"
%\ar@{~}"v2";"e3"
%\ar@{~}"v2";"e2"
%\end{xy} 
}
\end{align}

\section{Five-point open superstring amplitudes} \label{s:5point}
We now turn to the five-point amplitudes.  The five-vector amplitude has been tackled in a number of works, and a very complete treatment has been given in~\cite{Barreiro:2005hv}.  The remaining amplitudes involving gauginos can then also be determined from the knowledge of the bosonic amplitude and space-time supersymmetry~\cite{Medina:2006uf,Barreiro:2012aw}.  Indeed, using the pure spinor approach the full amplitudes can be computed in terms of SYM partial amplitudes~\cite{Mafra:2011nv}.  On the other hand, it is sometimes useful to have the explicit component expansions for these amplitudes, and they are sufficiently complicated that it may be useful to have an independent computation of these in the direct NSR approach.  In this section, we will first review the results of~\cite{Barreiro:2005hv} on the five-vector amplitude in our notation and then turn to amplitudes involving fermions.

The five-point massless kinematics is conveniently parametrized by the invariants
\begin{align}
q_{ij} \in \{ q_{12},~ q_{13},~ q_{23},~ q_{24},~ q_{34} \}~.
\end{align}
The remaining $q_{ij}$ are given by
\begin{align}
q_{14} &= -q_{12}-q_{13}-q_{24}-q_{34}-q_{23}~, \nonumber\\
q_{15} &= q_{24}+q_{34}+q_{23}, \nonumber\\
q_{25} &= -q_{12}-q_{23}-q_{24}, \nonumber\\
q_{35} &= -q_{13}-q_{23}-q_{34}, \nonumber\\
q_{45} &= q_{12}+q_{13}+q_{23}~.
\end{align}

\subsection{The 5-point vector amplitude} \label{ss:bbbbb}
Setting up this amplitude is not much more difficult now that we gained experience with the four-point functions.  We fix positions $z_{1,4,5}$ and denote the integration region $\mathfrak{T} = \{ (z_2,z_3) \in \mathbb{R}^2 ~|~ z_1 \leq z_2 \leq z_3 \leq z_4 \}$.  Leaving the flat measure $\ed z_3 \ed z_2$ implicit, we have
\begin{align}
A(1,2,3,4,5) & = \int_{\mathfrak{T}} \langle c_1 V_{v}^{0}[\xi^1,k_1] V_{v}^0[\xi^2,k_2] V_{v}^0[\xi^3,k_3] c_4 V_{v}^{-1}[\xi^4,k_4] c_5 V_{v}^{-1}[\xi^5, k_5] \rangle~,
\end{align}
The correlation function is then given by $\la c_1 c_4 c_5 \ra\DC{2}^5 = -z_{41} z_{51} z_{54} \DC{2}^5$ times the matter correlator
\begin{align}
\label{eq:bigmatter}
&\la \bU[\xi^4] \bU[\xi^5] \ra \cT_5^{(1,2,3)} \nonumber\\[2mm]
&
+\la \bJ[\mm^1]\bU[\xi^4] \bU[\xi^5] \ra \cT_5^{(2,3)}
+\la \bJ[\mm^2]\bU[\xi^4] \bU[\xi^5] \ra \cT_5^{(1,3)}
+\la \bJ[\mm^3]\bU[\xi^4] \bU[\xi^5] \ra \cT_5^{(1,2)}  \nonumber\\[2mm]
&
+\la \bJ[\mm^2]\bJ[\mm^3]\bU[\xi^4] \bU[\xi^5] \ra \cT_5^{(1)}
+\la \bJ[\mm^1]\bJ[\mm^3]\bU[\xi^4] \bU[\xi^5] \ra \cT_5^{(2)}
+\la \bJ[\mm^1]\bJ[\mm^2]\bU[\xi^4] \bU[\xi^5] \ra \cT_5^{(3)}\nonumber\\[2mm]
&
+\la \bJ[\mm^1] \bJ[\mm^2]\bJ[\mm^3] \bU[\xi^4] \bU[\xi^5]\ra \cT_5~.
\end{align}
The computation of these is now fairly mechanical and is reduced to the two-point function $\la \bU[\xi^4]\bU[\xi^5]\ra$ by the Ward identities.  To evaluate these it is extremely convenient to take the $z_5\to \infty$ limit as soon as possible, thereby reducing the number of terms that appear.  This is easily done since the ghost factor scales as $z_5^2$, while
\begin{align}
\cT_{5} &= \DC{3}  \prod_{1\leq i<j<5} z_{ji}^{q_{ji}}+ O(z_5^{-1})~, &
\la \bU[\xi^4] \bU[\xi^5] \ra = -\frac{\omega^{45}}{z_{5}^2} +O(z_5^{-3})~,
\end{align}
so we can just take the leading terms as $z_5\to \infty$ in all of the remaining terms.  As an example, we list the only correlator that is unfamiliar from the four-point computations:
\begin{align}\label{psicor4}
& \lim_{z_5\to\infty} z_{5}^2 \langle \bJ_{1}[\mm^1] \bJ_2[\mm^2] \bJ_3[\mm^3] \bU_4[\xi^4] \bU_5[\xi^5] \rangle \cr
&  = ~^t\xi^5 \bigg\{ 8 \frac{\tr( \mm^{12} \mm^3) }{z_{21} z_{31} z_{32}} + 4 \left( \frac{\tr (\mm^1 \mm^2) \mm^3}{z_{21}^2 z_{43}} + \frac{\tr (\mm^3 \mm^1) \mm^2}{z_{31}^2 z_{42}} + \frac{\tr(\mm^2 \mm^3) \mm^1}{z_{32}^2 z_{41}}  \right)+ \cr
& \quad + 8 \left[ \frac{\mm^1 \mm^2 \mm^3}{z_{21} z_{32} z_{43}} - \frac{\mm^2 \mm^1 \mm^3}{z_{21} z_{31} z_{43}} - \frac{ \mm^3 \mm^1 \mm^2}{z_{31} z_{21} z_{42}} + \frac{\mm^3 \mm^2 \mm^1}{z_{32} z_{21} z_{41}} - \frac{\mm^2 \mm^3 \mm^1}{z_{32} z_{31} z_{41}} - \frac{\mm^1 \mm^3 \mm^2}{z_{31} z_{32} z_{42}} \right] \bigg\} \xi^4~. \cr
\end{align}
As a check of this expression we note that it has the $S_3$ symmetry that permutes the $\bJ[\mm^i]$.

We now fix the remaining coordinates to $z_1 = 0$, $z_2 = x_2$, $z_3 = x_3$, $z_4 = 1$, and after some algebra obtain the amplitude as a sum of $26$ double integrals\footnote{We are not aware of the relation to either the critical bosonic string or to the English alphabet, but the relations, if any, may be at least as deep as those explored in~\cite{Siegel:1987gs}.}
\begin{align}
A(1,2,3,4,5) & = \DC{2}^5 \DC{3} \sum_{I \in \Sigma'}   N[I] \times I~.
\end{align}
With a few simple integral identities these may be reduced to the $19$ integrals presented in~\cite{Barreiro:2005hv}.  We review these $19$ integrals, the $17$ relations among them, as well as their momentum expansion in~Appendix~\ref{app:integrals}.~  For now, suffice it to to say that the index set for them is
\begin{align}\label{LsandKs}
\Sigma =  \{K_1,~\cdots,K_6,~K'_1,~K'_4,~K'_5,~L_1,~\cdots,L_7,~L'_1,~L'_3,~L'_4\}~,
\end{align}
and a typical integral is
\begin{align}
K_3 & = \int^1_0 \ed x_3 \int^{x_3}_0 \ed x_2 \,
x_3^{q_{13}} (1-x_3)^{q_{34}}x_2^{q_{12}} (1-x_2)^{q_{24}} (x_3-x_2)^{q_{23}} \times\frac{1}{(1-x_2) x_3} \nonumber\\
& = \int^1_0 \ed y_1 \int^1_0 \ed y_2~
y_1^{q_{45}} (1-y_1)^{q_{34}} y_2^{q_{12}} (1-y_2)^{q_{23}}(1-y_1y_2)^{q_{24}-1}~.
\end{align}
To obtain the last form we used $q_{13}+q_{12}+q_{23} = q_{45}$ and the change of variables $x_3 = y_1$ and $x_2 = y_1 y_2$.  This integral is finite as $q_{ij}\to 0$, and it may be expanded in powers of $q$, where the coefficients reduce to integrals of standard power series.  We obtain
\begin{align}
\label{eq:K3expand}
K_3 &  = \frac{\pi^2}{6} -\zeta(3) \left[ q_{12} + 2 q_{23} + q_{24} +2 q_{34} + q_{45} \right] + O (q^2)
\nonumber\\
& =
\frac{\pi^2}{6} -\zeta(3) \left[ q_{12} + q_{23} + q_{34} + q_{45} + q_{51} \right] + O (q^2)~.
\end{align}
In the second line we used kinematic relations to write the answer in an explicitly cyclic-invariant form.  In fact, as shown in~\cite{Barreiro:2005hv}, the full $K_3$ integral is $\Z_5$-invariant.  The remaining $18$ integrals can be reduced to $K_3$ and another cyclically-invariant and regular combination
\begin{align}
T = q_{12} q_{34} K_2 + (q_{51} q_{12} -q_{12} q_{34} + q_{34} q_{45}) K_3~,
\end{align}
which has the momentum expansion
\begin{align}
\label{eq:Texpand}
T = 1 -\zeta(3)\left[
q_{12} q_{23}q_{34} + q_{23} q_{34} q_{45} + q_{34} q_{45} q_{51} + q_{45} q_{51} q_{12} + q_{51} q_{12} q_{23}\right]+O(q^4)~.
\end{align}

Using these integrals, we find the following structure for the amplitude:
\begin{align}\label{5vecres}
A(1,2,3,4,5) & = \DC{2}^5 \DC{3} \times\left[ \cM_1 + \cM_2 + \cM_3 + \cM_4 + \cM_5\right]~,
\end{align}
where
\begin{align}\label{ourM1}
\cM_1 =&~ \omega^{12} \omega^{34} \left[ \Delta_{3}^5 (q_{12} - 1) L_5 - \Delta_{2}^5 q_{13} L_3 + \Delta_{1}^5 q_{23} L_2 \right] + \cr
&~ + \omega^{13} \omega^{24} \left[ \Delta_{2}^5 (q_{13} -1) L_6 - \Delta_{3}^5 q_{12} L_1 - \Delta_{1}^5 q_{23} L_7 \right] + \cr
&~ + \omega^{14} \omega^{23} \left[ \Delta_{1}^5 (q_{23} - 1) K_6 + \Delta_{3}^5 q_{12} K_4 - \Delta_{2}^5 q_{13} K_5 \right] + \cr
&~ + \omega^{23} \bigg\{ \Delta_{1}^5 \left[ \Delta_{2}^1 \Delta_{3}^4 L_2 - \Delta_{3}^1 \Delta_{2}^4 L_7 \right] + \cr
&~ \qquad \quad + \Delta_{2}^5 \left[ \Delta_{2}^1 \Delta_{3}^4 L_2 + \Delta_{3}^1 \Delta_{1}^4 K_5 + \Delta_{3}^1 \Delta_{3}^4 L_{4}' + \Delta_{4}^1 \Delta_{3}^4 K_{4}' \right] + \cr
&~ \qquad \quad + \Delta_{3}^5 \left[ - \Delta_{2}^1 \Delta_{1}^4 K_4 - \Delta_{3}^1 \Delta_{2}^4 L_7 - \Delta_{2}^1 \Delta_{2}^4 L_4 - \Delta_{4}^1 \Delta_{2}^4 K_{5}' \right] \bigg\} + \cr
&~ + \omega^{14} \bigg\{ \Delta_{2}^5 \left[ \Delta_{1}^2 \Delta_{4}^3 K_2 - \Delta_{4}^2 \Delta_{1}^3 K_3 \right] + \cr
&~ \qquad \quad + \Delta_{1}^5 \left[ \Delta_{1}^2 \Delta_{4}^3 K_2 - \Delta_{3}^2 \Delta_{4}^3 K_{4}' + \Delta_{4}^2 \Delta_{2}^3 K_{5}' - \Delta_{4}^2 \Delta_{4}^3 K_{1}' \right] + \cr
&~ \qquad \quad + \Delta_{4}^5 \left[  \Delta_{1}^2 \Delta_{2}^3 K_4 + \Delta_{1}^2 \Delta_{1}^3 K_1 - \Delta_{4}^2 \Delta_{1}^3 K_3 - \Delta_{3}^2 \Delta_{1}^3 K_5 \right] \bigg\}~,
\end{align}
and the remaining $\cM_{2},\ldots, \cM_{5}$ are obtained by cyclic permutations of $\cM_1$.  This matches~(4.1) of~\cite{Barreiro:2005hv}.  By using the relations among the integrals reviewed in Appendix \ref{app:integrals}, one can express the result in terms of just two members of the set \eqref{LsandKs}.  A convenient basis is $T$ and $K_3$, as these integrals are $\mathbbm{Z}_5$-invariant, resulting in an expression of the form
\begin{equation}\label{TK3form}
A(1,2,3,4,5) = A^{(i)}(1,2,3,4,5) T + A^{(ii)}(1,2,3,4,5) K_3~.
\end{equation}

This form of the amplitude was explored further in~\cite{Barreiro:2005hv}.  It was shown that the coefficient of $T$ is precisely the Yang--Mills five-point color-ordered amplitude, while the coefficient of $K_3$ includes poles that use the four-vector vertex in~(\ref{eq:newvertices4}), as well as a regular term.  The former originate from the $F^4$ terms in the low-energy effective action while the latter receives contributions from both the $F^4$ and $F^5$ terms of the action.

\subsection{Two fermions and three gauge bosons}
We now finally turn to a new computation, the string amplitude $A(\tilde{1},2,3,4,\tilde{5})$.  From the basic form
\begin{align}
A(\tilde{1},2,3,4,\tilde{5}) & = \int_{\mathfrak{T}} \langle c_1 V_{g}^{-1/2}[\zeta^1,k_1] V_{v}^0[\xi^2,k_2] V_{v}^0[\xi^3,k_3] c_4 V_{v}^{-1}[\xi^4,k_4] c_5 V_{g}^{-1/2}[\zeta^5, k_5] \rangle~,
\end{align}
we arrive at
\begin{align}\label{Afbbbf1}
A(\tilde{1},2,3,4,\tilde{5}) &= \frac{\DC{1}^2 \DC{2}^3}{\sqrt{2}} \int_{\mathfrak{T}} \la c_1 c_4c_5\ra \times
\left[X \cT_5^{(2,3)}+ Y_2\cT_5^{(3)} + Y_3\cT_5^{(2)} + Z\cT_5\right]~,
\end{align}
where
\begin{align}\label{XYZdefs}
X & \equiv \sqrt{2}\la \bV[\zeta^1] \bU[\xi^4]\bV[\zeta^5]\ra ~,&
Y_2 & \equiv \sqrt{2}\la \bV[\zeta^1] \bJ[\mm^2] \bU[\xi^4]\bV[\zeta^5]\ra ~,\nonumber\\[3mm]
Y_3 & \equiv \sqrt{2}\la \bV[\zeta^1] \bJ[\mm^3] \bU[\xi^4]\bV[\zeta^5]\ra~,&
Z & \equiv \sqrt{2}\la \bV[\zeta^1]\bJ[\mm^2]\bJ[\mm^3] \bU[\xi^4] \bV[\zeta^5]\ra~.
\end{align}
Of these the only new correlator is $Z$.  By applying the Ward identity first with $\bJ[\mm^2]$ and then with $\bJ[\mm^3]$ we obtain the result
\begin{align}\label{bigZcor}
\frac{Z}{\sqrt{2}} =&~  \frac{1}{4z_{21} z_{31}} \langle \bV_1[\mms^3 \mms^2 \zeta^1] \bU[\xi^4] \bV[\zeta^5]\rangle + \frac{1}{z_{21} z_{34}} \langle \bV_1[\mms^2 \zeta^1] \bU_4[\mm^3\xi^4] \bV[\zeta^5]\rangle + \cr
&~ +  \frac{1}{4z_{21} z_{35}} \langle \bV_1[\mms^2 \zeta^1] \bU[\xi^4] \bV_5[\mms^3\zeta^5]\rangle + \frac{2 \tr(\mm^2\mm^3)}{z_{23}^2} \langle \bV[\zeta^1] \bU[\xi^4] \bV[\zeta^5]\rangle + \cr
&~ + \frac{2}{z_{23} z_{31}} \langle \bV_1[\mms^{23} \zeta^1] \bU[\xi^4] \bV[\zeta^5]\rangle + \frac{8}{z_{23} z_{34}} \langle \bV[\zeta^1] \bU_4[\mm^{23}\xi^4] \bV[\zeta^5]\rangle + \cr
&~ + \frac{2}{z_{23} z_{35}} \langle \bV[\zeta^1] \bU[\xi^4] \bV_5[\mms^{23}\zeta^5]\rangle + \frac{1}{z_{24} z_{31}} \langle \bV_1[\mms^3\zeta^1] \bU_4[\mm^2\xi^4] \bV[\zeta^5]\rangle + \cr
&~ + \frac{4}{z_{24} z_{34}} \langle \bV[\zeta^1] \bU_4[\mm^3\mm^2\xi^4] \bV[\zeta^5]\rangle + \frac{1}{z_{24} z_{35}} \langle \bV[\zeta^1] \bU_4[\mm^2\xi^4] \bV_5[\mms^3\zeta^5]\rangle + \cr
&~ + \frac{1}{4z_{25} z_{31}} \langle \bV_1[\mms^3\zeta^1] \bU[\xi^4] \bV_5[\mms^2\zeta^5]\rangle + \frac{1}{z_{25} z_{34}} \langle \bV[\zeta^1] \bU_4[\mm^3\xi^4] \bV_5[\mms^2\zeta^5]\rangle + \cr
&~ + \frac{1}{4z_{25} z_{35}} \langle \bV[\zeta^1] \bU[\xi^4] \bV_5[\mms^3\mms^2 \zeta^5] \rangle~.
\end{align}
In fact it follows from the definition of $Z$, \eqref{XYZdefs}, that this result should be symmetric under the exchange of labels $2$ and $3$.  Although it takes a little work, it can be shown that this is indeed the case.

Every term in \eqref{bigZcor} is reduced to a three-point correlator, all of which have the same denominator structure as $X$.  Similarly, the Ward identity reduces $Y_{2,3}$ to a sum of terms with the same denominator structure as $X$---namely, every term scales as $(z_{14} z_{15} z_{45})^{-1}$, precisely the ghost amplitude.  So, we can once again easily take the $z_5 \to \infty$ limit, keeping the leading terms in the matter correlators.  These are the terms in \eqref{bigZcor}, for example, that do not already have explicit $z_{5}$'s in the prefactors of the three-point correlators.  This reduces the thirteen terms of \eqref{bigZcor} to seven.

Letting $\xis^{[i} \xis^{j]} \equiv \xi_{\mu}^i \xi_{\nu}^j \Gamma^{\mu\nu}$ and $\kiss_{[i} \kiss_{j]} \equiv k_{i}^\mu k_{j}^{\nu} \Gamma_{\mu\nu}$, we obtain\footnote{Here we exchanged labels 2 and 3 in \eqref{bigZcor} and then took the limit $z_5 \to \infty$.}
\begin{align}\label{limitZcor}
 \lim_{z_5\to\infty} z_{5}^2 Z & =
\frac{ 2\mm^{2\mu\nu}\mm^{3}_{\mu\nu} \zetab^5 \xis^4 \zeta^1}{z_{32}^2}
%%%
-\frac{\ff{1}{2}\zetab^5\xis^4 \left[ \Delta^3_2\kiss_3 \xis^2 +\omega^{23} \kiss_{[2} \kiss_{3]}+q_{23}\xis^{[2} \xis^{3]}  -\Delta^2_3 \kiss_2 \xis^3
\right]\zeta^1}{z_{32} z_{21}}  \nonumber\\
%%%%%%%%%%
%%%%%%%%%%
&\quad +\frac{ \zetab^5\left[
(\Delta^3_2 \omega^{24}-\Delta^4_2 \omega^{32}) \kiss_3+(\Delta^2_3\Delta^4_2-q_{32}\omega^{24}) \xis^3 -(2\leftrightarrow3)\right]\zeta^1}{z_{42}z_{32} }\nonumber\\
%%%%%%%%%%%%
&\quad
-\frac{\ff{1}{4}\zetab^5\xis^4 \mms^2\mms^3 \zeta^1}{ z_{31} z_{21}}
+\frac{\ff{1}{2}\zetab^5(-\Delta^4_2\xis^2+\omega^{24} \kiss_2)\mms^3\zeta^1}{z_{42}z_{31}}
-\frac{ \ff{1}{2}\zetab^5 (-\omega^{34}\kiss_3+\Delta^4_3\xis^3)\mms^2\zeta^1}{z_{43} z_{21} }\nonumber\\
%%%%%%%%%
%%%%%%%%%
&\quad
-\frac{\zetab^5[ (\omega^{34} \Delta^2_3 -\omega^{23}\Delta^4_3) \kiss_2+(\Delta^4_3\Delta^3_{2}-\omega^{34} q_{23}) \xis^2]\zeta^1}{z_{43}z_{42}}~.
\end{align}
Fixing positions as in the five-vector amplitude and combining the terms from the correlators we find
\begin{align}
A(\tilde{1},2,3,4,\tilde{5})  =  \sum_{I \in \Sigma_K} N[I] I~,
\end{align}
where the $9$ $K$ integrals in $\Sigma_K$ are of the form
\begin{align}\label{Kintegrals}
I =  \int^1_0 \ed x_3 \int^{x_3}_0 \ed x_2
x_3^{q_{13}} (1-x_3)^{q_{34}}x_2^{q_{12}} (1-x_2)^{q_{24}} (x_3-x_2)^{q_{23}} \frac{1}{ \kappa(I)}~,
\end{align}
and the denominators $\kappa(I)$, together with the contributing correlator, are listed in table~\ref{tab:ourK}.
\begin{table}[t!]
\centering
\begin{tabular}{c c c c c}
denominator $\kappa(I)$ & with $z_{1,4} = 0,1$ & name $I$ &  location$(fbbbf)$ & location$(fbbfb)$ \\[2mm]
%\hline
%\hline
  $z_{21} z_{31}$ & $x_2 x_3$ & $K_1$  &   $X$, $Y_2$, $Y_3$, $Z$  & $X'$, $Y_{2}'$, $Y_{3}'$, $Z'$  \\[2mm]
%\hline
 $z_{21} z_{43}$ & $x_2 (1-x_3)$ & $K_2$  & $X$, $Y_2$, $Y_3$, $Z$  & $X'$, $Y_{2}'$, $Y_{3}'$, $Z'$  \\[2mm]
%\hline
 $z_{31} z_{42}$ & $(1-x_2) x_3$ & $K_3$ &   $X$, $Y_2$, $Y_3$, $Z$  &  $X'$, $Y_{2}'$, $Y_{3}'$, $Z'$ \\[2mm]
%\hline
$z_{21} z_{32}$ & $x_2 (x_3-x_2)$ & $K_4$   &    $X$, $Y_2$, $Z$  & $X'$, $Y_{2}'$ \\[2mm]
%\hline
$z_{31} z_{32}$ & $x_{3} (x_3-x_2)$ & $K_5$  &    $X$, $Y_3$  & $X'$, $Y_{3}'$, $Z'$   \\[2mm]
%\hline
$z_{32}^2$ & $(x_3-x_2)^2$ & $K_6$    &    $X$, $Z$  & $X'$, $Z'$  \\[2mm]
%\hline
$z_{42} z_{43}$ & $(1-x_2) (1-x_3)$ & $K_1'$  &   $X$, $Y_2$, $Y_3$, $Z$  & $X'$, $Y_{2}'$, $Y_{3}'$, $Z'$  \\[2mm]
%\hline
 $z_{32} z_{43}$ & $(1-x_3) (x_3-x_2)$ & $K_4'$ &   $X$, $Y_3$  &  $X'$, $Y_{3}'$, $Z'$   \\[2mm]
%\hline
$z_{32} z_{42}$ & $(1-x_2) (x_3-x_2)$ & $K_{5}'$ &    $X$, $Y_2$, $Z$  & $X'$, $Y_{2}'$ \\[2mm]
\end{tabular}
\caption{The type $K$ denominators.  The fourth column lists the terms of \eqref{Afbbbf1} that contribute to the corresponding numerator.  The fifth column does the same for the second ordering of the amplitude, computed in subsection \ref{ssec:fbbfb} below; see \eqref{Afbbfb1}.}
\label{tab:ourK}
\end{table}

Using the determined values of the $\DC{i}$, the $9$ numerator factors are as follows:
\begin{align}
 N[K_1] &= ig^3
 \zetab^5 \xis^4 (\Delta^2_1+\ff{1}{2}\mms^2)(\Delta^3_1 +\ff{1}{2}\mms^3)\zeta^1~,\nonumber\\[2mm]
 %%%%%%%%%%%
 %%%%%%%%%%%
 N[K_2] &=ig^3 \zetab^5\left[\omega^{43}\kiss_3 +\Delta^3_4\xis^4-\Delta^4_3 \xis^3 \right](-\Delta^2_1-\ff{1}{2}\mms^2) \zeta^1~,\nonumber\\[2mm]
%%%%%%%%%%%%%
%%%%%%%%%%%%%
N[K_3]&= ig^3 \zetab^5 \left[\omega^{42}\kiss_2 +\Delta^2_4\xis^4-\Delta^4_2 \xis^2 \right](-\Delta^3_1-\ff{1}{2}\mms^3)\zeta^1 ~,\nonumber\\[2mm]
%%%%%%%%%%%%%
%%%%%%%%%%%%%
N[K_4] &= ig^3 \zetab^5\xis^4\left[\Delta^2_1\Delta^3_2 +\ff{1}{2}(\Delta^3_2 \mms^2
+ \Delta^3_2\kiss_3 \xis^2 +\omega^{23} \kiss_{[2} \kiss_{3]}+q_{23}\xis^{[2} \xis^{3]}  -\Delta^2_3 \kiss_2 \xis^3 )
\right]\zeta^1~, \nonumber\\[2mm]
%%%%%%%%%%%%
%%%%%%%%%%%%
N[K_5] & = ig^3 \zetab^5\xis^4\Delta^2_3\left[-\Delta^3_1-\ff{1}{2}\mms^3
\right]\zeta^1 ~,\nonumber\\[2mm]
%%%%%%%%%%%%%
%%%%%%%%%%%%%
N[K_6] &= ig^3 (1-q_{23})\omega^{23}\zetab^5 \xis^4\zeta^1~,
%%%%%%%%%%
%%%%%%%%%%
\end{align}

and
\begin{align}
N[K'_1] &= ig^3\zetab^5 \left[ \Delta^2_4 \Delta^3_4\xis^4
+\Delta^3_4(\omega^{42}\kiss_2 -\Delta^4_2 \xis^2 )
+\Delta^2_4(\omega^{43}\kiss_3 -\Delta^4_3 \xis^3 ) \right. \nonumber\\
& \qquad\qquad \left.+(\omega^{34} \Delta^2_3 -\omega^{23}\Delta^4_3) \kiss_2+(\Delta^4_3\Delta^3_{2}-\omega^{34} q_{23}) \xis^2
\right]\zeta^1~, \nonumber\\[2mm]
%%%%%%%%%%%%%
%%%%%%%%%%%%%
N[K'_4] &= ig^3\zetab^5\left[\Delta^2_3\Delta^3_4\xis^4
{+}\Delta^2_3(\omega^{43}\kiss_3 -\Delta^4_3 \xis^3 )\right] \zeta^1~, \nonumber\\[2mm]
%%%%%%%%%%%%%%
%%%%%%%%%%%%%
N[K'_5] &= ig^3\zetab^5\left[-\Delta^2_4 \Delta^3_2\xis^4
-\Delta^3_2(\omega^{42}\kiss_2 -\Delta^4_2 \xis^2 ) \right.\nonumber\\[2mm]
&  \qquad\qquad \left.
-[(\Delta^3_2 \omega^{24}-\Delta^4_2 \omega^{32}) \kiss_3+(\Delta^2_3\Delta^4_2-q_{32}\omega^{24}) \xis^3 -(2\leftrightarrow3)]\right]\zeta^1~~.
%%%%%%%%%%%%%%%
%%%%%%%%%%%%%%%
\end{align}
Finally, using~(\ref{eq:reducedK}) we can express the amplitude in the form
\begin{align}\label{ftform}
A(\tilde{1},2,3,4,\tilde{5}) = A^{(i)}(\tilde{1},2,3,4,\tilde{5}) T + A^{(ii)}(\tilde{1},2,3,4,\tilde{5}) K_3~.
\end{align}
As we will argue next, in parallel to the discussion for the five-vector amplitude given in~\cite{Barreiro:2005hv}, $A^{(i)}$ is precisely the super-Yang--Mills amplitude, while $A^{(ii)}$ arises from the fermionic couplings that accompany the $\tr F^4$ couplings in the open string effective action.

\subsection{Comparison with super-Yang--Mills}\label{sec:SYMcomp1}
From~(\ref{eq:reducedK}), we obtain the following form for $A^{(i)}$:
\begin{align}\label{AYMstrings}
A^{(i)} &=
\frac{N[K_1]+ N[K_4]}{q_{12} q_{45}}
+
\frac{N[K_2]}{q_{12} q_{34}}
+
\frac{N[K_5]+N[K_4]+q_{13} N[K_6]'}{q_{23} q_{45}} \nonumber\\
&
\qquad\qquad+
\frac{N[K'_5]+N[K'_4]-q_{34} N[K_6]' }{q_{15} q_{23}}
+
\frac{N[K'_1]+N[K'_4]}{q_{15}q_{34}}
-\frac{N[K_6]'}{q_{15}}~,
\end{align}
where $N[K_6]' = N[K_6] / (1-q_{23})$.  The displayed pole structure is just right to match the Yang--Mills computation, where we have the following diagrams for $\AYM(\tilde{1},2,3,4,\tilde{5})$:
\begin{align}
\label{eq:YMpolesdiagrams}
\xymatrix@R=4mm@C=15mm{
\begin{xy} <1.0mm,0mm>:
(10,-7)*{D_1,\quad\frac{1}{q_{15}}},
(0,0)*{5}="e5",
(10,0)*{\bullet}="v1",
(20,0)*{1}="e1",
(10,7)*{\bullet}="v2",
(0,7)*{4}="e4",
(10,14)*{3}="e3",
(20,7)*{2}="e2"
\ar@{-}|{<}"e5";"v1"
\ar@{-}|{<}"v1";"e1"
\ar@{~}"v1";"v2"
\ar@{~}"v2";"e2"
\ar@{~}"v2";"e3"
\ar@{~}"v2";"e4"
\end{xy} &
%%%%%%%%%%%%%
%%%%%%%%%%%%%
\begin{xy} <1.0mm,0mm>:
(10,-7)*{D_2,\quad\frac{1}{q_{15}q_{34}}},
(0,0)*{5}="e5",
(10,0)*{\bullet}="v1",
(20,0)*{1}="e1",
(10,14)*{\bullet}="v3",
(10,7)*{\bullet}="v2",
(0,14)*{4}="e4",
(20,14)*{3}="e3",
(20,7)*{2}="e2"
\ar@{-}|{<}"e5";"v1"
\ar@{-}|{<}"v1";"e1"
\ar@{~}"v1";"v2"
\ar@{~}"v2";"e2"
\ar@{~}"v3";"e3"
\ar@{~}"v3";"e4"
\ar@{~}"v2";"v3"
\end{xy}  &
%%%%%%%%%%%%%%
%%%%%%%%%%%%%%
\begin{xy} <1.0mm,0mm>:
(10,-7)*{D_3,\quad\frac{1}{q_{15}q_{23}}},
(0,0)*{5}="e5",
(10,0)*{\bullet}="v1",
(20,0)*{1}="e1",
(10,14)*{\bullet}="v3",
(10,7)*{\bullet}="v2",
(0,7)*{4}="e4",
(0,14)*{3}="e3",
(20,14)*{2}="e2"
\ar@{-}|{<}"e5";"v1"
\ar@{-}|{<}"v1";"e1"
\ar@{~}"v1";"v2"
\ar@{~}"v2";"e4"
\ar@{~}"v3";"e3"
\ar@{~}"v3";"e2"
\ar@{~}"v2";"v3"
\end{xy} \\
%%%%%%%%%%%%%%
%%%%%%%%%%%%%%
%%%%%%%%%%%%%%
\begin{xy} <1.0mm,0mm>:
(15,-7)*{D_4,\quad\frac{1}{q_{45} q_{23}}},
(0,0)*{5}="e5",
(10,0)*{\bullet}="v1",
(20,0)*{\bullet}="v2",
(30,0)*{1}="e1",
(5,10)*{4}="e4",
(20,7)*{\bullet}="v3",
(15,14)*{3}="e3",
(25,14)*{2}="e2",
\ar@{-}|{<}"e5";"v1"
\ar@{-}|{<}"v1";"v2"
\ar@{-}|{<}"v2";"e1"
\ar@{~}"v1";"e4"
\ar@{~}"v2";"v3"
\ar@{~}"v3";"e3"
\ar@{~}"v3";"e2"
\end{xy} &
%%%%%%%%%%%%%%
%%%%%%%%%%%%%%
\begin{xy} <1.0mm,0mm>:
(15,-7)*{D_5,\quad\frac{1}{q_{12} q_{34}}},
(0,0)*{5}="e5",
(10,0)*{\bullet}="v1",
(20,0)*{\bullet}="v2",
(30,0)*{1}="e1",
(25,10)*{2}="e2",
(10,7)*{\bullet}="v3",
(5,14)*{4}="e4",
(15,14)*{3}="e3",
\ar@{-}|{<}"e5";"v1"
\ar@{-}|{<}"v1";"v2"
\ar@{-}|{<}"v2";"e1"
\ar@{~}"v3";"e4"
\ar@{~}"v1";"v3"
\ar@{~}"v3";"e3"
\ar@{~}"v2";"e2"
\end{xy} &
%%%%%%%%%%%%%%
%%%%%%%%%%%%%%
\begin{xy} <1.0mm,0mm>:
(20,-7)*{D_6,\quad\frac{1}{q_{45} q_{12}}},
(0,0)*{5}="e5",
(10,0)*{\bullet}="v1",
(20,0)*{\bullet}="v2",
(30,0)*{\bullet}="v3",
(40,0)*{1}="e1",
(10,9)*{4}="e4",
(20,9)*{3}="e3",
(30,9)*{2}="e2",
\ar@{-}|{<}"e5";"v1"
\ar@{-}|{<}"v1";"v2"
\ar@{-}|{<}"v2";"v3"
\ar@{-}|{<}"v3";"e1"
\ar@{~}"v1";"e4"
\ar@{~}"v2";"e3"
\ar@{~}"v3";"e2"
\end{xy}
}
\end{align}
Of course there is no reason for the terms to line up as fortuitously as these naive poles suggest.  The numerators in \eqref{AYMstrings} are generally cubic in the momenta of the external particles and thus can, and do, contain terms in which one of the two $q_{ij}$ factors in the denominator is canceled.  Such terms can contribute to $D_1$ or cancel against each other.  It is only the sum of terms on the right of \eqref{AYMstrings} that must match the sum of diagrams, and indeed it does.  To demonstrate how this comes about, we will first compare $D_5$ with the $N[K_2]$ term in $A^{(i)}$.  Using our rules we have
\begin{align}
D_5 = -ig^3 \frac{1}{4q_{12}q_{34}} \times \zetab^5
 \left[\omega^{34}\kiss_{34} +2 \Delta^3_4\xis^4 -2\Delta^4_3 \xis^3 \right] (\kiss_1+\kiss_2) \xis^2 \zeta^1~.
\end{align}
Since $\zetab^5\kiss_5 = 0$ and $(\kiss_1+\kiss_2)^2 = 2q_{12}$, we may write this as
\begin{align}
D_5 = -ig^3 \frac{1}{2q_{12}q_{34}} \times \zetab^5
 \left[\omega^{34}\kiss_{3} + \Delta^3_4\xis^4 -\Delta^4_3 \xis^3 \right] (\kiss_1+\kiss_2) \xis^2 \zeta^1
 - ig^3\frac{1}{2q_{34}} \times \zetab^5 \xis^2\zeta^1~.
\end{align}
On the other hand, we also have the simple relation
\begin{align}
(\Delta^i_1 + \ff{1}{2}\mms^i) \zeta^1 =\frac{1}{2} (\kiss_1\xis^i+\xis^i\kiss_1 + \kiss_i\xis^i)\zeta^1= \frac{1}{2} (\kiss_1 + \kiss_i)\xis^i\zeta^1~,
\end{align}
and this allows us to write
\begin{align}
N[K_2] & = -\frac{ig^3}{2 q_{12} q_{34}} \zetab^5 \left[\omega^{34} \kiss_3 + \Delta^3_4\xis^4 -\Delta^4_3\xis^3\right] (\kiss_1+\kiss_2) \xis^2\zeta^1~,
\end{align}
so that evidently
\begin{align}
N[K_2] & = D_5 + \frac{ig^3 \omega^{34}\zetab^5\xis^2\zeta^1}{q_{34}}~.
\end{align}
By similar manipulations, we obtain the following relations
\begin{align}
\label{eq:YMresults}
0& = D_1-\frac{ig^3}{2q_{15}} \zetab^5\left[2\omega^{24}\xis^3 -\omega^{34}\xis^2 -\omega^{23}\xis^4\right]\zeta^1~,\nonumber\\[2mm]
%%%%%%%%%
%%%%%%%%%
\frac{(N[K'_4]+N[K'_1])}{q_{15}q_{34}} & =D_2
- \frac{ig^3}{2q_{34}} \zetab^5 \xis^2\zeta^1 \omega^{34}
+\frac{ig^3}{2 q_{15}} \zetab^5 \xis^2 \zeta^1 \omega^{34}~,\nonumber\\[2mm]
%%%%%%%%%
%%%%%%%%%
\frac{(N[K'_4]+N[K'_5]-q_{34}N[K_6]')}{q_{15}q_{23}} & = D_3
-\frac{ig^3}{2q_{23}} \zetab^5 \xis^4\zeta^1 \omega^{23}
+\frac{ig^3}{2q_{15}} \zetab^5\left[  \omega^{23}\xis^4+2\omega^{24} \xis^3 -2\omega^{34} \xis^2\right]\zeta^1~,\nonumber\\[2mm]
%%%%%%%%
 \frac{N[K_4]+N[K_5]+q_{13}N[K_6]'}{q_{23}q_{45}} & = D_4
- ig^3 \frac{\zetab^5\xis^4\xis^3 \xis^2\zeta^1}{2q_{45}}
+ig^3 \frac{\omega^{23} \zetab^5 \xis^4 \zeta^1}{2q_{23}}~,\nonumber\\
%%%%%%%%
 \frac{N[K_2]}{q_{12}q_{34}} &= D_5
+\frac{i g^3}{2q_{34}} \zetab^5 \xis^2 \zeta^1  \omega^{34}~,\nonumber\\[2mm]
%%%%%%%%%%
 \frac{N[K_1]+N[K_4]}{q_{12} q_{45}} &= D_6 + ig^3 \frac{\zetab^5 \xis^4\xis^3\xis^2 \zeta^1}{2 q_{45}}~,\nonumber\\[2mm]
 -\frac{N[K_6]'}{q_{15}} &= -\frac{i g^3}{q_{15}} \zetab^5 \xis^4 \zeta^1 \omega^{23}~.
\end{align}
Taking the sum, we therefore obtain the promised
\begin{align}
A^{(i)}(\tilde{1},2,3,4,\tilde{5}) = \sum_{i=1}^6 D_i = \AYM(\tilde{1},2,3,4,\tilde{5})~.
\end{align}
This is of course precisely in line with the results of~\cite{Barreiro:2005hv} and expectations from space-time supersymmetry.

We can also extend these observations to the coefficient of the K3 integral.  We expect these terms to arise from the superpersymmetric completion of the $F^4$ term in the effective Lagrangian.  Using the vertices from~(\ref{eq:newvertices4}), we already expect a number of single pole contributions such as~(\ref{eq:reducedK}) would suggest, as well as a local vertex.  That is, we express $A^{(ii)}$ as
\begin{align}
A^{(ii)} = & =
\frac{1}{q_{15}} B_{15} +
\frac{1}{q_{23}} B_{23} +
\frac{1}{q_{45}} B_{45} +
\frac{1}{q_{12}} B_{12} +
\frac{1}{q_{34}} B_{34} +
B_0~,
\end{align}
where
\begin{align}
B_{34} & = - q_{12}q_{15}q_{34}(D_2+D_5)~,\nonumber\\
B_{12} & = -q_{12}q_{34}q_{45}(D_5+D_6)~,\nonumber\\
B_{45} & = -q_{12}q_{23}q_{45}(D_4+D_6)~,\nonumber\\
B_{23} & = -q_{15}q_{23}q_{45}(D_3+D_4)~,\nonumber\\
B_{15} & = -q_{15}q_{23}q_{34} (D_1+D_2+D_3)~, \nonumber\\
B_0 & = N[K_3] - q_{15} q_{45} D_1 +q_{15}q_{23}(D_1+D_3)+q_{12}q_{34} D_5 + q_{23} q_{45} D_4 \nonumber\\
&\qquad -\ff{1}{2} ig^3 (q_{13}+q_{24}) \zetab^5\xis^4\zeta^1
+\ff{1}{2} q_{24} ig^3\zetab^5 \xis^4\xis^3\xis^2\zeta^1~,
\end{align}
are all polynomial in momenta and can be given the following diagrammatic interpretation:
\begin{align}
\label{eq:F4polesdiagrams}
\xymatrix@R=4mm@C=15mm{
\begin{xy} <1.0mm,0mm>:
(10,-7)*{B_{15},\quad 1/q_{15}},
(0,0)*{5}="e5",
(10,0)*{\bullet}="v1",
(20,0)*{1}="e1",
(10,7)*{\blacksquare}="v2",
(0,7)*{4}="e4",
(10,14)*{3}="e3",
(20,7)*{2}="e2"
\ar@{-}|{<}"e5";"v1"
\ar@{-}|{<}"v1";"e1"
\ar@{~}"v1";"v2"
\ar@{~}"v2";"e2"
\ar@{~}"v2";"e3"
\ar@{~}"v2";"e4"
\end{xy} &
%%%%%%%%%%%%%
%%%%%%%%%%%%%
\begin{xy} <1.0mm,0mm>:
(10,-7)*{B_{34},\quad 1/q_{34}},
(0,0)*{5}="e5",
(10,0)*{\blacksquare}="v1",
(20,0)*{1}="e1",
%(10,14)*{\bullet}="v3",
(10,7)*{\bullet}="v2",
(0,14)*{4}="e4",
(20,14)*{3}="e3",
(20,7)*{2}="e2"
\ar@{-}|{<}"e5";"v1"
\ar@{-}|{<}"v1";"e1"
\ar@{~}"v1";"v2"
\ar@{~}"v1";"e2"
\ar@{~}"v2";"e3"
\ar@{~}"v2";"e4"
%\ar@{~}"v2";"v3"
\end{xy}  &
%%%%%%%%%%%%%%
%%%%%%%%%%%%%%
\begin{xy} <1.0mm,0mm>:
(10,-7)*{B_{23},\quad 1/q_{23}},
(0,0)*{5}="e5",
(10,0)*{\blacksquare}="v1",
(20,0)*{1}="e1",
%(10,14)*{\bullet}="v3",
(10,7)*{\bullet}="v2",
(0,7)*{4}="e4",
(0,14)*{3}="e3",
(20,14)*{2}="e2"
\ar@{-}|{<}"e5";"v1"
\ar@{-}|{<}"v1";"e1"
\ar@{~}"v1";"v2"
\ar@{~}"v1";"e4"
\ar@{~}"v2";"e3"
\ar@{~}"v2";"e2"
%\ar@{~}"v2";"v3"
\end{xy} \\
%%%%%%%%%%%%%%
%%%%%%%%%%%%%%
%%%%%%%%%%%%%%
\begin{xy} <1.0mm,0mm>:
(15,-7)*{B_{45},\quad 1/q_{45}},
(0,0)*{5}="e5",
(10,0)*{\bullet}="v1",
(20,0)*{\blacksquare}="v2",
(30,0)*{1}="e1",
(5,10)*{4}="e4",
%(20,7)*{\bullet}="v3",
(15,10)*{3}="e3",
(25,10)*{2}="e2",
\ar@{-}|{<}"e5";"v1"
\ar@{-}|{<}"v1";"v2"
\ar@{-}|{<}"v2";"e1"
\ar@{~}"v1";"e4"
%\ar@{~}"v2";"v3"
\ar@{~}"v2";"e3"
\ar@{~}"v2";"e2"
\end{xy} &
%%%%%%%%%%%%%%
%%%%%%%%%%%%%%
\begin{xy} <1.0mm,0mm>:
(15,-7)*{B_{12},\quad 1/q_{12}},
(0,0)*{5}="e5",
(10,0)*{\blacksquare}="v1",
(20,0)*{\bullet}="v2",
(30,0)*{1}="e1",
(25,10)*{2}="e2",
%(10,7)*{\bullet}="v3",
(5,10)*{4}="e4",
(15,10)*{3}="e3",
\ar@{-}|{<}"e5";"v1"
\ar@{-}|{<}"v1";"v2"
\ar@{-}|{<}"v2";"e1"
\ar@{~}"v1";"e4"
%\ar@{~}"v1";"v3"
\ar@{~}"v1";"e3"
\ar@{~}"v2";"e2"
\end{xy} &
%%%%%%%%%%%%%%
%%%%%%%%%%%%%%
\begin{xy} <1.0mm,0mm>:
(10,-7)*{B_0~.},
(0,0)*{5}="e5",
(10,0)*{\blacksquare}="v1",
%(20,0)*{\bullet}="v2",
%(30,0)*{\bullet}="v3",
(20,0)*{1}="e1",
(0,9)*{4}="e4",
(10,9)*{3}="e3",
(20,9)*{2}="e2",
\ar@{-}|{<}"e5";"v1"
%\ar@{-}|{>}"v1";"v2"
%\ar@{-}|{>}"v2";"v3"
\ar@{-}|{<}"v1";"e1"
\ar@{~}"v1";"e4"
\ar@{~}"v1";"e3"
\ar@{~}"v1";"e2"
\end{xy}
}
\end{align}
%

%%%%%%%%%%%%%%%%%
\subsection{The second color ordering}\label{ssec:fbbfb}
%%%%%%%%%%%%%%%%%

As in the case of the four-point amplitude, there is a second, cyclically-inequivalent amplitude with two gauginos:
\begin{align}
A(\tilde{1},2,3,\tilde{4},5) =&~ \int_{\mathfrak{T}} \langle c_1 V_{g}^{-1/2}[\zeta^1,k_1] V_{v}^0[\xi^2,k_2] V_{v}^0[\xi^3,k_3] c_4V_{g}^{-1/2}[\zeta^4,k_4] c_5V_{v}^{-1}[\xi^5,k_5] \rangle~.
\end{align}
This factorizes to
\begin{align}\label{Afbbfb1}
A(\tilde{1},2,3,\tilde{4},5) =&~ - \frac{\DC{1}^2 \DC{2}^3}{\sqrt{2}} \int_{\mathfrak{T}} \langle c_1 c_4 c_5\rangle \times \left[ X' \cT_{5}^{(2,3)} + Y_{2}' \cT_{5}^{(3)} + Y_{3}' \cT_{5}^{(2)} + Z' \cT_5 \right]~,
\end{align}
where
\begin{align}
X' & \equiv \sqrt{2}\la \bV[\zeta^1] \bU[\xi^5]\bV[\zeta^4]\ra ~,&
Y_{2}' & \equiv \sqrt{2}\la \bV[\zeta^1] \bJ[\mm^2] \bU[\xi^5]\bV[\zeta^4]\ra ~,\nonumber\\[3mm]
Y_{3}' & \equiv \sqrt{2}\la \bV[\zeta^1] \bJ[\mm^3] \bU[\xi^5]\bV[\zeta^4]\ra~,&
Z' & \equiv \sqrt{2}\la \bV[\zeta^1]\bJ[\mm^2]\bJ[\mm^3] \bU[\xi^5] \bV[\zeta^4]\ra~.
\end{align}

The $\psi$-sector correlators in the $z_5 \to \infty$ limit are
\begin{align}
X' =&~ \frac{\zetab^1 \xis^5 \zeta^4}{z_{41} z_{51} z_{54}}~, \cr
Y_{2}' =&~ -\frac{1}{z_{41} z_{51} z_{54}}\bigg\{ \frac{\zetab^1 \mms^2 \xis^5 \zeta^4}{2z_{21}} + \frac{\zetab^1 \xis^5 \mms^2 \zeta^4}{2z_{42}} + O(1/z_5) \bigg\}~, \cr
Y_{3}' =&~ -\frac{1}{z_{41} z_{51} z_{54}} \bigg\{ \frac{\zetab^1 \mms^3 \xis^5 \zeta^4}{2z_{31}} + \frac{\zetab^1 \xis^5 \mms^3 \zeta^4}{2z_{43}} + O(1/z_5) \bigg\}~,
\end{align}
and
\begin{align}
Z' =&~ \frac{1}{z_{41} z_{51} z_{54}} \bigg\{ \frac{\zetab^1 \mms^2 \mms^3 \xis^5 \zeta^4}{4z_{21} z_{31}} + \frac{\zetab^1 \mms^2 \xis^5 \mms^3 \zeta^4}{4z_{21} z_{43}} + \frac{2 \tr(\mm^2\mm^3) \zetab^1 \xis^5 \zeta^4}{z_{32}^2} +  \frac{2 \zetab^1\mms^{23} \xis^5 \zeta^4}{z_{32} z_{31}}   +  \cr
&~ \qquad \qquad \qquad + \frac{2 \zetab^1 \xis^5 \mms^{23} \zeta^4}{2 z_{32} z_{43}} + \frac{\zetab^1 \mms^3 \xis^5 \mms^2 \zeta^4}{4z_{42} z_{31}} + \frac{\zetab^1 \xis^5 \mms^3 \mms^2 \zeta^4}{4z_{42} z_{43}} + O(1/z_5) \bigg\}~.
\end{align}
The prefactors of $(z_{41} z_{51} z_{54})^{-1}$ cancel the ghost correlator up to a sign.  The result for $Z'$ is easily obtained by first exchanging the $4$ and $5$ labels in \eqref{bigZcor} and then sending $z_5 \to \infty$.  This selects a subset of seven terms from \eqref{bigZcor}, different from the seven obtained in \eqref{limitZcor}.

Making use of these correlators together with the relevant Koba--Nielsen ones, we find that \eqref{Afbbfb1} reduces to a sum over same type $K$ integrals, \eqref{Kintegrals}:
\begin{equation}
A(\tilde{1},2,3,\tilde{4},5) = \sum_{I \in \Sigma_K} N'[I] I~.
\end{equation}
The numerator factors receive contributions from the terms in \eqref{Afbbfb1} as listed in the last column of Table \ref{tab:ourK}.  They are
\begin{align}
N'[K_1] =&~ ig^3 \, \zetab^1 \left( \Delta_{1}^2 - \ff{1}{2} \mms^2 \right)\left( \Delta_{1}^3 - \ff{1}{2} \mms^3 \right) \xis^5 \zeta^4~, \cr
N'[K_2] =&~ - ig^3 \, \zetab^1 \left(\Delta_{1}^2 - \ff{1}{2} \mms^2\right) \xis^5 \left(\Delta_{4}^3 + \ff{1}{2} \mms^3\right) \zeta^4~, \cr
N'[K_3] =&~ - ig^3 \, \zetab^1 \left( \Delta_{1}^3 - \ff{1}{2} \mms^3 \right) \xis^5 \left( \Delta_{4}^2 + \ff{1}{2} \mms^2\right) \zeta^4~, \cr
N'[K_4] =&~  ig^3  \Delta_{2}^3 \, \zetab^1 \left(\Delta_{1}^2 - \ff{1}{2} \mms^2\right) \xis^5 \zeta^4~, \cr
N'[K_5] =&~  ig^3 \, \zetab^1 \left[ 2 \mms^{23} - \Delta_{3}^2 \left(\Delta_{1}^3 - \ff{1}{2} \mms^3\right) \right] \xis^5 \zeta^4~, \cr
N'[K_6] =&~  ig^3(1-q_{23}) \omega^{23} \, \zetab^1\xis^5 \zeta^4~,
\end{align}
and
\begin{align}
N'[K_{1}'] =&~  ig^3 \,\zetab^1 \xis^5 \left(\Delta_{4}^3 + \ff{1}{2} \mms^3\right) \left(\Delta_{4}^2 + \ff{1}{2}\mms^2\right) \zeta^4~, \cr
N'[K_{4}'] =&~  ig^3 \,\zetab^1 \xis^5 \left[ 2 \mms^{23} + \Delta_{3}^2 \left( \Delta_{4}^3 + \ff{1}{2} \mms^3\right) \right] \zeta^4 ~, \cr
N'[K_{5}'] =&~ - ig^3 \Delta_{2}^3 \, \zetab^1 \xis^5 \left( \Delta_{4}^2 + \ff{1}{2} \mms^2 \right) \zeta^4~.
\end{align}
Using \eqref{eq:reducedK} this amplitude can also be put in the form \eqref{TK3form}.  Since the same set of $K$ integrals appear here as in the first color ordering, the form of $A^{(i)}(\tilde{1},2,3,\tilde{4},5)$ will be the same as \eqref{AYMstrings} with $N[\cdot] \to N'[\cdot]$.  One can check that the factor associated with $T$ is the corresponding color-ordered Yang--Mills amplitude.

%%%%%%%%%%%%%%%%%%
\subsection{Four fermions and a gauge boson}
%%%%%%%%%%%%%%%%%%

Our final five-point open string tree amplitude has four gauginos and one gauge boson as external states.  There is a single amplitude up to cyclic permutations.  Placing the boson in position three,
\begin{align}
A(\tilde{1},\tilde{2},3,\tilde{4},\tilde{5}) =& \DC{1}^4 \DC{2} \int_{\mathfrak{T}} \langle c_1 V_{g}^{-1/2}[\zeta^1,k_1] V_{g}^{-1/2}[\zeta^2,k_2] V_{v}^0[\xi^3,k_3] c_4V_{g}^{-1/2}[\zeta^4,k_4] c_5V_{g}^{-1/2}[\zeta^5,k_5] \rangle \cr
=& - \frac{\DC{1}^4 \DC{2}}{2} \int_{\mathfrak{T}} \left( F \cT_{5}^{(3)} + G \cT_5 \right) \langle c_1 c_4 c_5 \rangle~,           \raisetag{20pt}
\end{align}
where the $\psi$-sector correlators are
\begin{align}
& F \equiv 2 \langle \bV[\zeta^1] \bV[\zeta^2] \bV[\zeta^4] \bV[\zeta^5] \rangle~, \qquad G \equiv 2 \langle \bV[\zeta^1] \bV[\zeta^2] \bJ[\mm^3] \bV[\zeta^4] \bV[\zeta^5] \rangle~.
\end{align}

A simple relabeling of indices of \eqref{4Vcor} gives
\begin{equation}\label{Feval}
F = - \frac{ \zetab^1 \Gamma^\mu \zeta^2 \,\zetab^4 \Gamma_\mu \zeta^5 }{z_{21} z_{42} z_{54} z_{52}} - \frac{\zetab^1 \Gamma^\mu \zeta^4 \,\zetab^2 \Gamma_\mu \zeta^5 }{z_{41} z_{42} z_{52} z_{54}} - \frac{\zetab^1 \Gamma^\mu \zeta^5 \,\zetab^2 \Gamma_\mu \zeta^4  }{z_{51} z_{52} z_{42} z_{54}} ~.
\end{equation}
With $\langle c_1 c_4 c_5 \rangle = - z_{41} z_{51} z_{54}$ as usual, it is clear that only the leading $O(1/z_{5}^2)$ terms of $F$ will contribute to the amplitude in the $z_5 \to \infty$ limit.  We take this limit immediately and set the remaining $z$ values to $z_1 = 0$, $z_{2,3} = x_{2,3}$, and $z_4 = 1$.  Then
\begin{align}
\langle c_1 c_4 c_5 \rangle F =&~ \frac{ \zetab^1 \Gamma^\mu \zeta^2 \,\zetab^4 \Gamma_\mu \zeta^5}{x_2 (1-x_2)} + \frac{\zetab^1 \Gamma^\mu \zeta^4 \,\zetab^2 \Gamma_\mu \zeta^5}{(1-x_2)} + O(1/z_5)~.
\end{align}

Now for $G$ we use the Ward identity:
\begin{align}\label{Gward}
G =&~ \frac{1}{2 z_{31}} \langle \bV_1[\mms^3 \zeta^1] \bV[\zeta^2] \bV[\zeta^4] \bV[\zeta^5] \rangle + \frac{1}{2 z_{32}} \langle \bV[\zeta^1] \bV_2[\mms^3\zeta^2] \bV[\zeta^4] \bV[\zeta^5] \rangle + \cr
&~ + \frac{1}{2 z_{34}} \langle \bV[\zeta^1] \bV[\zeta^2] \bV_4[\mms^3\zeta^4] \bV[\zeta^5] \rangle + \frac{1}{2 z_{35}} \langle \bV[\zeta^1] \bV[\zeta^2] \bV[\zeta^4] \bV_5[\mms^3 \zeta^5] \rangle~.
\end{align}
The $z$-dependences of each four-point correlator will be exactly the set of denominators appearing in \eqref{Feval}.  Hence, considering the $z_5 \to \infty$ limit, we can ignore the fourth term in \eqref{Gward}, and when we plug in the four-point correlators for the remaining three we only get the first two type of terms in \eqref{Feval}.  This will give a total of six terms:
\begin{align}
\langle c_1 c_4 c_5 \rangle G =&~ \frac{1}{2 x_3} \left\{ -\frac{ \zetab^1 \mms^3 \Gamma^\mu \zeta^2 \,\zetab^4 \Gamma_\mu \zeta^5}{x_2 (1-x_2)} - \frac{\zetab^1\mms^3 \Gamma^\mu \zeta^4 \,\zetab^2 \Gamma_\mu \zeta^5}{(1-x_2)} \right\} + \cr
&~ + \frac{1}{2(x_3-x_2)} \left\{ \frac{ \zetab^1 \Gamma^\mu \mms^3 \zeta^2 \,\zetab^4 \Gamma_\mu \zeta^5}{x_2 (1-x_2)} - \frac{\zetab^1 \Gamma^\mu \zeta^4 \,\zetab^2 \mms^3 \Gamma_\mu \zeta^5}{(1-x_2)}  \right\} + \cr
&~ - \frac{1}{2(1-x_3)} \left\{ - \frac{ \zetab^1 \Gamma^\mu \zeta^2 \,\zetab^4 \mms^3 \Gamma_\mu \zeta^5}{x_2 (1-x_2)} + \frac{\zetab^1 \Gamma^\mu \mms^3 \zeta^4 \,\zetab^2 \Gamma_\mu \zeta^5}{(1-x_2)} \right\} + O(1/z_5)~. \qquad
\end{align}

In terms of these two results, the full amplitude is
\begin{align}\label{Affbff}
A(\tilde{1},\tilde{2},3,\tilde{4},\tilde{5}) =&~ \frac{\DC{1}^4 \DC{2}}{2} \int_{\mathfrak{T}} \left( \langle c_1c_4c_5\rangle F \left( -\frac{\Delta_{1}^3}{x_3} - \frac{\Delta_{2}^3}{x_3-x_2} + \frac{\Delta_{4}^3}{1-x_3} \right) - \langle c_1 c_4 c_5 \rangle G \right) \cT_5~,
\end{align}
where $\cT_5$ is understood to be evaluated on its leading behavior as $z_5 \to \infty$.  We encounter both type $L$ and type $K$ integrals.  The full list of denominators appearing in the amplitude, together with the corresponding integral names, is given in Table \ref{tab:denlist}.
\begin{table}[t!]
\centering
\begin{tabular}{c c c}
name   & denominator \\[2mm]
%\hline
%\hline
$K_3$ &  $(1-x_2) x_3$   \\[2mm]
%\hline
$K_1'$  &  $(1-x_2) (1-x_3)$    \\[2mm]
%\hline
$K_{5}'$ & $(1-x_2) (x_3-x_2)$   \\[2mm]
%\hline
$L_1$ & $x_2 (1-x_2) x_3$   \\[2mm]
%\hline
$L_4$ & $x_2 (1-x_2) (x_3-x_2)$   \\[2mm]
%\hline
$L_{3}'$ & $x_2 (1-x_2) (1-x_3)$
%\hline
%\hline
\end{tabular}
\caption{The denominators appearing in \eqref{Affbff}.}
\label{tab:denlist}
\end{table}
The $F$ and $G$ terms contribute one term to each of these.  Denoting the collection of integrals by
\begin{equation}
\Sigma' = \{ K_3, K_{1}', K_{5}', L_1, L_4, L_{3}' \}~,
\end{equation}
we have
\begin{equation}
A(\tilde{1},\tilde{2},3,\tilde{4},\tilde{5}) =  \sum_{I \in \Sigma'} N''[I] I  ~.
\end{equation}
The corresponding numerators are
\begin{align}
N''[K_3] =&~ \frac{i g^3}{2} \zetab^1 \left( \Delta_{1}^3 - \ff{1}{2} \mms^3\right) \Gamma^\mu \zeta^4 \, \zetab^2 \Gamma_\mu \zeta^5~, \cr
N''[K_{1}'] =&~ - \frac{i g^3}{2} \zetab^1 \Gamma^\mu \left( \Delta_{4}^3 + \ff{1}{2} \mms^3 \right) \zeta^4 \, \zetab^2 \Gamma_\mu \zeta^5~, \cr
N''[K_{5}'] =&~ \frac{i g^3}{2} \zetab^1 \Gamma^\mu \zeta^4 \, \zetab^2 \left( \Delta_{2}^3 - \ff{1}{2} \mms^3\right) \Gamma_\mu \zeta^5~, \cr
N''[L_1] =&~ \frac{i g^3}{2} \zetab^1 \left( \Delta_{1}^3 - \ff{1}{2} \mms^3\right) \Gamma^\mu \zeta^2 \, \zetab^4 \Gamma_\mu \zeta^5~, \cr
N''[L_4] =&~ \frac{i g^3}{2} \zetab^1 \Gamma^\mu \left( \Delta_{2}^3 + \ff{1}{2} \mms^3\right) \zeta^2 \, \zetab^4 \Gamma_\mu \zeta^5~, \cr
N''[L_{3}'] =&~ - \frac{i g^3}{2} \zetab^1 \Gamma^\mu \zeta^2 \, \zetab^4 \left( \Delta_{4}^3 - \ff{1}{2} \mms^3 \right) \Gamma_\mu \zeta^5~.
\end{align}

We could of course use \eqref{LtoK} to exchange the type $L$ integrals for type $K$ ones, but there is no particular reason to do so.  Rather one can use \eqref{LtoK} and \eqref{eq:reducedK} to write the amplitude directly in the form \eqref{ftform}:
\begin{equation}\label{ffbffamp}
A(\tilde{1},\tilde{2},3,\tilde{4},\tilde{5}) =  A^{(i)}(\tilde{1},\tilde{2},3,\tilde{4},\tilde{5}) T + A^{(ii)}(\tilde{1},\tilde{2},3,\tilde{4},\tilde{5}) K_3~,
\end{equation}
with
\begin{equation}\label{ffbffampT}
A^{(i)} = \frac{N''[K_{1}'] + N''[L_{3}']}{q_{34} q_{15}} + \frac{N''[K_{5}'] + N''[L_{4}']}{q_{23} q_{15}}  +  \frac{N''[L_1] + N''[L_4]}{q_{12} q_{45}} + \frac{N''[L_4]}{q_{23} q_{45}} + \frac{N''[L_{3}']}{q_{12} q_{34}}~, \quad
\end{equation}
and where $A^{(ii)}$ will be given below.  The results of \cite{Barreiro:2005hv} combined with  supersymmetry dictate that $A^{(i)}$ must be the tree-level color-ordered super-Yang--Mills amplitude, while $A^{(ii)}$ is the correction to the Yang--Mills amplitude due to the higher derivative vertices \eqref{eq:newvertices4}.  In the next subsection we perform field theory checks of these statements analogous to those we did for the three gauge boson -- two gaugino amplitude in subsection \ref{sec:SYMcomp1}.

%%%%%%%%%%%%%%%%%
\subsection{Comparison with super-Yang--Mills}
%%%%%%%%%%%%%%%%%

The tree-level diagrams contributing to $A_{\rm YM}(\tilde{1},\tilde{2},3,\tilde{4},\tilde{5})$, along with their associated pole structures, are the following:
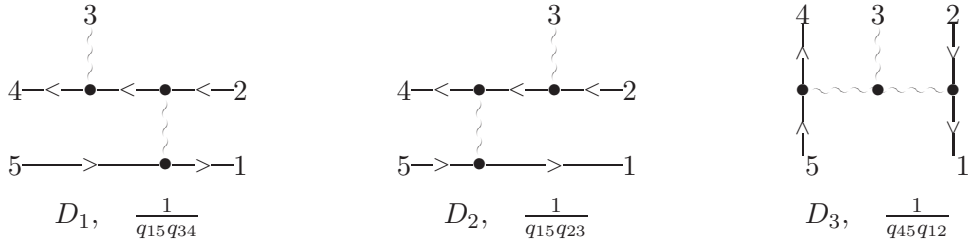
\begin{align}
\xymatrix@R=0mm@C=1cm{
%%%%%%%
%%%%%%%D1
\begin{xy} <1.0mm,0mm>:
(5,-7)*{D_1,\quad\frac{1}{q_{15}q_{34}}},
(-10,0)*{5}="e5";
(10,0)*{\bullet}="v1";
(20,0)*{1} ="e1";
(-10,10)*{4}="e4";
(10,10)*{\bullet}="v2";
(20,10)*{2} = "e2";
(0,20)*{3}= "e3";
(0,10)*{\bullet}="v3";
\ar@{-}|{>}"e5";"v1";
\ar@{-}|{<}"e4";"v3";
\ar@{-}|{>}"v1";"e1";
\ar@{-}|{<}"v3";"v2";
\ar@{-}|{<}"v2";"e2";
\ar@{~}"v1";"v2";
\ar@{~}"v3";"e3";
\end{xy}  \qquad &
%%%%%%%%D2
\begin{xy} <1.0mm,0mm>:
(5,-7)*{D_2,\quad\frac{1}{q_{15}q_{23}}},
(-10,0)*{5}="e5";
(0,0)*{\bullet}="v1";
(20,0)*{1} ="e1";
(-10,10)*{4}="e4";
(0,10)*{\bullet}="v2";
(20,10)*{2} = "e2";
(10,20)*{3}= "e3";
(10,10)*{\bullet}="v3";
\ar@{-}|{>}"e5";"v1";
\ar@{-}|{<}"e4";"v2";
\ar@{-}|{>}"v1";"e1";
\ar@{-}|{<}"v2";"v3";
\ar@{-}|{<}"v3";"e2";
\ar@{~}"v1";"v2";
\ar@{~}"v3";"e3";
\end{xy} \qquad &
%%%%%%%%D3
\begin{xy} <1.0mm,0mm>:
(0,-7)*{D_3,\quad\frac{1}{q_{45}q_{12}}},
(-10,20)*{4}="e4";
(0,20)*{3} ="e3";
(10,20)*{2}="e2";
(-10,10)*{\bullet}="v1";
(0,10)*{\bullet}="v2";
(10,10)*{\bullet}="v3";
(-10,0)*{~~5}="e5";
(10,0)*{~~1} ="e1";
\ar@{-}|{\wedge} "e5";"v1";
\ar@{-}|{\wedge} "v1";"e4";
\ar@{~}"v1";"v3";
\ar@{-}|{\vee}"e2";"v3";
\ar@{-}|{\vee}"v3";"e1";
\ar@{~}"v2";"e3";
\end{xy}
%\\
%%%%%%%%%% names and poles
%%%%%%%%%%
%D_1,\quad \frac{1}{q_{15}q_{34}} &
%D_2, \quad \frac{1}{q_{15}q_{23}} &
%D_3, \quad \frac{1}{q_{45}q_{12}}
} \nonumber
\end{align}
and
\begin{align}\label{ffbffdiagrams}
\xymatrix@R=4mm@C=1cm{
%%%%%%%
%%%%%%%D4
\begin{xy} <1.0mm,0mm>:
(0,-18)*{D_4,\quad\frac{1}{q_{45}q_{23}}},
(-10,10)*{4}="e4";
(10,10)*{3}="e3";
(-10,0)*{\bullet}="v1";
(0,0)*{\bullet}="v2";
(10,0)*{\bullet}="v3";
(-10,-10)*{~~5}="e5";
(0,-10)*{~~1}="e1";
(20,0)*{2}="e2";
\ar@{-}|{\wedge}"e5";"v1";
\ar@{-}|{\wedge}"v1";"e4";
\ar@{~}"v1";"v2";
\ar@{-}|{\vee}"e1";"v2";
\ar@{-}|{<}"v2";"v3";
\ar@{~}"v3";"e3";
\ar@{-}|{<}"e2";"v3";
\end{xy}  \qquad &
%%%%%%%D5
\begin{xy} <1.0mm,0mm>:
(5,-18)*{D_5,\quad\frac{1}{q_{34}q_{12}}~.},
(0,10)*{3}="e3";
(20,10)*{2}="e2";
(-10,0)*{4}="e4";
(0,0)*{\bullet}="v1";
(10,0)*{\bullet}="v2";
(20,0)*{\bullet}="v3";
(10,-10)*{~~5}="e5";
(20,-10)*{~~1}="e1";
\ar@{~}"e3";"v1";
\ar@{~}"v2";"v3";
\ar@{-}|{<}"e4";"v1";
\ar@{-}|{<}"v1";"v2";
\ar@{-}|{\wedge}"e5";"v2";
\ar@{-}|{\vee}"e2";"v3";
\ar@{-}|{\vee}"v3";"e1";
\end{xy}
%%%%%%%%  names and poles
%%%%%%%%
%\\
%D_4~,1/q_{45}q_{23} &
%D_5~,1/q_{34}q_{12} ~.
}
\end{align}
The double poles match precisely with those appearing in \eqref{ffbffampT}.  

In contrast to the amplitude involving three gauge bosons and two gauginos, there are no explicit single pole terms in \eqref{ffbffampT}, nor are there implicit single pole terms.  Indeed, the numerators of \eqref{ffbffampT} are linear in the external momenta and therefore cannot contain terms in which the double pole structure indicated by the denominator is reduced because one of the $q_{ij}$ factors is canceled.  This simplifies the comparison: each of the five diagrams above must match the corresponding term in \eqref{ffbffampT} on the nose.  

Explicit evaluation of the diagrams using the rules \eqref{Feynrules1}, \eqref{Feynrules2} confirms this.  The case of $D_1$ provides a good illustration.  On the one hand the color-ordered rules straightforwardly yield
\begin{align}
D_1 =&~ i g \zetab^1 \Gamma^\mu \zeta^5 \left( \frac{-i \eta_{\mu\nu}}{(k_1+k_5)^2} \right) i g (\zetab^4 \xis^3)_a \frac{i (-\kiss_3 - \kiss_4)_{ab}}{(k_3+k_4)^2} ig (\Gamma^\nu \zeta^2)_b \cr
=&~ \frac{i g^3}{4 q_{15} q_{34}} \zetab^1 \Gamma^\mu \zeta^5 \, \zetab^4 \xis^3 (\kiss_3 + \kiss_4) \Gamma_\mu \zeta^2 \cr
=&~ \frac{i g^3}{2 q_{15} q_{34}} \zetab^1 \Gamma^\mu \zeta^5 \, \zetab^4 \left( \Delta_{4}^3 - \ff{1}{2} \mms^3 \right) \Gamma_\mu \zeta^2~,
\end{align}
where in the last step we invoked the physical state conditions $\zetab^4 \kiss_4 = 0$ and $\Delta_{3}^3 = 0$.  On the other hand we can employ the Fierz identity \eqref{eq:Fierz} to write
\begin{align}
N''[K_{1}'] + N''[L_{3}'] =&~ -\frac{i g^3}{2} \bigg\{ \zetab^1 \Gamma^\mu \left( \Delta_{4}^3 + \ff{1}{2} \mms^3 \right) \zeta^4 \, \zetab^2 \Gamma_\mu \zeta^5 +  \zetab^1 \Gamma^\mu \zeta^2 \, \zetab^4 \left( \Delta_{4}^3 - \ff{1}{2} \mms^3 \right) \Gamma_\mu \zeta^5 \bigg\} \cr
=&~  -\frac{i g^3}{2} \bigg\{  \Delta_{4}^3 \left( \zetab^2 \Gamma^{\mu} \zeta^5 \, \zetab^1 \Gamma_\mu \zeta^4 + \zetab^4 \Gamma^\mu \zeta^5 \, \zetab^1 \Gamma_\mu \zeta^2 \right) + \cr
&~ \qquad  \qquad +  \half \left( \zetab^2 \Gamma^{\mu} \zeta^5 \, \zetab^1 \Gamma_\mu (\mms^3 \zeta^4) + (\overline{\mms^3 \zeta^4}) \Gamma^\mu \zeta^5 \, \zetab^1 \Gamma_\mu \zeta^2 \right) \bigg\} \cr
=&~  \frac{i g^3}{2}  \zetab^1 \Gamma^\mu \zeta^5 \, \zetab^4 \left( \Delta_{4}^3 - \ff{1}{2} \mms^3\right) \zeta^2~, \raisetag{20pt}
\end{align}
whence
\begin{equation}\label{ffbffDrel1}
D_1 = \frac{N''[K_{1}'] + N''[L_{3}']}{q_{15}q_{34}} ~.
\end{equation}
Similar manipulations show that 
\begin{align}\label{ffbffDrel2}
& D_2 = \frac{N''[K_{5}']+N''[L_{4}']}{q_{23} q_{15}}~, \quad D_3 = \frac{N''[L_1] + N''[L_4]}{q_{12} q_{45}}~, \quad D_4 = \frac{N''[L_4]}{q_{23} q_{45}}~, \quad D_5 = \frac{N''[L_{3}']}{q_{12} q_{34}}~,
\end{align}
and therefore
\begin{equation}
A^{(i)}(\tilde{1},\tilde{2},3,\tilde{4},\tilde{5}) = \sum_{i=1}^{5} D_i = A_{\rm YM}(\tilde{1},\tilde{2},3,\tilde{4},\tilde{5})~.
\end{equation}
We note that diagrams $D_{3,4,5}$ received an extra sign relative to $D_{1,2}$ for the same reason as described under equation \eqref{AYM4gaugino}.

We can also give a diagrammatic interpretation of $A^{(ii)}$, the coefficient of the $K_3$ integral in \eqref{ffbffamp}.  By making use of \eqref{LtoK} and \eqref{eq:reducedK} together with the relations \eqref{ffbffDrel1}, \eqref{ffbffDrel2}, one finds
\begin{equation}\label{ffbffampK3}
A^{(ii)} = \frac{1}{q_{23}} B_{23} + \frac{1}{q_{34}} B_{34} + \frac{1}{q_{45}} B_{45} + \frac{1}{q_{12}} B_{12} + \frac{1}{q_{15}} B_{15} + B_0~.
\end{equation}
with
\begin{align}\label{ffbffBtoDs}
B_{23} =&~ - (q_{12} + q_{13}) q_{23} q_{15} D_2 - (q_{24} + q_{34}) q_{23} q_{45} D_4~, \cr
B_{34} =&~ -q_{12} q_{34} q_{15} D_1 - (q_{23} + q_{24}) q_{12} q_{34} D_5~, \cr
B_{45} =&~ -q_{34} q_{12} q_{45} D_3 - q_{12} q_{23} q_{45} D_4~, \cr
B_{12} =&~ -q_{12} q_{45} q_{34} D_3 - (q_{13} + q_{23}) q_{12} q_{34} D_5~, \cr
B_{15} =&~ - q_{23} q_{34} q_{15} D_1 - q_{34} q_{23} q_{15} D_2~,
\end{align}
and
\begin{equation}
B_0 = N[L_{3}'] - N[K_3] - N[L_1] ~.
\end{equation}

The $B_{ij}$ correspond to field theory diagrams with a single pole in $q_{ij}$ and constructed using one of the higher derivative vertices \eqref{eq:newvertices4}, while $B_0$ is a five-point contact interaction.  The diagrams are
\begin{align}
\xymatrix@R=0mm@C=1cm{
%%%%%%%
%%%%%%%B1
\begin{xy} <1.0mm,0mm>:
(-10,0)*{5}="e5";
(0,0)*{\blacksquare}="v1";
(10,0)*{\bullet}="v2";
(0,-10)*{~~1}="e1";
(20,0)*{2}="e2";
(10,10)*{3}="e3";
(0,10)*{4}="e4";
\ar@{-}|{>}"e5";"v1";
\ar@{-}|{\vee}"v1";"e1";
\ar@{-}|{<}"v2";"v1";
\ar@{-}|{\wedge}"v1";"e4";
\ar@{~}"v2";"e3";
\ar@{-}|{<}"v2";"e2";
\end{xy} \quad &
%%%%%%B2
\begin{xy} <1.0mm,0mm>:
(-10,0)*{5}="e5";
(0,0)*{\blacksquare}="v1";
(0,8)*{\bullet}="v2";
(0,-10)*{~~1}="e1";
(10,0)*{2}="e2";
(10,8)*{3}="e3";
(0,16)*{4}="e4";
\ar@{-}|{>}"e5";"v1";
\ar@{-}|{\vee}"v1";"e1";
\ar@{-}|{\wedge}"v2";"v1";
\ar@{-}|{\wedge}"v2";"e4";
\ar@{~}"v2";"e3";
\ar@{-}|{<}"v1";"e2";
\end{xy} \quad &
%%%%%%B3
\begin{xy} <1.0mm,0mm>:
(-10,10)*{4}="e4";
(-10,0)*{\bullet}="v1";
(-10,-10)*{~~5}="e5";
(0,0)*{\blacksquare}="v2";
(0,-10)*{~~1}="e1";
(10,0)*{2}="e2";
(0,10)*{3}="e3";
\ar@{-}|{\wedge}"e5";"v1";
\ar@{-}|{\wedge}"v1";"e4";
\ar@{~}"v1";"v2";
\ar@{~}"v2";"e3";
\ar@{-}|{\vee}"e1";"v2";
\ar@{-}|{<}"v2";"e2";
\end{xy}
\quad &
%%%%%%%B4
\begin{xy}<1.0mm,0mm>:
(-10,10)*{4}="e4";
(-10,0)*{\blacksquare}="v1";
(-10,-10)*{~~5}="e5";
(0,10)*{3}="e3";
(0,0)*{\bullet}="v2";
(0,-10)*{~~1}="e1";
(10,0)*{2}="e2";
\ar@{-}|{\wedge}"e5";"v1";
\ar@{-}|{\wedge}"v1";"e4";
\ar@{~}"v1";"e3";
\ar@{-}|{<}"v2";"e2";
\ar@{~}"v1";"v2";
\ar@{-}|{\vee}"e1";"v2";
\end{xy}
%%%%%%%%
%%%%%%%%
\\
B_1,~~1/q_{23} & B_2,~~1/q_{34}	& B_3,~~1/q_{45} & B_4,~~1/q_{12}
} \nonumber
\end{align}
and
\begin{align}
\xymatrix@R=0mm@C=1cm{
%%%%%%%
%%%%%%%B5
\begin{xy} <1.0mm,0mm>:
(-10,0)*{4}="e4";
(0,0)*{\blacksquare}="v1";
(10,0)*{2}="e2";
(-10,-10)*{5}="e5";
(0,-10)*{\bullet}="v2";
(10,-10)*{1} ="e1";
(0,10)*{3}="e3";
\ar@{-}|{>}"e5";"v2";
\ar@{-}|{>}"v2";"e1";
\ar@{-}|{<}"e4";"v1";
\ar@{~}"v1";"e3";
\ar@{-}|{<}"v1";"e2";
\ar@{~}"v1";"v2";
\end{xy} \qquad &
%%%%%%% B0
\begin{xy} <1.0mm,0mm>:
(0,0)*{\blacksquare}="v";
(-10,0)*{5}="e5";
(0,10)*{4}="e4";
(10,10)*{3}="e3";
(10,0)*{2}="e2";
(0,-10)*{~~1}="e1";
\ar@{-}|{>}"e5";"v";
\ar@{-}|{<}"v";"e2";
\ar@{-}|{\vee}"v";"e1";
\ar@{-}|{\wedge}"v";"e4";
\ar@{~}"v";"e3";
\end{xy} \\
%%%%%%%
B_{5},~~1/{q_{15}}	& B_0~.
}
\end{align}

%%%%%%%%%%%%%%%%%
%%%%%%%%%%%%%%%%%
\section{KLT relations for the closed amplitudes}
%%%%%%%%%%%%%%%%%
%%%%%%%%%%%%%%%%%

\subsection{Closed string vertex operators}

Now we turn to the computation of tree-level closed string amplitudes.  The closed-string vertex operators are simply products of open string vertex operators.  For example, an NS-NS sector operator in the $(-1,-1)$ picture would be given by
\begin{equation}
V_{\mathrm{NS-NS}}^{(-1,-1)}[\xi,k]=\xi_{\mu\nu}\mathcal{U}^\mu\widetilde{\mathcal{U}}^\nu e^{ik\cdot X}.
\end{equation}
This represents a dilaton, a graviton, or a $B$-field, depending on whether the polarization tensor $\xi_{\mu\nu}$ is a trace\footnote{More precisely, the polarization for a dilaton is constructed by first choosing a null vector $\ell^\mu$ that satisfies $k\cdot\ell=1$.  Then we choose
\begin{equation}
\xi_{\mu\nu}=\Xi\left(\eta_{\mu\nu}-k_\mu\ell_\nu-\ell_\mu k_\nu\right),
\end{equation}
where $\Xi$ is the scalar ``polarization" of the dilaton.}, traceless symmetric, or antisymmetric, respectively.  As with the open string vertex operators, there are physical state conditions
\begin{equation}
\xi_{\mu\nu}k^\mu=0,\qquad\xi_{\mu\nu}k^\nu=0,\qquad k^2=0,
\end{equation}
and gauge invariances,
\begin{equation}
\xi_{\mu\nu}\sim\xi_{\mu\nu}+\lambda_\mu k_\nu+k_\mu\widetilde{\lambda}_\nu,
\end{equation}
where the vectors $\lambda_\mu$ and $\widetilde{\lambda}_\mu$ satisfy $k\cdot\lambda=k\cdot\widetilde{\lambda}=0$.

The gravitini vertex operators in the $(-1,-1/2)$ and $(-1/2,-1)$ pictures (the $(0,-1/2)$ and $(-1/2,0)$ pictures are similar) are
\begin{equation}
V_{\mathrm{NS-R}}^{(-1,-1/2)}=\Psi_{\mu a}\mathcal{U}^\mu\widetilde{\mathcal{V}}^ae^{ik\cdot X},\qquad V_{\mathrm{R-NS}}^{(-1/2,-1)}=\Psi'_{\mu a}\mathcal{V}^a\widetilde{\mathcal{U}}^\mu e^{ik\cdot X}.
\end{equation}
In type IIB string theory the two gravitini are spinors of the same chirality, while for type IIA $\Psi_\mu$ and $\Psi'_\mu$ have opposite chirality.

The physical state conditions on the polarizations $\Psi_\mu$ are that (suppressing the spinor indices)
\begin{equation}
\Psi_\mu k^\mu=0,\qquad \kiss\Psi_\mu=0,\qquad k^2=0.
\end{equation}
Additionally there is a gauge symmetry,
\begin{equation}
\Psi_{\mu a}\sim\Psi_{\mu a}+k_\mu\zeta_a,
\end{equation}
where $\kiss\zeta=0$.
Of course $\Psi'_{\mu a}$ has the same physical state conditions and gauge symmetry.

Finally, for states in the Ramond-Ramond sector we will use the $(-1/2,-1/2)$ picture, and we can write
\begin{equation}
V_{\mathrm{R-R}}^{(-1/2,-1/2)}=\zeta_{ab}\mathcal{V}^a\widetilde{\mathcal{V}}^be^{ik\cdot X}.
\end{equation}
The polarization $\zeta_{ab}$ is a bispinor, and we can expand it in terms of differential forms in ten dimensions.  For IIA, the left and right spinors have opposite chirality and the forms will have even degree, while for IIB they have the same chirality and the forms will have odd degree.  Explicitly,
\begin{equation}
\zeta^{(IIA)}_{ab}=\sum_{i=0}^2F^{(2i)}_{\mu_1\cdots\mu_{2i}}\left(\mathcal{C}\Gamma^{\mu_1\cdots\mu_{2i}}\right)_{ab},\qquad\zeta^{(IIB)}_{ab}=\sum_{i=0}^2F^{(2i+1)}_{\mu_1\cdots\mu_{2i+1}}\left(\mathcal{C}\Gamma^{\mu_1\cdots\mu_{2i+1}}\right)_{ab}.
\end{equation}
For the purposes of amplitude calculations, it is more convenient to leave the polarization as $\zeta_{ab}$, in either theory, rather than breaking it up into a sum of forms.

The correlation functions now roughly factorize into holomorphic and anti-holomorphic parts, where unlike in the open string case we are not constrained to the real line $z=\bar{z}$.  Finally, the amplitudes are computed by fixing the positions of three of the vertex operators and integrating over the positions of the remaining vertex operators in the complex plane.  This means in particular that a closed string $N$-point amplitude involves integration over $2N-6$ real variables, compared to $N-3$ for an $N$-point open string amplitude.  Fortunately, by situating some of these real integrations as contours in complex planes, and by judicious contour rotations and careful analysis of the resulting phases, it was shown by Kawai, Lewellen, and Tye~\cite{Kawai:1985xq}, that the closed string amplitudes can in fact be written as weighted sums (over a subset of the possible permutations of the operators) of products of open string amplitudes.

\subsection{KLT relations}

We will simply quote the results of \cite{Kawai:1985xq} for three-, four-, and five-point amplitudes.
\begin{equation}
A_{\mathrm{closed}}^{(3)}=\frac{i\kappa}{2g^2}A_{\mathrm{open}}(1,2,3)\widetilde{A}_{\mathrm{open}}(1,2,3)~,
\end{equation}
\begin{equation}\label{4ptKLT}
A_{\mathrm{closed}}^{(4)}=\frac{i\kappa^2}{2\pi g^4}\sin(\pi q_{12})A_{\mathrm{open}}(1,2,3,4)\widetilde{A}_{\mathrm{open}}(1,2,4,3)~,
\end{equation}
and
\begin{multline}
\label{eq:KLT5point}
A_{\mathrm{closed}}^{(5)}=\frac{i\kappa^3}{2\pi^2g^6}\left[\sin(\pi q_{12})\sin(\pi q_{34})A_{\mathrm{open}}(1,2,3,4,5)\widetilde{A}_{\mathrm{open}}(2,1,4,3,5)\right.\\
\left.+\sin(\pi q_{13})\sin(\pi q_{24})A_{\mathrm{open}}(1,3,2,4,5)\widetilde{A}_{\mathrm{open}}(3,1,4,2,5)\right]~.
\end{multline}
In each case, gauge invariance of the closed string amplitude is guaranteed by the gauge invariance of the open string amplitudes.

These relations are of course somewhat schematic.  The tildes on the second open string amplitude in each term above just means that we compute the amplitude using the right-moving sectors of our vertex operators.  If one wishes to restore dimensions to these amplitudes, one must use the conventional closed string choice, $\alpha'=2$, in contrast to $\alpha'=1/2$ which was used for the open string amplitudes.
Also, when we expand the right-hand side we should replace each product of left and right polarizations by a closed string polarization tensor.  For example, if particle 1 is in the NS-NS sector, then we should make the replacement
\begin{equation}
\label{eq:KLTReplacement}
\xi^1_\mu\widetilde{\xi}^1_\nu\quad\longrightarrow\quad\xi^1_{\mu\nu}~.
\end{equation}
Let's see how this works for the three-point functions.

For three NS-NS sector fields, we use (\ref{eq:3vAmplitude}) and get
\begin{multline}
A_{\mathrm{closed}}^{(3)}(\xi^1,k_1;\xi^2,k_2;\xi^3,k_3)\\
=-\frac{i}{2}\kappa\xi^1_{\mu\mu'}\xi^2_{\nu\nu'}\xi^3_{\rho\rho'}\left[k_{23}^\mu\eta^{\nu\rho}+k_{31}^\nu\eta^{\rho\mu}+k_{12}^\rho\eta^{\mu\nu}\vphantom{k_{23}^{\mu'}}\right]\left[k_{23}^{\mu'}\eta^{\nu'\rho'}+k_{31}^{\nu'}\eta^{\rho'\mu'}+k_{12}^{\rho'}\eta^{\mu'\nu'}\right]~.
\end{multline}
Recall that the original open string three-vector amplitude was invariant under $\Z_3$ cyclic symmetry, but was antisymmetric under exchange of any two vectors, and hence was not invariant under the full $S_3$ symmetry group.  The closed string amplitude is $S_3$-invariant; it inherits the cyclic symmetry from the open string amplitudes, and under exchange of two vectors we get a minus sign from each factor, leaving the product invariant.  This enhancement from cyclic symmetry to the full permutation symmetry occurs in each of the cases we will examine.

For one NS-NS sector field and two NS-R sector gravitini, we plug in (\ref{eq:3vAmplitude}) and (\ref{eq:v2fAmplitude}) to find
\begin{equation}
A_{\mathrm{closed}}^{(3)}(\xi^1,k_1;\Psi^2,k_2;\Psi^3,k_3)=\frac{i}{2}\kappa\xi^1_{\mu\mu'}\overline{\Psi}^2_\nu\Gamma^{\mu'}\Psi^3_\rho\left[k_{23}^\mu\eta^{\nu\rho}+k_{31}^\nu\eta^{\rho\mu}+k_{12}^\rho\eta^{\mu\nu}\vphantom{k_{23}^{\mu'}}\right]~.
\end{equation}
The case of two R-NS sector gravitini is completely analogous.

Two more amplitudes remain, and they are both products of two $bff$ (boson-fermion-fermion) open string amplitudes.  First we have one NS-NS and two R-R fields,
\begin{equation}
A_{\mathrm{closed}}^{(3)}(\xi^1,k_1;\zeta^2,k_2;\zeta^3,k_3)=-\frac{i}{2}\kappa\xi^1_{\mu\nu}\Tr(\zeta^2\mathcal{C}\Gamma^\mu\zeta^{3\,T}\mathcal{C}\Gamma^\nu)~.
\end{equation}
The other possibility is one R-R, one NS-R, and one R-NS field,
\begin{equation}
A_{\mathrm{closed}}^{(3)}(\zeta^1,k_1;\Psi^2,k_2;\Psi^{\prime\,3},k_3)=-\frac{i}{2}\kappa\overline{\Psi}^{\prime\,3}_\mu\Gamma^\nu\zeta^1\mathcal{C}\Gamma^\mu\Psi^2_\nu~.
\end{equation}

All of these amplitudes match the results that would be obtained (at tree level) from the effective action of type IIA or IIB supergravity.  For graviton fields in particular, we would expand the Einstein-Hilbert Lagrangian around flat space, with fluctuations
\begin{equation}
G_{\mu\nu}(x)=\eta_{\mu\nu}+2\kappa\xi_{\mu\nu}e^{ik\cdot x}~,
\end{equation}
with traceless symmetric $\xi_{\mu\nu}$.

Next we move on to four-point functions.  For four NS-NS sector fields we have, using \eqref{4ptKLT} and \eqref{4vstrings},
\begin{multline}
A_{\mathrm{closed}}^{(4)}(\xi^1,k_1;\xi^2,k_2;\xi^3,k_3;\xi^4,k_4)\\
=-\frac{2i}{\pi}\kappa^2\sin(\pi q_{12})\frac{B(q_{12},q_{23})}{q_{13}}\frac{B(q_{12},q_{24})}{q_{14}}K(1,2,3,4)\widetilde{K}(1,2,4,3)\\
=-2i\kappa^2\frac{\Gamma(q_{12})\Gamma(q_{13})(\Gamma(q_{14})}{\Gamma(1-q_{12})\Gamma(1-q_{13})\Gamma(1-q_{14})}K(1,2,3,4)\widetilde{K}(1,2,4,3)~,
\end{multline}
where we have used the identity
\begin{equation}
\frac{\pi}{\sin(\pi q_{12})}=\Gamma(q_{12})\Gamma(1-q_{12})~,
\end{equation}
and it is understood that we make the replacement (\ref{eq:KLTReplacement}).  $K(1,2,3,4)$ or $\widetilde{K}(1,2,4,3)$ are both given by (\ref{K4vec}), with the right-moving polarizations inserted in the latter case (as an intermediate step before we substitute (\ref{eq:KLTReplacement})).  Note that the prefactor,
\begin{equation}
\mathcal{P}=\frac{\Gamma(q_{12})\Gamma(q_{13})\Gamma(q_{14})}{\Gamma(1-q_{12})\Gamma(1-q_{13})\Gamma(1-q_{14})}~,
\end{equation}
is now invariant under the full $S_4$ symmetry.  Since the kinematic factor $K(1,2,3,4)$ also has $S_4$ symmetry, the full amplitude has the appropriate permutation symmetry.

Similarly, for the amplitude with two NS-NS fields and two NS-R gravitini, we would compute
\begin{align}
A_{\mathrm{closed}}^{(4)}(\xi^1,k_1;\xi^2,k_2;\Psi^3,k_3;\Psi^4,k_4)=& \frac{i\kappa^2}{2\pi g^4}\sin(\pi q_{12})A_{\mathrm{open}}^{(4)}(1,2,3,4)\widetilde{A}_{\mathrm{open}}^{(4)}(1,2,\widetilde{4},\widetilde{3})\nonumber\\
=& -\frac{i\kappa^2}{2\pi g^4}\sin(\pi q_{12})A_{\mathrm{open}}^{(4)}(1,2,3,4)\widetilde{A}_{\mathrm{open}}^{(4)}(\widetilde{3},1,2,\widetilde{4})\nonumber\\
=& -\frac{i}{2}\kappa^2\mathcal{P}K(1,2,3,4)\widetilde{K}(\widetilde{3},1,2,\widetilde{4})~,
\end{align}
where now $\widetilde{K}(\tilde{3},1,2,\tilde{4})$ is given by (\ref{ourK2v2g}).
In the third line we have used the cyclicity of $\widetilde{A}_{\mathrm{open}}^{(4)}$, with an extra sign since we moved the fermions through each other.  In the last line we made the substitutions from (\ref{eq:FourPointSummary}).  This result has the expected symmetry under exchange of the two bosons and antisymmetry under exchange of the two fermions.  Note, however, that we could have put the states in a different order to start, writing
\begin{align}
A_{\mathrm{closed}}^{(4)}(\xi^1,k_1;\Psi^3,k_3;\Psi^4,k_4;\xi^2,k_2)=& \frac{i\kappa^2}{2\pi g^4}\sin(\pi q_{13})A_{\mathrm{open}}^{(4)}(1,3,4,2)\widetilde{A}_{\mathrm{open}}^{(4)}(1,\widetilde{3},2,\widetilde{4})\nonumber\\
=& \frac{i\kappa^2}{2\pi g^4}\sin(\pi q_{13})A_{\mathrm{open}}^{(4)}(1,3,4,2)\widetilde{A}_{\mathrm{open}}^{(4)}(\widetilde{3},2,\widetilde{4},1)\nonumber\\
=& -\frac{i}{2}\kappa^2\mathcal{P}K(1,3,4,2)\widetilde{K}(\widetilde{3},2,\widetilde{4},1)~,
\end{align}
which uses (\ref{K2v2gB}).
The equivalence of these two computations relies on the symmetries of the kinematic factors, as well as the equality (\ref{eq:KequalsK}) between the kinematic factors of the two open string amplitudes and the relative sign appearing in (\ref{eq:FourPointSummary}) between the two types of ordering.

Similar stories can be told for all other possible four-point functions of massless closed strings.  In each case the leading term in the momentum expansion matches the result derived from the type IIA or IIB effective action at tree-level, while the higher terms in the momentum expansion originate from corrections (in fact the corrections are eight-derivative or higher, or $(\alpha')^3$) to those actions.

Finally we turn to the five-point function.  The procedure is the same as for the three- and four-point functions, so we will be brief.  As above, the closed-string amplitudes are obtained by simply substituting our open string amplitudes (\ref{5vecres}) or the appropriate version of (\ref{ftform}) into the KLT formula (\ref{eq:KLT5point}).  

As with the four-point amplitude, the closed string amplitude can have enhanced symmetry compared to the open string.  For example, the amplitude for five identical NS-NS sector states has a full $S_5$ permutation symmetry, which is not manifest from (\ref{eq:KLT5point}) (other than the transposition of particles labeled $2$ and $3$, which clearly just exchanges the two lines of the formula).  Indeed, verifying the full symmetry involves certain nontrivial identities, quadratic in generalized hypergeometric functions, which can be checked numerically.  

Similarly, the fact that the closed string amplitudes are independent of the distribution order of particles implies certain relations between different open string orderings.  For example, consider the amplitude with three NS-NS and two NS-R states.  If we put the fermions in positions $1$ and $5$, then the closed string result is constructed purely from the second ordering computed in section \ref{ssec:fbbfb},
\begin{multline}
A_{\mathrm{closed}}^{(5)}=\frac{i\kappa^3}{2\pi^2g^6}\left[\sin(\pi q_{12})\sin(\pi q_{34})A_{\mathrm{open}}(1,2,3,4,5)\widetilde{A}_{\mathrm{open}}(\tilde{1},4,3,\tilde{5},2)\right.\\
\left.+\sin(\pi q_{13})\sin(\pi q_{24})A_{\mathrm{open}}(1,3,2,4,5)\widetilde{A}_{\mathrm{open}}(\tilde{1},4,2,\tilde{5},3)\right]~,
\end{multline}
where we used cyclic symmetry to put the antiholomorphic sector amplitudes into the form we computed.  On the other hand, if we assign the fermions to positions $1$ and $4$, then it is the first ordering which appears.  The equivalence of these two choices implies a relation between the two orderings, though not one that is expressible in a simple way.  

We should mention that another way to uncover the emergence of the full permutation symmetry is through the use of linear relations (with momentum-dependent coefficients) between the open string $n$-point ordered amplitudes that reduce the number of independent open sub-amplitudes to $(n-3)!$~\cite{Stieberger:2009hq,BjerrumBohr:2009rd}.  

%%%%%%%%%%%%%%%%%
%%%%%%%%%%%%%%%%%
\section*{Acknowledgements}
%%%%%%%%%%%%%%%%%
%%%%%%%%%%%%%%%%%
IVM would like to thank O.~Schlotterer for correspondence regarding pure spinor computations of amplitudes with space-time fermions.
K.~Becker and M.~Becker were supported by PHY-1214333 and NSF Focussed Research grant DMS-1159404. D.~Robbins was supported by PHY-1214333, NSF Focussed Research grant DMS-1159404 and by the Mitchell Family Foundation. I.~V.~Melnikov was supported by NSF Focussed Research grant DMS-1159404 and the University of Texas A\&M. A.~B.~Royston was supported by the Mitchell Family Foundation.

\appendix

\section{A study of integrals}\label{app:integrals}
In this appendix we discuss the integrals relevant for the five-point amplitudes.  All of this structure is already given in~\cite{Barreiro:2005hv}, and we merely include this for ease of reference.  First, recall that in~(\ref{eq:bigmatter}) we have the following correlation functions:
\begin{align}
Z & = \la \bU[\xi^4] \bU[\xi^5] \ra \cT_5^{(1,2,3)}~,&
Y_{i} & = \la \bJ[\mm^i]\bU[\xi^4] \bU[\xi^5] \ra \cT_5^{(2,3)}~, \nonumber\\
X_{ij} & = \la \bJ[\mm^i]\bJ[\mm^j]\bU[\xi^4] \bU[\xi^5] \ra \cT_5^{(1)} ~,&
W_{123} & = \la \bJ[\mm^1] \bJ[\mm^2]\bJ[\mm^3] \bU[\xi^4] \bU[\xi^5]\ra \cT_5~.
\end{align}
After fixing the positions as indicated in section~\ref{ss:bbbbb}, we find that these correlators give rise to the denominators collected in tables~\ref{tab:L}, \ref{tab:K}, and \ref{tab:D}.  These integrals are of the hypergeometric form, and with some manipulations can be reduced to a product of Beta functions and $~_{3} F_2$.   The latter is a sufficiently complicated object that for the most part it is best to work with the integral representations directly.  In each case, the integral corresponding to a denominator $\kappa$ is given by
\begin{align}
I[\kappa] & = \int^1_0 \ed x_3 \int^{x_3}_0 \ed x_2 \frac{ \tau}{\kappa}~, &
\tau =
x_3^{q_{13}} (1-x_3)^{q_{34}}x_2^{q_{12}} (1-x_2)^{q_{24}} (x_3-x_2)^{q_{23}}~.
\end{align}

\subsection{Reduction to $K_3$ and $T$}
The $26$ integrals satisfy $24$ relations.  These are obtained by a combination of integration by parts and partial fractions.

For instance, we have
\begin{align}\label{newrln1}
& \int_{0}^1 \ed x_3 \int_{0}^{x_3} \ed x_2 \, \tau \left( \frac{q_{12} - 1}{x_{2}^2 x_3} - \frac{q_{23}}{x_2 x_3 (x_3-x_2)} \right) =  \cr
& \qquad = \int_{0}^1 \ed x_3 \int_{0}^{x_3} \ed x_2 \left( \frac{q_{12} - 1}{x_2} - \frac{q_{23}}{x_3 - x_2} \right) x_{2}^{q_{12}-1} (1-x_2)^{q_{24}} x_{3}^{q_{13}-1} (1-x_3)^{q_{34}} (x_3 - x_2)^{q_{23}} \cr
& \qquad = \int_{0}^1 \ed x_3 \int_{0}^{x_3} \ed x_2 \left\{ \frac{\p}{\p x_2} \left[ x_{2}^{q_{12} - 1} (x_3-x_2)^{q_{23}} \right] \right\} (1-x_2)^{q_{24}} x_{3}^{q_{13} - 1} (1-x_3)^{q_{34}} \cr
& \qquad = -  \int_{0}^1 \ed x_3 \int_{0}^{x_3} \ed x_2 \left( -\frac{q_{24}}{1-x_2} \right) x_{2}^{q_{12}-1} (1-x_2)^{q_{24}} x_{3}^{q_{13}-1} (1-x_3)^{q_{34}} (x_3 - x_2)^{q_{23}} \cr
& \qquad =  \int_{0}^1 \ed x_3 \int_{0}^{x_3} \ed x_2 \,  \frac{q_{24}\tau}{x_2(1-x_2)x_3}~,
\end{align}
which leads to the relation
\begin{align}
(q_{12} - 1) M_2 - q_{23} M_1 = q_{24} L_1~.
\end{align}
In a similar fashion we obtain
\begin{align}
(q_{23} - 1) M_4 + q_{13} M_1 &= q_{34} L_2 ~, \nonumber\\
- (q_{23} - 1) M_5 + q_{12} M_1 &= q_{24} L_7 ~, \nonumber\\
(q_{13} - 1) M_6 + q_{23} M_1 &=  q_{34} L_3 ~.
\end{align}
For a partial fraction example, we have
\begin{align}\label{nr5}
\frac{1}{x_2 x_3 (x_3 - x_2)} = \frac{1}{x_{2}^2 (x_3 - x_2)} - \frac{1}{x_{2}^2 x_3}~,
\end{align}
which implies
\begin{align}
 M_1 = M_3 - M_2~.
\end{align}
Similarly, we obtain
\begin{align}
M_1 &= M_4 - M_5~,&
M_1 &= M_6 + M_7~.
\end{align}
Thus, we have $7$ linear relations that allow us to determine the $M$ integrals in terms of the $K$ and $L$ integrals.

\begin{table}[t!]
\centering
\begin{tabular}{c c c c}
denominator & with $z_1 =0$, $z_4 = 1$ & name   & location  \\[2mm]
%\hline
%\hline
$z_{21} z_{31} z_{42}$ & $x_2 x_3 (1-x_2)$ & $L_1$ & $W_{123}$, $X_{12}$, $Y_2$, $Z$    \\[2mm]
%\hline
$z_{21} z_{32} z_{43}$ & $x_2 (x_3-x_2)(1-x_3)$ & $L_2$   & $W_{123}$, $X_{23}$, $Y_3$, $Z$   \\[2mm]
%\hline
$z_{21} z_{31} z_{43}$ & $x_2 x_3 (1-x_3)$ & $L_3$  & $W_{123}$, $X_{31}$, $Y_3$, $Z$   \\[2mm]
%\hline
$z_{21} z_{32} z_{42}$ & $x_2 (1-x_2) (x_3-x_2)$ & $L_4$  & $X_{12}$, $X_{23}$, $Y_2$, $Z$  \\[2mm]
%\hline
$z_{21}^2 z_{43}$ & $x_{2}^2 (1-x_3)$ & $L_5$     & $W_{123}$, $X_{12}$, $Y_3$, $Z$   \\[2mm]
%\hline
$z_{31}^2 z_{42}$ & $x_{3}^2 (1-x_2)$ & $L_6$ & $W_{123}$, $X_{31}$, $Y_2$, $Z$  \\[2mm]
%\hline
$z_{31} z_{32} z_{42}$ & $(1-x_2) x_3 (x_3 -x_2)$ & $L_7$ & $W_{123}$, $X_{23}$, $Y_2$, $Z$   \\[2mm]
%\hline
$z_{31} z_{42} z_{43}$ & $(1-x_2) x_3 (1-x_3)$ & $L_1'$ & $X_{31}$, $Y_2$, $Y_3$, $Z$  \\[2mm]
%\hline
$z_{21} z_{42} z_{43}$ & $x_2 (1-x_2) (1-x_3)$ & $L_3'$  & $X_{12}$, $Y_2$, $Y_3$, $Z$  \\[2mm]
%\hline
$z_{31} z_{32} z_{43}$ & $x_3 (1-x_3) (x_3-x_2)$ & $L_4'$ & $X_{23}$, $X_{31}$, $Y_3$, $Z$   \\[2mm]
%\hline \hline
\end{tabular}
\caption{The type $L$ denominators and their origins in the $five$-vector amplitude.}
\label{tab:L}
\end{table}
\begin{table}[t!]
\centering
\begin{tabular}{c c c c}
denominator & with $z_1 =0$, $z_4 = 1$ &  name   & location  \\[2mm]
%\hline
%\hline
$z_{21} z_{31} z_{41}$ & $x_2 x_3$ & $K_1$  & $X_{12}$, $X_{31}$, $Y_1$, $Z$   \\[2mm]
%\hline
$z_{21} z_{41} z_{43}$ & $x_2 (1-x_3)$ & $K_2$  &$X_{12}$ $Y_1$, $Y_3$, $Z$  \\[2mm]
%\hline
$z_{31} z_{41} z_{42}$ & $(1-x_2) x_3$ & $K_3$ & $X_{31}$, $Y_1$, $Y_2$, $Z$  \\[2mm]
%\hline
$z_{21} z_{32} z_{41}$ & $x_2 (x_3-x_2)$ & $K_4$   & $W_{123}$, $X_{12}$, $Y_1$, $Z$   \\[2mm]
%\hline
$z_{31} z_{32} z_{41}$ & $x_{3} (x_3-x_2)$ & $K_5$  & $W_{123}$, $X_{31}$, $Y_1$, $Z$   \\[2mm]
%\hline
$z_{32}^2 z_{41}$ & $(x_3-x_2)^2$ & $K_6$    & $W_{123}$, $X_{23}$, $Y_1$, $Z$   \\[2mm]
%\hline
$z_{41} z_{42} z_{43}$ & $(1-x_2) (1-x_3)$ & $K_1'$  & $Y_1$, $Y_2$, $Y_3$, $Z$  \\[2mm]
%\hline
$z_{32} z_{41} z_{43}$ & $(1-x_3) (x_3-x_2)$ & $K_4'$ & $X_{23}$, $Y_1$, $Y_3$, $Z$  \\[2mm]
%\hline
$z_{32} z_{41} z_{42}$ & $(1-x_2) (x_3-x_2)$ & $K_{5}'$ & $X_{23}$, $Y_1$, $Y_2$, $Z$  \\[2mm]
\end{tabular}
\caption{The type $K$ denominators and their origins in the five-vector amplitude.}
\label{tab:K}
\end{table}
\begin{table}[t!]
\centering
\begin{tabular}{c c c c}
denominator & with $x_1 =0$, $x_4 = 1$ & name   & location \\[2mm]
$z_{21} z_{31} x_{32}$ & $x_{2} x_{3} (x_3 - x_2)$ & $M_1$ & $W_{123}$, $Z$ \\[2mm]
%\hline
$z_{21}^2 z_{31}$ & $x_{2}^2 x_3$ & $M_2$ & $X_{12}$, $Z$ \\[2mm]
%\hline
$z_{21}^2 z_{32}$ & $x_{2}^2 (x_3-x_2)$ & $M_3$ & $X_{12}$, $Z$ \\[2mm]
%\hline
$z_{21} z_{32}^2$ & $x_2 (x_3-x_2)^2$ & $M_4$ &  $X_{23}$, $Z$ \\[2mm]
%\hline
$z_{31} z_{32}^2$ & $x_3 (x_3-x_2)^2$ & $M_5$ & $X_{23}$, $Z$ \\[2mm]
%\hline
$z_{21} z_{31}^2$ & $x_2 x_{3}^2$ & $M_6$ & $X_{31}$, $Z$ \\[2mm]
%\hline
$z_{31}^2 z_{32}$ & $x_{3}^2 (x_3-x_2)$ & $M_7$ & $X_{31}$, $Z$ \\[2mm]
%\hline \hline
\end{tabular}
\caption{Type $M$ denominators and their origins in the five-vector amplitude.}
\label{tab:D}
\end{table}

The remaining integrals are then reduced according to~\cite{Barreiro:2005hv}.  Integration by parts yields
\begin{align}
q_{43} K_2 - q_{31} K_1 - q_{32} K_4 & = 0~,\nonumber\\
q_{42} K_3 - q_{21} K_1 + q_{32} K_5 & = 0~, \nonumber\\
(q_{32}-1) K_6 - q_{43} K'_4 + q_{31} K_5 & = 0~, \nonumber\\
q_{21} K_2 - q_{42} K'_1 -q_{32} K'_4 & = 0~, \nonumber\\
q_{31} K_3 -q_{43} K'_1 + q_{32} K'_5 & = 0~.
\end{align}
while from partial fractions we learn that
\begin{align}
K_1-K_4+K_5 & = 0~, &
K'_1-K_4'+K'_5 & = 0~.
\end{align}
These $7$ relations allow us to express the nine $K$ integrals in terms of just two independent ones, which we take to be $K_3$, and the linear combination ingeniously  identified in~\cite{Barreiro:2005hv}:
\begin{align}
T = q_{21} q_{43} K_2 + (q_{51}q_{21}-q_{21} q_{43} + q_{43} q_{54}) K_3~.
\end{align}
The cyclic invariance of $T$ is highly non-trivial and requires the integral relations.  There is a good reason that the relations for the $K$ integrals close without involving any of the $L$ integrals:  the two-fermion, three-vector amplitude only involves the $K$ integrals.

\subsection{Reduction of the $K$ integrals to $K_3$ and $T$}
Here we give the expressions for all the integrals in terms of our basis $K_3$ and $T$ with a composition according to the pole structure.  We use
\begin{align}
q_{15} &= q_{23} + q_{24} + q_{34}~,&
q_{45} &=q_{12} + q_{13} + q_{23}~.
\end{align}
\begin{align}
\label{eq:reducedK}
K_1
%& = \frac{ -(q_{12}q_{23}+q_{12}q_{34}+q_{13}q_{34}+q_{23}q_{34}) K_3 + \cT}{q_{12} (q_{12}+q_{13}+q_{23})}~\nonumber\\
& = - \left[\frac{q_{23}}{q_{45}}  + \frac{q_{34}}{q_{12}}\right] K_3  + \frac{ T}{q_{12} q_{45}}~,
\nonumber\\[2mm]
K_2
%& =\frac{ -(q_{12}q_{23}+q_{12}q_{24}+q_{12}q_{34}+q_{13}q_{34}+q_{23}q_{34}) K_3 + \cT}{q_{12} q_{34}} \nonumber\\
&= -\left[\frac{ q_{23} + q_{24}}{q_{34}} + \frac{q_{13} + q_{23}}{q_{12}} +1 \right]K_3 + \frac{T}{q_{12} q_{34}}~,\nonumber\\[2mm]
K_4
&= -\left[ \frac{q_{12}+q_{23}}{q_{45}}  + \frac{q_{34}}{q_{12}} +\frac{q_{24}+q_{34}}{q_{23}}\right] K_3 + \frac{1}{q_{45}}\left[ \frac{1}{q_{23}} + \frac{1}{q_{12}} \right]T ~,\nonumber\\
K_5
& = -\left[\frac{q_{24}+q_{34}}{q_{23}} + \frac{q_{12}}{q_{45}} \right] K_3 + \frac{1}{q_{23} q_{45}} T~,
\nonumber\\[2mm]
K'_1
&= - \left[\frac{q_{12}}{q_{34}} + \frac{q_{23}}{q_{15}} \right] K_3 + \frac{1}{q_{34} q_{15}} T~,
\nonumber\\[2mm]
K'_4
&= -\left[ \frac{q_{23} + q_{34}}{q_{15}}+\frac{q_{12}}{q_{34}} + \frac{q_{12}+q_{13}}{q_{23}}  \right] K_3
+\frac{1}{q_{15}} \left[\frac{1}{q_{34}} + \frac{1}{q_{23}} \right] T
\nonumber\\[2mm]
K'_5
& = -\left[\frac{q_{34}}{q_{15}}+\frac{q_{12}+q_{13}}{q_{23}} \right] K_3 + \frac{1}{q_{23}q_{15}} T
\nonumber\\[2mm]
(q_{23}-1)K_6
& = -\left[ q_{13} K_5 - q_{34} K'_4\right]~ \nonumber\\
&= - \left[ q_{12} + q_{34} \left( \frac{q_{23}+q_{34}}{q_{15}} + \frac{q_{12}+q_{13}}{q_{23}} \right) -q_{13} \left(\frac{q_{24}+q_{34}}{q_{23}} + \frac{q_{12}}{q_{45}}\right) \right] K_3 \nonumber\\
&\qquad + \left[ \frac{1}{q_{15}} + \frac{1}{q_{23}} \left(\frac{q_{34}}{q_{15}} -\frac{q_{13}}{q_{45}}\right) \right] T~.
 \end{align}
Note that terms with double poles only show up as coefficients of $T$.

The remaining $L$ integrals can also be expressed in terms of $K_3$ and $T$.  From partial fractions we obtain
\begin{align}\label{LtoK}
L_1 & = K_1+K_3~,&
L'_1 &= K'_1 +K_3~,\nonumber\\
L_3 & = K_1+K_2~,&
L'_3 &=K'_1+K_2~,\nonumber\\
L_4 &=K_4+K'_5~,&
L'_4 &=K'_4+K_5~,\nonumber\\
L_2 &= L_3+L'_4~,\nonumber\\
L_7 &= L_4 -L_1~,
\end{align}
while integration by parts yields
\begin{align}
(q_{21} -1) L_5 & = q_{42} L'_3 + q_{32} L_2~,&
(q_{31}-1) L_6 & = q_{43} L'_1- q_{32} L_7~.
\end{align}

\subsection{The small momentum expansion}
We now tackle the small momentum expansion of the integrals.   This is most easily done for integrals that are finite as $q_{ij} \to 0$, such as $K_3$.  Here, after setting $x_3 = y_1$ and $x_2 = y_1y_2$ we obtain
\begin{align}
K_3 = \int^1_0 \ed y_1 \int^1_0 \ed y_2~
y_1^{q_{45}} (1-y_1)^{q_{34}} y_2^{q_{12}} (1-y_2)^{q_{23}}(1-y_1y_2)^{q_{24}-1}~.
\end{align}
In what follows we will not explicitly specify the integration domain or the obvious  measure $\ed y_1 \ed y_2$.
Expanding in the $q_{ij}$ leads to the form claimed in~(\ref{eq:K3expand}) :
\begin{align}
K_3 = C_0 + (q_{45} + q_{12}) C_1 + (q_{34}+q_{23})C_2 + q_{24} C_3 + O(q^2)~,
\end{align}
with
\begin{align}
C_0 & = \int  \sum_{k=0}^\infty (y_1y_2)^k = \frac{\pi^2}{6}~,\nonumber\\
C_1 & = \int  \frac{ \log y_1}{1-y_1y_2} = -\zeta(3)~,\nonumber\\
C_2 & = \int \frac{ \log (1-y_1)}{1-y_1y_2} = -2 \zeta(3)~,\nonumber\\
C_3 & = \int  \frac{ \log (1-y_1y_2)}{1-y_1y_2} = -\zeta(3)~.
\end{align}
All of these integrals are evaluated by expanding the logarithms and integrating term by term.  We give an example below.

The $K_2$ integral presents a little more of a challenge because it is singular in the $q_{ij} \to 0$ limit.  However, a close examination of the integration region shows that the poles arise from different regions, and thus we can easily subtract them off to find a finite remainder.  We begin with
\begin{align}
K_{2} &=
\int^1_0 \ed y_1 \int^1_0 \ed y_2~
y_1^{q_{45}} (1-y_1)^{q_{34}-1} y_2^{q_{12}-1} (1-y_2)^{q_{23}}(1-y_1y_2)^{q_{24}}~.
\end{align}
With a little effort, we see that we can write this as
\begin{align}
K_2  & = \frac{1}{q_{12}q_{34}} + K'_2~,\nonumber\\
K'_2 & = \int (1-y_1)^{q_{34}-1} y_2^{q_{12}-1}  \left[ y_1^{q_{45}}  (1-y_2)^{q_{23}}(1-y_1y_2)^{q_{24}}-1\right]~,\nonumber\\
K'_2 & =
\frac{1}{q_{34}}\left[ B(q_{12},1+q_{23}+q_{24}) -\frac{1}{q_{12}} \right]
+\frac{1}{q_{12}}\left[  B(q_{34},1+q_{45}) -\frac{1}{q_{34}} \right] + K''_2~,\nonumber\\
K''_2 & = \int (1-y_1)^{q_{34}-1} y_2^{q_{12}-1} \left[ y_1^{q_{45}} ( (1-y_2)^{q_{23}} (1-y_1y_2)^{q_{24}} -1) - (1-y_2)^{q_{23}+q_{24}} +1 \right]~.
\end{align}

The remaining non-trivial integral $K''_2$ is regular as $q_{ij} \to 0$ and has expansion
\begin{align}
K''_2 & = q_{24} C_4 +O(q^2)~, &
C_4 & = \int \frac{\log(1-y_1y_2) - \log(1-y_2)}{(1-y_1) y_2}~.
\end{align}
We evaluate $C_4$ is by power-series expansion:
\begin{align}
C_4 = \sum_{k=1}^{\infty} \frac{1}{k} \int \frac{ (1-y_1^k)y_2^{k-1}}{1-y_1} =
\sum_{k=1}^{\infty} \frac{1}{k^2} \sum_{l=1}^{\infty} \left[ \frac{1}{l} - \frac{1}{l+k}\right]
=\sum_{k,l=1}^\infty \frac{1}{kl(k+l)} = 2\zeta(3)~.
\end{align}
The last equality can be obtained as follows.  Observe that the sum can be written as
%Let $\Psi(x) = \Gamma'(x)/ \Gamma(x)$ be the digamma function.  Observe that
%\begin{align}
%\sum_{k=1}^\infty \frac{ \Psi(k+1) -\Psi(1)}{k^2} = \sum_{k=1}^\infty \frac{1}{k^2} \sum_{n=1}^\infty \frac{k}{n(n+k)} =
%\sum_{k=1}^\infty \sum_{n=1}^\infty \frac{1}{nk(n+k)}~.
%\end{align}
\begin{align}
\int^1_0 \ed x \sum_{l=1}^{\infty} \sum_{k=1}^\infty \frac{x^{l+k-1}}{lk} &= \int^1_0 \ed x \frac{\log(1-x)^2}{x} = 2 \int^1_0 \ed x \frac{\log(1-x)\log(x)}{x} \nonumber\\
 & = -2 \sum_{k=1}^\infty \frac{1}{k} \int^1_0 \ed x ~ x^{k-1} \log(x) = 2\sum_{k=1}^\infty \frac{1}{k^3} = 2\zeta(3)~.
\end{align}
Using this in $T$ we find the form claimed in~(\ref{eq:Texpand}).

The integrals $K_{2}$ and $K_3$ also have closed-form expressions in terms of Gamma functions and ${}_3F_2$ hypergeometric functions---and hence, via \eqref{eq:reducedK} and \eqref{LtoK}, so do all of the other $K$'s and $L$'s.  We have not found these expressions useful in this work, but for completeness we give them here:
\begin{align}
K_3 &= \frac{\Gamma(1+q_{23})\Gamma(1+q_{34})\Gamma(1+q_{15})\Gamma(1+q_{45})}{\Gamma(2+q_{23}+q_{45})\Gamma(2+q_{15}+q_{45})}\ \vphantom{F}_3F_2\left[\left.\begin{matrix}1+q_{23},\ \ 1-q_{35},\ \ 1+q_{15} \\ 2+q_{23}+q_{34},\ \ 2+q_{15}+q_{45}\end{matrix}\right| 1\right],\\
K_2 &= \frac{\Gamma(1+q_{23})\Gamma(q_{34})\Gamma(1+q_{15})\Gamma(1+q_{45})}{\Gamma(1+q_{23}+q_{45})\Gamma(2+q_{15}+q_{45})}\ \vphantom{F}_3F_2\left[\left.\begin{matrix}1+q_{23},\ \ 1-q_{35},\ \ 1+q_{15} \\ 1+q_{23}+q_{34},\ \ 2+q_{15}+q_{45}\end{matrix}\right| 1\right].
\end{align}

\bibliographystyle{./utphys}
\bibliography{./bigref}

\end{document}

%% file: basedefs2.tex
\DeclareFontFamily{U}{rsf}{}
\DeclareFontShape{U}{rsf}{m}{n}{
  <5> <6> rsfs5 <7> <8> <9> rsfs7 <10-> rsfs10}{}
\DeclareMathAlphabet\Scr{U}{rsf}{m}{n}

\def\CO#1#2{{[#1,#2]}}
\def\AC#1#2{{\{#1,#2\}}}

\def\iden{{\mathbbm 1}}

\def\rep#1{{{\boldsymbol{#1}}}}

\def\R{{\mathbb R}}
\def\Z{{\mathbb Z}}

\def\Sym{\operatorname{Sym}}
\def\Tr{\operatorname{Tr}}

\def\tr{\operatorname{tr}}

\def\PSL{\operatorname{PSL}}

\def\so{\operatorname{\mathfrak{so}}}
\def\Lsl{\operatorname{\mathfrak{sl}}}

\def\spin{\operatorname{\mathfrak{spin}}}

\def\Lg{\operatorname{\mathfrak{g}}}

\def\p{\partial}

\def\slash{\!\!\!/}

\def\la{\langle}
\def\ra{\rangle}

\def\ff#1#2{{\textstyle\frac{#1}{#2}}}
\def\half{\frac{1}{2}}

\def\cC{{\cal C}}

\def\cM{{\cal M}}

\def\cT{{\cal T}}
\def\cU{{\cal U}}
\def\cV{{\cal V}}

\newcommand\zetab{\overline{\zeta}}

\newcommand\psit{\widetilde{\psi}}

\newcommand\vphi{\varphi}

\newcommand\vphit{\widetilde{\vphi}}

\newcommand\Thetat{\widetilde{\Theta}}

\newcommand\zb{\overline{z}}

\newcommand\ct{\widetilde{c}}